\documentclass[12pt]{article}
\usepackage{lscape}
\usepackage{graphicx}
\usepackage{bm}
\usepackage{amsmath}
\usepackage{amssymb}
\usepackage{amsfonts}
\usepackage{mathtools}
\usepackage{hyperref}
\usepackage{braket}
\usepackage{slashed}
\usepackage{mathrsfs} 
\usepackage{anyfontsize}
\usepackage[normalem]{ulem}
\usepackage{wrapfig}
\usepackage{comment}
\usepackage[left=1in,right=1in,top=1in,bottom=1in]{geometry}

\usepackage[ascii]{inputenc}
\usepackage{bbm,braket,microtype,mathtools}
\usepackage[T1]{fontenc}
\usepackage{lmodern}
\usepackage[USenglish]{babel}

\hypersetup{colorlinks=true,urlcolor=[rgb]{0,0,0.5},citecolor=[rgb]{0.5,0,0},linkcolor=[rgb]{0,0,0.4}}

\renewcommand{\title}[1]{\vbox{\center\bf{\Large{#1}}}\vspace{5mm}}
\renewcommand{\author}[1]{\vbox{\center#1}\vspace{5mm}}
\newcommand{\address}[1]{\vbox{\center\em#1}}

\newcommand\emails[1]{\begingroup
	\renewcommand\thefootnote{}\footnote{#1}
	\addtocounter{footnote}{-1}\endgroup}

\newcommand{\be}{\begin{equation}}
\newcommand{\bea}{\begin{eqnarray}}
\newcommand{\eea}{\end{eqnarray}}
\newcommand{\ba}{\begin{array}}
\newcommand{\ea}{\end{array}}
\newcommand{\ee}{\end{equation}}
\newcommand{\bes}{\begin{equation*}}
\newcommand{\beas}{\begin{eqnarray*}}
\newcommand{\eeas}{\end{eqnarray*}}
\newcommand{\bas}{\begin{array*}}
\newcommand{\eas}{\end{array*}}
\newcommand{\ees}{\end{equation*}}

\def\1den{\hbox{$1\hskip -1.2pt\vrule depth 0pt height 1.53ex width 0.7pt
                  \vrule depth 0pt height 0.3pt width 0.12em$}}

\def\and{\quad {\rm and} \quad}

\begin{document}

\begin{titlepage}
\begin{flushright}
IPM/P-2018/027\\
\end{flushright}

\begin{center}
\vspace*{2cm}
\title{{\fontsize{0.73cm}{0.73cm}\selectfont Renormalization Group Circuits for \\ \vspace*{2mm} Weakly Interacting Continuum
Field Theories}}
\author{Jordan Cotler${}^a$, M. Reza Mohammadi Mozaffar,${}^b$ Ali Mollabashi,${}^b$ and Ali Naseh{${}^c$}
}
\address{{\fontsize{0.4cm}{.4cm}\selectfont
${}^a$ Stanford Institute for Theoretical Physics,\\ Stanford University, Stanford, California 94305, USA
\\
\vspace*{2mm}
${}^b$ School of Physics, ${}^c$ School of Particles and Accelerators
\\Institute for Research in Fundamental Sciences (IPM), \\ P.O. Box 19395-5531, Tehran, Iran
\\
}}

\emails{ \hspace*{-8mm}
\href{mailto:jcotler@stanford.edu}{\tt jcotler@stanford.edu},
\href{mailto:m_mohammadi@ipm.ir}{\tt m$\_$mohammadi@ipm.ir},
\href{mailto:mollabashi@ipm.ir}{\tt mollabashi@ipm.ir},
\href{mailto:naseh@ipm.ir}{\tt naseh@ipm.ir}}

\end{center}

\begin{abstract}
We develop techniques to systematically construct local unitaries which map scale-invariant, product state wavefunctionals to the ground states of weakly interacting, continuum quantum field theories.  More broadly, we devise a ``quantum circuit perturbation theory'' to construct local unitaries which map between any pair of wavefunctionals which are each Gaussian with arbitrary perturbative corrections.  Further, we generalize cMERA to interacting continuum field theories, which requires reworking the existing formalism which is tailored to non-interacting examples.  Our methods enable the systematic perturbative calculation of cMERA circuits for weakly interacting theories, and as a demonstration we compute the 1-loop cMERA circuit for scalar $\varphi^4$ theory and analyze its properties.  In this case, we show that Wilsonian renormalization of the spatial momentum modes is equivalent to a local position space cMERA circuit.  This example provides new insights into the connection between position space and momentum space renormalization group methods in quantum field theory.  The form of cMERA circuits derived from perturbation theory suggests useful ansatzes for numerical variational calculations.

\end{abstract}

\end{titlepage}

\tableofcontents
\newpage

\section{Introduction}

Quantum field theory is traditionally treated in the Heisenberg or interaction pictures which focus on the evolution of operators and relegate the structure of quantum states to an often implicit, supporting role. Although Schr\"{o}dinger picture and wavefunctional methods for QFT's have been developed, they have historically been a niche subject since for many applications they do not offer advantage over more conventional and streamlined approaches. In recent years, the incorporation of quantum information theoretic techniques into high energy and condensed matter physics has punctuated the importance of studying the entanglement structure of quantum states.

In particular, ground states of interacting theories encode a pattern of entanglement between subsystems which is structurally dependent on the characteristic length scales of those subsystems (for an overview, see \cite{EEQFT1}--\cite{EEQFT5}).  In other words, entanglement looks different between subsystems of different sizes.  This structure is intimately connected with the framework of the renormalization group (RG) -- a broad collection of technical and conceptual tools that allows one to study the effective physics of a system which can only be probed at distance scales larger than a specified small distance scale.  The new insight afforded by studying entanglement is that ground states of interacting systems have \textit{structurally similar patterns of scale-dependent entanglement}.  This implies that there are structural commonalities between their renormalization group flows, or more colloquially, the procedure by which one coarse-grains the description of a quantum state to study an effective description at larger distance scales.

This insight has been used to construct tensor networks (i.e., a representation of quantum circuits), which efficiently implement RG flows \cite{ER1}.  These ``entanglement renormalization'' tensor networks are comprised of sequences of \textit{spatially local} quantum gates.  Concretely, suppose we have a quantum state which captures correlations above a distance scale $\ell$, and that we want to perform RG to create a new state which only captures the correlations of the original state above a larger distance scale $b \cdot \ell$, where $b > 1$.  Then an entanglement renormalization tensor network transforms the original quantum state by \textit{disentangling} all subsystems with distance scales between $\ell$ and $b \cdot \ell$, so that only entanglement and hence correlations above scale $b \cdot \ell$ remain.

For many states, we can efficiently continue this procedure and disentangle subsystems of increasing size until the resulting final ``IR'' state is separable with respect to all spatial subsystems.\footnote{For some entanglement renormalization schemes, one can obtain a separable ``IR'' state even from ground states of UV theories with interacting IR fixed points (in the sense of Wilsonian RG).  Thus, not all entanglement renormalization schemes correspond to Wilsonian RG schemes for all theories -- we will have more to say about this later.  For the moment, we remark in passing that massive theories have their masses diverge in the IR in Wilsonian RG, causing their ground states to progressively become product states as they flow into the IR (in the sense of Wilsonian RG).}  But then running the tensor network in reverse, we can start with a completely disentangled state and entangle it at successively smaller distance scales to build up the correct short-distance correlations of the ``UV'' state.  Hence, by finding tensor network architectures that perform RG flow on ground states of interacting theories, we can parametrize these architectures and run them in reverse on simple IR states to yield a parametric family of UV states which capture viable short-distance correlations of ground states of interacting theories.  More broadly, if one has knowledge of the entanglement structure of the UV state,
one can construct a parametric family of tensor networks which generate a family of UV states that have the desired entanglement structure.

Such parametric families of tensor networks have been utilized as variational ansatzes, and are tremendously successful in capturing correlations in ground states of interacting lattice systems in 1+1 dimensions.  Some parametric families of tensor networks like the Density Matrix Renormalization Group (DMRG) and its variants \cite{DMRG1, DMRG2, DMRG3}, as well as the Multiscale Entanglement Renormalization Ansatz (MERA) and its variants \cite{ER1, ER2, ER3}, are based on the RG intuition described above.  Other parametric families like Matrix Product States (MPS) and its variants \cite{MPS1, MPS2, MPS3} are based on direct knowledge of the entanglement structure of the UV state, such as area-law entanglement.\footnote{In one spatial dimension, ``area-law'' entanglement is a misnomer since it conventionally means that contiguous subsystems of length $L$ have von Neumann entropy that goes as $\log L$. However, in one spatial dimension the boundary of a contiguous region of length $L$ is simply its endpoints, which have $\mathcal{O}(L^0)$ area instead of $\mathcal{O}(\log L)$.}  There are various technical and computational obstructions to generalizing tensor networks to higher dimensional lattice systems, and crucially to continuum field theories.  For lattice systems, higher-dimensional tensor networks are hard to implement on computers since known methods require an almost intractably large number of tensor contractions, even for systems of modest size \cite{Contract1}.  This obstruction is referred to as the ``contraction problem.''

For continuum field theories in 1+1 dimensions, a continuum analog of the MPS tensor network architecture, called cMPS has been proposed \cite{Verstraete:2010ft}. cMPS is successful for studying non-relativistic continuum field theories in 1+1 dimensions, but its higher-dimensional generalizations suffer from the contraction problem mentioned above.  Furthermore, it is difficult to generalize cMPS to generic \textit{relativistic} field theories in 1+1 dimensions, since there are difficulties in treating both long-distance (IR) and short-distance (UV) divergences that arise \cite{cMPSrel1, cMPSrel2}.  Some progress has been made on 1+1 relativistic \textit{fermionic} theories by leveraging features of the Dirac sea \cite{cMPSfermionic1}, although applying cMPS to bosonic theories or gauge theories seems to require new ideas.

A more promising tensor network architecture is the continuum analog of MERA, called cMERA \cite{cMERA1}. This ansatz comprises of a continuum quantum circuit which acts on a scale-invariant product state in the IR, and entangles it as a function of decreasing distance scale down into the short-distance UV regime.  In other words, it performs RG flow in reverse, in the manner previously described.  As a matter of principle, the cMERA ansatz has no difficulty treating continuum relativistic or non-relativistic theories in \textit{any} spacetime dimension.  Unfortunately, cMERA has previously only been understood in the context of free field theories, for which the ground states are exactly solvable using standard methods \cite{cMERA1, cMERA2, cMERA3, cMERA4}.

To apply cMERA to interacting continuum field theories, a first approach is to observe that for the free field cases which have been previously understood, the cMERA ansatz produces Gaussian wavefunctionals.  While this ``mean-field theory'' approximation is exact for free theories, one can still apply it as an ansatz to interacting theories using techniques developed several decades ago \cite{GEP1}--\cite{GEP6}.  This approach was carried out in \cite{CMM1}, and it was found that the Gaussian cMERA ansatz captures some features of the true ground state of the interacting system.  However, the Gaussian ansatz has trivial RG flow, and so the scale-dependent correlations it can capture are fundamentally limited.

In principle, the correct way to generalize cMERA for interacting theories is to consider circuits which produce \textit{non}-Gaussian wavefunctionals, since the ground states of interacting theories are generically non-Gaussian.  But this requires working with cMERA unitaries that have non-quadratic Hermitian generators. In other words, we have unitaries of the form $U = e^{i \, Q}$ where $Q$ is a non-quadratic Hermitian operator.  For example, if we consider a scalar field with $\widehat{\phi}(\vec{x})$ and $\widehat{\pi}(\vec{x})$ operators, we might have
\begin{align}
\label{generatorExpand1}
Q = c_0 \, \textbf{1} &+ \int d^d \vec{x} \, \left(c_{1}^{(0)}(\vec{x}) \, \widehat{\phi}(\vec{x}) + c_{1}^{(1)}(\vec{x}) \, \widehat{\pi}(\vec{x}) \right) \nonumber \\ \nonumber \\
&+ \int d^d \vec{x}_1 \, d^d \vec{x}_2 \, \big(c_{2}^{(0)}(\vec{x}_1, \vec{x}_2) \,\widehat{\phi}(\vec{x}_1) \widehat{\phi}(\vec{x}_2) +  c_{2}^{(1)}(\vec{x}_1, \vec{x}_2) \, \widehat{\phi}(\vec{x}_1) \widehat{\pi}(\vec{x}_2) \nonumber \\
& \qquad \qquad \qquad \quad + c_{2}^{(1)}(\vec{x}_1, \vec{x}_2) \,\widehat{\pi}(\vec{x}_2) \widehat{\phi}(\vec{x}_1) + c_{2}^{(2)}(\vec{x}_1, \vec{x}_2) \, \widehat{\pi}(\vec{x}_1) \widehat{\pi}(\vec{x}_2) \big) \nonumber \\ \nonumber \\
&+ \int d^d \vec{x}_1 \, d^d \vec{x}_2 \, d^d \vec{x}_3 \, \big(c_{3}^{(0)}(\vec{x}_1, \vec{x}_2, \vec{x}_3) \,\widehat{\phi}(\vec{x}_1) \widehat{\phi}(\vec{x}_2) \widehat{\phi}(\vec{x}_3) +  c_{3}^{(1)}(\vec{x}_1, \vec{x}_2,\vec{x}_3) \, \widehat{\phi}(\vec{x}_1) \widehat{\phi}(\vec{x}_2) \widehat{\pi}(\vec{x}_3) \nonumber \\
& \qquad \qquad \qquad \qquad \quad + c_{3}^{(1)}(\vec{x}_1, \vec{x}_2,\vec{x}_3) \,\widehat{\pi}(\vec{x}_3) \widehat{\phi}(\vec{x}_2) \widehat{\phi}(\vec{x}_1) + c_{3}^{(2)}(\vec{x}_1, \vec{x}_2,\vec{x}_3) \, \widehat{\phi}(\vec{x}_1) \widehat{\pi}(\vec{x}_2) \widehat{\pi}(\vec{x}_3) \nonumber \\
& \qquad \qquad \qquad \qquad \quad + c_{3}^{(2)}(\vec{x}_1, \vec{x}_2,\vec{x}_3) \, \widehat{\pi}(\vec{x}_3) \widehat{\pi}(\vec{x}_2) \widehat{\phi}(\vec{x}_1)  + c_{3}^{(3)}(\vec{x}_1, \vec{x}_2, \vec{x}_3) \, \widehat{\pi}(\vec{x}_1) \widehat{\pi}(\vec{x}_2) \widehat{\pi}(\vec{x}_3) \big) \nonumber \\ \nonumber \\
&+ \cdots
\end{align}
where all the $c_n^{(k)}$ are real-valued functions or distributions.  Eqn.~\eqref{generatorExpand1} is a completely general expansion of a Hermitian operator in scalar field theory, and was mentioned in \cite{CMM1} in terms of creation and annihilation operators.
 While a nice class of unitaries with quadratic Hermitian generators called squeezing operators and displacement operators (i.e., in the context of squeezed coherent states) is well-understood in field theory, unitaries with Hermitian generators higher than quadratic order as in Eqn.~\eqref{generatorExpand1} are less studied.
 
In this paper, we study unitaries with Hermitian generators higher than quadratic order for continuum field theories.  By analyzing the Lie algebraic structure of higher-order (i.e., higher than quadratic) Hermitian generators, we identify:
\begin{enumerate}
\item Why unitaries with quadratic generators are special;
\item How to systematically treat unitaries with higher-order generators perturbatively in the higher order generators (i.e., ``quantum circuit perturbation theory''); and
\item Special circumstances required for us to treat unitaries with higher-order generators non-perturbatively in the higher order generators.
\end{enumerate}
By developing quantum circuit perturbation theory, we can systematically construct unitaries which map scale-invariant product states to the ground states of weakly interacting quantum field theory, to any fixed order in perturbation theory.  We can also construct unitaries which map between any pair of states that are each Gaussian with arbitrary perturbative corrections.  In very special cases, unitaries between non-Gaussian states (or between a Gaussian state and a non-Gaussian state) can be constructed non-perturbatively, although we will not discuss such cases here.

Next, we generalize cMERA to interacting fields, which involves generalizing our methods above to unitaries created by path-ordered exponentials of Hermitian generators.  We discover new features of cMERA which are absent for the free field examples which have previously been studied.  Our techniques enable systematic perturbative calculations of cMERA circuits for interacting field theories, and we calculate the 1-loop cMERA circuit for $\varphi^4$ theory as an example.
Furthermore, the disentangler is constructed to disentangle spatial momentum modes in a manner \textit{exactly} corresponding to perturbative Wilsonian RG (on spatial momentum modes), and yet in position space disentangles \textit{spatially local} subsystems.  This relationship between disentangling degrees of freedom in momentum space and position space draws a sharp connection between momentum space and position space RG methods in QFT.

Our perturbative calculations also suggest useful ansatzes for the numerical non-perturbative variational calculations.  Although we focus on scalar field theories in this paper, our techniques can be adapted to fermionic theories and gauge theories.  Our results are also summarized in a companion paper \cite{shortversion}. 

The paper is organized as follows:
\begin{itemize}
\item In Section 2, we develop ``quantum circuit perturbation theory'' for unitary circuits with higher-order Hermitian generators in field theories.  We begin with an analysis of the Lie algebraic structure of the higher-order Hermitian generators, and then develop perturbation theory in the higher order generators.  As a simple case, we review how to construct unitaries between Gaussian states, followed by the more interesting case of a unitary which maps between a Gaussian state and the ground state of $\varphi^4$ theory to first order in perturbation theory.  We then outline more general perturbative examples.
\item In Section 3, we consider cMERA for interacting fields.  After giving an overview of the cMERA formalism with several modifications pertinent to the interacting case, we revisit the cMERA for a free scalar field from our new perspective.  Next, we perform a perturbative calculation of the 1-loop cMERA circuit for $\varphi^4$ theory.  We discuss how the disentangler acts perturbatively as a dilatation operator, point out novel features of the 1-loop circuit, and show how the form of the circuit elucidates the connection between momentum space and position space RG.  Finally, we use our perturbative analysis to suggest new numerical ansatzes for variational calculations.
\item In Section 4, we summarize the paper and discuss future directions.
\item In the Appendix, we collect useful derivations, formulas, and mathematical proofs.
\end{itemize}

\newpage

\section{Circuits for wavefunctionals}\label{Sec:2}

Here we develop tools for analyzing quantum circuits acting on infinite-dimensional Hilbert spaces.  The setting of infinite-dimensional Hilbert space includes as a special case the standard quantum mechanics of point particles (i.e., a fixed number of particles in $\mathbb{R}^n$), which we conventionally call $(0+1)$--dimensional field theory.
We will begin with $(0+1)$--dimensional field theory because our methods are particularly transparent in this context.  However, we quickly generalize to $(d+1)$--dimensional field theory, which is our main interest.  We will focus on scalar field theory in particular, although our approach generalizes naturally to fermionic theories and gauge theories.

Before proceeding, it is useful to clarify what we mean by a ``quantum circuit.''  From our perspective, a quantum circuit is a unitary $U$ or sequence of unitaries $U_1, U_2,...,U_n$ which acts on a Hilbert space $\mathcal{H}$.  When we have a sequence of unitaries in mind, we imagine the unitaries act on the Hilbert space in order multiplicatively as $U_n \cdots U_2 U_1$.  Sometimes we may want that each $U_i$ to have a restricted form, for example $U_i = e^{i Q_i}$ where $Q_i$ is a finite sum of \textit{local} Hermitian operators.  Physically, this means we can think of each $U_i$ as implementing \textit{local} Hamiltonian evolution (i.e., $e^{i Q_i} = e^{-i t H}$ for some $t$ and local $H$).  Even when each $U_i = e^{Q_i}$ has a local Hermitian generator $Q_i$, it may not be the case that $U_n \cdots U_2 U_1$ itself can be written as $e^{i Q}$ where $Q$ is a finite sum of local Hermitian operators.

Quantum circuits are central objects in quantum algorithms.  In particular, a quantum circuit $U$ transforms an input state $|\psi_i\rangle$ into a final state $|\psi_f\rangle = U |\psi_i\rangle$ which is the output of the quantum algorithm.  Hence we can think of a unitary as a quantum circuit which \textit{performs computation on quantum states}.  This perspective is conceptually useful, since it evokes the idea that quantum computation operates by rearranging correlations and entanglement in quantum states.

\subsection{Algebra of operators and continuum circuits}
\subsubsection{Standard quantum mechanics in infinite--dimensional Hilbert space}
Let us begin with a $(0+1)$--dimensional quantum field theory, in particular a single quantum particle in one spatial dimension.  We denote the Hilbert space by $\mathcal{H}$, and we will utilize $\widehat{x}$ and $\widehat{p}$ operators which satisfy $[\widehat{x},\widehat{p}] = i$, where we have set $\hbar = 1$.  Now we ask: what is the form of the most general unitary $U$ acting on $\mathcal{H}$?  Since we can always write $U = e^{i Q}$ where $Q$ is Hermitian, we can instead ask: what is the most general Hermitian operator $Q$ acting on $\mathcal{H}$?

This question is tricky because Hermitian operators $Q$ are infinite-dimensional.  However, we can still construct a basis of Hermitian operators on $\mathcal{H}$.  A convenient choice is the Bender-Dunne basis \cite{BD1, BD2},\footnote{For more formal properties of the Bender-Dunne basis, see \cite{FormalBD1, FormalBD2}.} given by the set of Hermitian operators $\{T_{m,n}\}_{n,m=-\infty}^\infty$ where
\begin{equation}
\label{BD1}
T_{m,n} := \frac{1}{2^n} \sum_{k=0}^\infty \frac{\Gamma(n+1)}{k! \, \Gamma(n-k+1)} \, \widehat{x}^k \widehat{p}^m \widehat{x}^{n-k} = \frac{1}{2^m} \sum_{j=0}^\infty \frac{\Gamma(m+1)}{j! \, \Gamma(m-j+1)}\,\widehat{p}^j \widehat{x}^n \widehat{p}^{m-j}
\end{equation}  
where the operators are presented in Weyl-ordered form.  For example, we have:
\begin{align}
& T_{0,0} = 1\,,\qquad T_{0,1} = \widehat{x}\,, \qquad T_{1,0} = \widehat{p}\,, \qquad T_{1,1} = \frac{1}{2}(\widehat{x}\widehat{p} + \widehat{p}\widehat{x})\,,\qquad T_{0,2} = \widehat{x}^2\,, \qquad T_{2,0} = \widehat{p}^2\,, \nonumber \\ \nonumber \\
&T_{1,2} = \frac{1}{3}(\widehat{p}\widehat{x}^2 + \widehat{x}\widehat{p}\widehat{x} + \widehat{x}^2 \widehat{p})\,, \qquad T_{2,2} = \frac{1}{6}\left(\widehat{p}^2 \widehat{x}^2 + \widehat{x}^2 \widehat{p}^2 + \widehat{p}\widehat{x}\widehat{p}\widehat{x} + \widehat{x}\widehat{p}\widehat{x}\widehat{p} + \widehat{p} \widehat{x}^2 \widehat{p} + \widehat{x} \widehat{p}^2 \widehat{x} \right)\,,\nonumber \\ \nonumber \\
& T_{-1,1} = \frac{1}{2}\left(\frac{1}{\widehat{p}} \, \widehat{x} + \widehat{x} \, \frac{1}{\widehat{p}} \right)\,, \qquad T_{-2,1} = \frac{1}{2}\left(\frac{1}{\widehat{p}^2} \, \widehat{x} + \widehat{x} \, \frac{1}{\widehat{p}^2} \right)\,,\nonumber \\ \nonumber \\
& T_{-2,4} = \frac{1}{16}\left(\frac{1}{\widehat{p}^2} \, \widehat{x}^4 + 4 \, \widehat{x} \, \frac{1}{\widehat{p}^2}\,\widehat{x}^3 + 6 \, \widehat{x}^2 \, \frac{1}{\widehat{p}^2}\, \widehat{x}^2 + 4 \, \widehat{x}^3 \, \frac{1}{\widehat{p}^2}\,\widehat{x} + \widehat{x}^4 \, \frac{1}{\widehat{p}^2} \right)\,.\nonumber
\end{align}
The operators $\{T_{m,n}\}_{m,n=-\infty}^\infty$ form an algebra \cite{BD1, BD2}.  For our purposes, we would like to restrict to operators that are non-singular,\footnote{We do not in principle need to make such a restriction, although for the purposes of this paper we do not gain utility in our later QFT examples by including the singular operators (e.g., in the QFT scalar field case, the singular operators would contain combinations of $1/\widehat{\phi}(\vec{x})$ and $1/\widehat{\pi}(\vec{x})$).  In short, the reason is that we do not consider any theories whose Hamiltonians have singular operators.  Such terms do not arise in the standard examples of relativistic or even non-relativistic quantum field theory.
In any case, the analysis in this paper can also be carried out without making the restriction $m,n \geq 0$, although we will not discuss this further here.} and so consider $\{T_{mn}\}_{m,n=0}^{\infty}$,\, i.e., $T_{m,n}$ for which $m,n \geq 0$.  In this case, Eqn.~\eqref{BD1} simplifies to
\begin{equation}
\label{BD2}
T_{m,n} := \frac{1}{2^n} \sum_{k=0}^n \binom{n}{k} \, \widehat{x}^k \widehat{p}^m \widehat{x}^{n-k} = \frac{1}{2^m} \sum_{j=0}^m \binom{m}{j}\,\widehat{p}^j \widehat{x}^n \widehat{p}^{m-j}\,.
\end{equation}  
This set $\{T_{mn}\}_{m,n=0}^{\infty}$ forms a subalgebra of $\{T_{m,n}\}_{n,m=-\infty}^\infty$ and so is closed.  So we will consider (non-singular) unitaries of the form
\begin{equation}
\label{unitary1}
U = \exp\left\{i \, \sum_{m,n = 0}^\infty  c_{m,n} \, T_{m,n}\right\}
\end{equation}
where the $c_{mn}$'s are real constants.  In our notation from before, we are saying that $Q = \sum_{m,n = 0}^\infty  c_{m,n} \, T_{m,n}$.

We would like to understand how unitaries $U$ of the form in Eqn.~\eqref{unitary1} transform a quantum state.  Conceptually, we would like to start with a simple state $|\psi\rangle$ (like the input state to a quantum algorithm) and output a more interesting state $U|\psi\rangle$ that we want to characterize.  To make this precise, we pose the following question: \\ \\
\textbf{Question 1:} Given an initial state $|\psi\rangle$ such that we can explicitly compute any correlation function $\langle \psi| T_{m,n} |\psi\rangle$ for $m,n \geq 0$, for which unitaries $U$ can we explicitly compute $\langle \psi|U^\dagger \, T_{m,n} \, U|\psi\rangle$ for $m,n \geq 0$\,? \\ \\
In other words, if we can explictly compute any finite-point (i.e., a finite number of $\widehat{x}$'s and $\widehat{p}$'s) correlation function of $|\psi\rangle$, then for which unitaries $U$ can we explicitly compute any finite-point correlation function of the transformed state $U|\psi\rangle$\,?

Leveraging the fact that $\{T_{m,n}\}_{m,n=0}^\infty$ is a closed algebra, we can write the correlation function $\langle \psi|U^\dagger \, T_{m,n} \, U|\psi\rangle$ as
\begin{align}
\langle \psi| \, e^{-i \sum_{p,q=0}^\infty c_{p,q} \, T_{p,q}} \, T_{m,n} \, e^{i \sum_{p,q=0}^\infty c_{p,q} \, T_{p,q}} \, |\psi\rangle &= \langle \psi|\,\sum_{r,s=0}^\infty \widetilde{c}_{r,s}^{\,m,n} \, T_{r,s} \, |\psi\rangle \nonumber \\
&= \sum_{r,s=0}^\infty \widetilde{c}_{r,s}^{\,m,n} \,\langle \psi|\, T_{r,s} \, |\psi\rangle
\end{align}
for some real constants $\widetilde{c}_{r,s}^{\,m,n}$.  So to answer Question 1, we would like to know for which unitaries $U$ we can explicitly compute the sum $\sum_{r,s=0}^\infty \widetilde{c}_{r,s}^{\,m,n} \,\langle \psi|\, T_{r,s} \, |\psi\rangle$ for any fixed $m,n \geq 0$, given that we can explicitly compute each $\langle \psi|\, T_{r,s} \, |\psi\rangle$.  Given a particular $U$, there are several possibilities for what can happen:
\begin{enumerate}
\item Each $U^\dagger T_{m,n} U$ can be written as a \textit{finite} sum of Bender-Dunne operators.  This means that for any fixed $m,n \geq 0$, the coefficients $\widetilde{c}_{r,s}^{\,m,n}$ are nonzero for only \textit{finitely} many values of $r,s$.  In this case, for any fixed $m,n$ we can compute the non-zero $\widetilde{c}_{r,s}^{\,m,n}$'s, and then explicitly compute the \textit{finite} sum $\sum_{r,s} \widetilde{c}_{r,s}^{\,m,n} \,\langle \psi|\, T_{r,s} \, |\psi\rangle$ by including only the non-zero terms.
\item There are some $m,n$ for which $U^\dagger T_{m,n} U$ is an \textit{infinite} sum of Bender-Dunne operators.  In other words, for some fixed $m,n$, there are non-zero coefficients $\widetilde{c}_{r,s}^{\,m,n}$ for arbitrarily large values of $r,s$.  Either:
\vspace{.1cm}
\subitem (a) For \textit{some} $m,n$ such that $U^\dagger T_{m,n} U$ is an \textit{infinite} sum of Bender-Dunne operators,
\subitem \,\,\,\,\,\,\, we cannot compute all of the infinitely many $\widetilde{c}_{r,s}^{\,m,n}$, in which case we cannot \subitem \,\,\,\,\,\,\, compute $\langle \psi|U^\dagger \, T_{m,n} \, U|\psi\rangle$; or
\vspace{.2cm}
\subitem (b) For \textit{every} $m,n$ such that $U^\dagger T_{m,n} U$ is an \textit{infinite} sum of Bender-Dunne operators, 
\subitem \,\,\,\,\,\,\,\, we can take advantage of special structure to compute all of the infinitely many 
\subitem \,\,\,\,\,\,\,\, $\widetilde{c}_{r,s}^{\,m,n}$, but we do not know how to perform some sum $\sum_{r,s=0}^\infty \widetilde{c}_{r,s}^{\,m,n} \,\langle \psi|\, T_{r,s} \, |\psi\rangle$\,; or
\vspace{.2cm}
\subitem (c) For \textit{every} $m,n$ such that $U^\dagger T_{m,n} U$ is an \textit{infinite} sum of Bender-Dunne operators, 
\subitem \,\,\,\,\,\,\, we can take advantage of special structure to compute all of the infinitely many 
\subitem \,\,\,\,\,\,\, $\widetilde{c}_{r,s}^{\,m,n}$ \textit{and} take advantage of additional special structure to perform 
\subitem \,\,\,\,\,\,\, \textit{any} of the sums $\sum_{r,s=0}^\infty \widetilde{c}_{r,s}^{\,m,n} \,\langle \psi|\, T_{r,s} \, |\psi\rangle$.
\end{enumerate}
If either 1. or 2.(c) occurs, then we can compute each $\langle \psi|U^\dagger \, T_{m,n} \, U|\psi\rangle$ explicitly.  However, if 2.(a) or 2.(b) occurs, then we cannot compute some $\langle \psi|U^\dagger \, T_{m,n} \, U|\psi\rangle$ explicitly.  As we will see, for most $U$ we will fall into 2.(a) or 2.(b).  There are some known special classes of $U$'s that fall into 2.(c), but we will not discuss them here.  Finally, there is a very small class of $U$ which fall into 1., and we will now characterize them exactly.

In order to find all unitaries $U$ that fall into possibility 1., we need to explore several aspects of the Bender-Dunne basis.  First we recall the following useful formula which is closely related to the Baker-Campbell-Hausdorff (BCH) formula:
\begin{align}
\label{BCHlemma1}
e^{-i \, A} \, B \, e^{i \, A} &= B - i [A,B] - \frac{1}{2!}\,[A,[A,B]] +  \frac{i}{3!} \, [A,[A,[A,B]]] + \cdots \\
&= e^{- i \,\text{ad}_{A}}B
\end{align} 
where $\text{ad}_A B = [A,B]$, $\text{ad}_A^2 B = [A,[A,B]]$, and so on.  Then we can write
\begin{equation}
\label{unitary2}
\langle \psi| U^\dagger \, T_{m,n} \, U|\psi\rangle = \langle \psi| \exp\left\{-i \, \text{ad}_{\sum_{r,s=0}^\infty c_{r,s}\, T_{p,q}}\right\} T_{m,n} \, |\psi\rangle\,,
\end{equation}
so it becomes clear that we should study commutators of the form $[T_{r,s}, T_{m,n}]$.  In $\cite{BD2}$ Bender and Dunne give the formula
\begin{align}
\label{BDformula1}
[T_{r,s}, T_{m,n}] = 2 \sum_{j=0}^\infty \frac{\left(\frac{i}{2}\right)^{2j+1}}{(2j+1)!} \sum_{\ell = 0}^{2j+1}(-1)^\ell \binom{2j+1}{\ell} \frac{r! \, s! \, m! \, n!}{(r-\ell)! \,(s+\ell -2j-1)! \, (m+\ell - 2j-1)! \, (n-\ell)! }& \nonumber \\
\times \, T_{r+m-2j-1,\,s+n-2j-1}& \nonumber \\
\end{align}  
which holds for\footnote{In $\cite{BD2}$, Bender and Dunne also give formulas for the more generic case in which $r,s,m,n$ can be any integers.} $r,s,m,n \geq 0$.  A concise derivation is given in \cite{FormalBD1}.  While Eqn.~\eqref{BDformula1} may look daunting, the infinite sum always truncates to a finite sum for any $r,s,m,n \geq 0$.  This can be seen by noticing that there are factorial terms in the denominator in the sum which become infinite (causing the quotient to go to zero) when the argument of any of those factorials becomes a negative integer.

Let us define the \textit{order} of a sum of Bender-Dunne operators $\sum_{m,n \geq 0} c_{m,n}\, T_{m,n}$ by the largest value of $m+n$ such that $c_{m,n} \not= 0$. (If we allowed negative $m,n$, then the order would be defined as the largest value of $|m|+|n|$ such that $c_{m,n} \not= 0$.)  The order is the largest total number of $\widehat{x}$ and $\widehat{p}$ operators that appear multiplicatively in any term of $\sum_{m,n \geq 0} c_{m,n}\, T_{m,n}$.  For example, $T_{0,1} + T_{1,2} = \widehat{x} + \frac{1}{2}(\widehat{x}^2 \widehat{p} + \widehat{p}\widehat{x}^2)$ has order $3$.  Upon examining Eqn.~\eqref{BDformula1}, we notice that operators of order $2$ and lower, namely linear combinations of the Bender-Dunne operators
\begin{equation*}
T_{0,0} = 1\,,\qquad T_{0,1} = \widehat{x}\,, \qquad T_{1,0} = \widehat{p}\,, \qquad T_{1,1} = \frac{1}{2}(\widehat{x}\widehat{p} + \widehat{p}\widehat{x})\,,\qquad T_{0,2} = \widehat{x}^2\,, \qquad T_{2,0} = \widehat{p}^2\,,
\end{equation*} 
have a special property.  We have the commutation relations
\begin{align}
[1, T_{m,n}] &= [T_{0,0}, T_{m,n}] = 0 \\ \nonumber \\
[\widehat{x}, T_{m,n}] &= [T_{0,1}, T_{m,n}] = i m \, T_{m-1, n} \\\nonumber \\
[\widehat{p}, T_{m,n}] &= [T_{1,0}, T_{m,n}] = -i n \, T_{m, n-1} \\\nonumber \\ 
\label{T11comm1} [\frac{1}{2}(\widehat{x}\widehat{p} + \widehat{p}\widehat{x}), T_{m,n}] &= [T_{1,1}, T_{m,n}] = i (m-n) \, T_{m,n} 
\end{align}
\begin{align}
[\widehat{x}^2, T_{m,n}] &= [T_{0,2}, T_{m,n}] = 2 i m \, T_{m-1, n+1} \\\nonumber \\
[\widehat{p}^2, T_{m,n}] &= [T_{2,0}, T_{m,n}] = - 2 i n \, T_{m+1, n-1}
\end{align}
from which we see that taking the commutator between (a) an order $2$ or lower order Bender-Dunne operators and (b) an arbitrary $T_{m,n}$ yields an operator which has the same order as $T_{m,n}$ or is lower order.  By taking linear combinations of the order $2$ and lower order Bender-Dunne operators, we find that in general
\begin{equation}
\label{wordEq1}
[\text{order }2, \, \text{order }k] \leq \text{order }k
\end{equation}
in evident notation.

Then if $Q_2$ is an operator of order $2$ and $Q_k$ is an operator of order $k$, it follows by Eqn.~\eqref{BCHlemma1} that
\begin{equation}
e^{-i Q_2} \, Q_k \, e^{i Q_2}\quad \text{has at most order }k\,.
\end{equation}
Therefore, we have the result
\begin{equation}
\label{result1}
\langle \psi | e^{- i \sum_{p,q = 0}^2 c_{r,s} \, T_{r,s}} T_{m,n} \,  e^{i \sum_{p,q = 0}^2 c_{r,s} \, T_{r,s}}|\psi\rangle = \sum_{r,s=0}^{m+n} \widetilde{c}_{r,s}^{\,m,n} \, \langle \psi | T_{r,s} |\psi\rangle
\end{equation}
where the reader should look closely at the lower and upper bounds of the indices of summation.  In words, Eqn.~\eqref{result1} expresses that if we choose a unitary $U$ with a Hermitian generator of at most order $2$, then $\langle \psi|U^\dagger \,T_{m,n}\,U|\psi\rangle$ is a sum of at most $m+n$ calculable terms.  Furthermore, unitaries $U$ with a Hermitian generator of at most order $2$ fall into possibility 1. listed on page 8.  We also see that Eqn.~\eqref{result1} provides a partial answer to Question 1, namely: \\ \\
\textbf{Partial answer to Question 1:} Given an initial state $|\psi\rangle$ such that we can explicitly compute any correlation function $\langle \psi| T_{m,n} |\psi\rangle$ for $m,n\geq 0$, for unitaries $U$ with Hermitian generators of at most order $2$, we can explicitly compute $\langle \psi|U^\dagger \,T_{m,n}\,U|\psi\rangle$ for $m,n \geq 0$. \\

It is natural to ask what works differently if a $U$ has a Hermitian generator \textit{strictly greater} than order $2$.  Firstly, there is \textit{no} analogous version of Eqn.~\eqref{wordEq1} for operators with order strictly greater than $2$.  In particular, if $Q_j$ is any operator of order $j > 2$, then there exists some operator $Q_k$ of order $k$ such that $[Q_j, Q_k]$ has order \textit{greater than} $k$.  For example,
\begin{equation}
[\widehat{x}^4, T_{m,n}] = [T_{0,4}, T_{m,n}] = 4im \, T_{m-1,n+3} - im(m-1)(m-2) \, T_{m-3,n+1}
\end{equation}
is an operator of order $m+n+2$, which is greater than order $m+n$.  More dramatically, $e^{-i Q_j} Q_k e^{i Q_{j}}$ is typically an operator of infinite order, since the nested commutator terms which come from applying Eqn.~\eqref{BCHlemma1} can proliferate out of control since each new nesting can produce operators of progressively higher orders.  It follows that unitaries $U$ with a Hermitian generator strictly greater than order $2$ falls into possibilities 2.(a), 2.(b) and 2.(c) on page 8.  All of this goes to show that operators of order $2$ and below are truly special in the sense that they are the unique set of operators such that their commutators with any operator are ``order non-increasing.''

We also remark\footnote{We thank William Donnelly for pointing this out.} that appropriately rescaled Bender-Dunne operators of at most order $2$, namely
$$\{-i\,T_{0,0}, i\,T_{1,0}, i\,T_{0,1}, \frac{i}{2}\,T_{1,1}, \frac{i}{2}\,T_{2,0}, \frac{i}{2}\,T_{0,2}\}\,$$
form a Lie algebra which is isomorphic to $\mathfrak{h}_3 \rtimes \mathfrak{sl}_2$.  Here $\mathfrak{h}_3$ is the Lie algebra of the Heisenberg group $H_3(\mathbb{R})$, and $\mathfrak{sl}_2$ is the Lie algebra of $SL_2(\mathbb{R})$.  The Lie algebra $\mathfrak{h}_3$ corresponds to $\{-i\,T_{0,0}, i\,T_{1,0}, i\,T_{0,1}\}$ whereas $\mathfrak{sl}_2$ corresponds to $\{\frac{i}{2}\,T_{1,1}, \frac{i}{2}\,T_{2,0}, \frac{i}{2}\,T_{0,2}\}$.


\subsubsection{Generalization to $(d+1)$-dimensional QFT}

Here we generalize our analysis in the previous section to $(d+1)$--dimensional quantum field theory, and specifically the case of a scalar field in Minkowski space.  Suppose we have a scalar field in $d$ spatial dimensions.  The field operator is denoted by $\widehat{\phi}(x)$ and the conjugate momentum operator is denoted by $\widehat{\pi}(\vec{x})$.  We have the usual commutation relation $[\widehat{\phi}(\vec{x}), \widehat{\pi}(\vec{y})] = i \, \delta^{(d)}(\vec{x}-\vec{y})$.

The new feature is that our field and conjugate momentum operators furnish position labels, and our canonical commutation relation accordingly has a delta function.  This new feature makes it cumbersome to work with a direct generalization of the Bender-Dunne basis since much of the nice structure of the Bender-Dunne operators would only apply to combinations of $\widehat{\phi}$'s and $\widehat{\pi}$'s that are symmetric in all position labels which furnish $\widehat{\phi}$'s, and also symmetric in all position labels which furnish $\widehat{\pi}$'s.  Then we would need to keep track of all of the symmetrization factors.  Instead, we use the basis of field operators
\begin{equation}
\label{fieldbasis1}
S_{n}^{(k)}(\vec{x}_1,...,\vec{x}_n) = \widehat{\phi}(\vec{x}_1)\cdots \widehat{\phi}(\vec{x}_k) \widehat{\pi}(\vec{x}_{k+1}) \cdots \widehat{\pi}(\vec{x}_n) + \widehat{\pi}(\vec{x}_{k+1}) \cdots \widehat{\pi}(\vec{x}_n) \widehat{\phi}(\vec{x}_1)\cdots \widehat{\phi}(\vec{x}_k)
\end{equation}
for all $n \geq 0$, $0 \leq k \leq n$, and all $\vec{x}_1,...,\vec{x}_n \in \mathbb{R}^d$. These operators are manifestly Hermitian.  The subscript of $S_{n}^{(k)}$ counts the number of total position labels, whereas the superscript counts the number of position labels belonging to $\widehat{\phi}$'s.  Indeed, we can write any non-singular Hermitian operator $Q$ as
\begin{equation}
\label{generatorExpand2}
Q = \sum_{n=0}^\infty \sum_{k=0}^n \int d^d \vec{x}_1 \cdots d^d \vec{x}_n \, c_{n}^{(k)}(\vec{x}_1,...,\vec{x}_n) \, S_{n}^{(k)}(\vec{x}_1,...,\vec{x}_n)
\end{equation}
where all the $c_n^{(k)}$ are real-valued functions or distributions.  The above equation is in fact the same as Eqn.~\eqref{generatorExpand1} in the introduction.

It is worth pointing out a possible source of confusion.  It is tempting to think that a Hermitian operator like
\begin{equation}
\label{possibleConfusion1}
c(\vec{x}_1,\vec{x}_2) \, \widehat{\phi}(\vec{x}_1) \widehat{\pi}(\vec{x}_2) + c^*(\vec{x}_1,\vec{x}_2) \, \widehat{\pi}(\vec{x}_2) \widehat{\phi}(\vec{x}_1)
\end{equation}
where $c(\vec{x}_1,\vec{x}_2)$ is \textit{complex} is not of the form given by Eqn.~\eqref{generatorExpand2}, which only contains \textit{real} functions or distributions.  However, writing $c(\vec{x}_1, \vec{x}_2) = a(\vec{x}_1, \vec{x}_2) + i \, b(\vec{x}_1, \vec{x}_2)$ where both $a(\vec{x}_1, \vec{x}_2)$ and $b(\vec{x}_1, \vec{x}_2)$ are real, we find that Eqn.~\eqref{possibleConfusion1} can be written as
\begin{align}
& a(\vec{x}_1, \vec{x}_2) \, \big(\widehat{\phi}(\vec{x}_1) \widehat{\pi}(\vec{x}_2) + \widehat{\pi}(\vec{x}_2) \widehat{\phi}(\vec{x}_1) \big) + i \, b(\vec{x}_1,\vec{x}_2) \, \big(\widehat{\phi}(\vec{x}_1) \widehat{\pi}(\vec{x}_2) - \widehat{\pi}(\vec{x}_2) \widehat{\phi}(\vec{x}_1) \big) \nonumber \\
=\,& a(\vec{x}_1, \vec{x}_2) \, \big(\widehat{\phi}(\vec{x}_1) \widehat{\pi}(\vec{x}_2) + \widehat{\pi}(\vec{x}_2) \widehat{\phi}(\vec{x}_1) \big) - b(\vec{x}_1, \vec{x}_2) \, \delta^{(d)}(\vec{x}_1-\vec{x}_2)
\end{align} 
where we have used the canonical commutation relation.  We see that the above equation is indeed of the form given by Eqn.~\eqref{generatorExpand2}.  By using the same trick, one can express any operator
\begin{equation}
\label{possibleConfusion2}
c(\vec{x}_1,...,\vec{x}_n) \, \widehat{\phi}(\vec{x}_1)\cdots\widehat{\phi}(\vec{x}_k) \widehat{\pi}(\vec{x}_{k+1}) \cdots \widehat{\pi}(\vec{x}_n) + c^*(\vec{x}_1,...,\vec{x}_n) \, \widehat{\pi}(\vec{x}_{k+1}) \cdots \widehat{\pi}(\vec{x}_n) \cdots \widehat{\phi}(\vec{x}_1)\cdots\widehat{\phi}(\vec{x}_k)
\end{equation}
in the form given by Eqn.~\eqref{generatorExpand2}.  The commutator term will similarly produce lower-order operators multiplied by real-valued functions or distributions.

Similar to before, we would like to understand how unitaries $U = e^{iQ}$, where $Q$ of the form in Eqn.~\eqref{generatorExpand2}, transform a quantum state $|\Psi\rangle$ of the field theory.  We can ask a modified version of Question 1 as follows: \\ \\
\textbf{Question 2:} Given an initial state $|\Psi\rangle$ such that we can explicitly compute any equal-time correlation function $\langle \Psi| S_{n}^{(k)}(\vec{x}_1,...,\vec{x}_n) |\Psi\rangle$ for $n \geq 0$ and $0 \leq k \leq n$, for which unitaries $U$ can we explicitly compute $\langle \Psi|U^\dagger \, S_{n}^{(k)}(\vec{x}_1,...,\vec{x}_n) \, U|\Psi\rangle$ for $n \geq 0$ and $0 \leq k \leq n$\,? \\ \\
The story is similar to the $(0+1)$--dimensional case.  In the case of the scalar field, we define the order a Hermitian operator $\sum_{n=0}^\infty \sum_{k=0}^n \int d^d \vec{x}_1 \cdots d^d \vec{x}_n \, c_{n}^{(k)}(\vec{x}_1,...,\vec{x}_n) \, S_{n}^{(k)}(\vec{x}_1,...,\vec{x}_n)$ as the largest value of $n$ for which there is some $c_{n}^{(k)}(\vec{x}_1,...,\vec{x}_n) \not = 0$.  For example, $S_n^{(k)}(\vec{x}_1,...,\vec{x}_n)$ by itself is order $n$.  As before, operators of order $2$ and lower, namely linear combinations of
\begin{align*}
&S_{0}^{(0)} = \textbf{1}\,,\qquad S_{1}^{(0)}(\vec{x}_1) = \widehat{\pi}(\vec{x}_1)\,, \qquad S_{1}^{(1)}(\vec{x}_1) = \widehat{\phi}(\vec{x}_1)\,, \qquad S_{2}^{(0)}(\vec{x}_1,\vec{x}_2) = \widehat{\pi}(\vec{x}_1) \widehat{\pi}(\vec{x}_2) \\ \\
&S_{2}^{(1)}(\vec{x}_1,\vec{x}_2) = \widehat{\phi}(\vec{x}_1) \widehat{\pi}(\vec{x}_2) + \widehat{\pi}(\vec{x}_2) \widehat{\phi}(\vec{x}_1)\,, \qquad S_{2}^{(2)}(\vec{x}_1) = \widehat{\phi}(\vec{x}_1)\widehat{\phi}(\vec{x}_2)\,,
\end{align*} 
are special.

We will now compute the commutation relations of these operators with the operator $S_{n}^{(k)}(\vec{y}_1,...,\vec{y}_n)$ for generic $n$ and $k$.  It is convenient to write $S_{n}^{(k)}(\vec{y}_1,...,\vec{y}_k \, ; \, \vec{y}_{k+1},...,\vec{y}_n)$, where the variables to the left of the semicolon are position labels belonging to $\widehat{\phi}$ operators, and the variables to the right of the semicolon are position labels belonging to $\widehat{\pi}$ operators.  We will sometimes include an additional semicolon, as in $S_{n-1}^{(\widetilde{k})}(\vec{y}_1,...,\vec{y}_k \, ; \, \vec{y}_{k+1},...,\vec{y}_n\,;\,\vec{y}_j)$, to denote that the $\vec{y}_j$ variable is to be omitted.  (Also notice that $k$ has been replaced by $\widetilde{k}$ in the superscript.)  Thus there are $n-1$ remaining variables.  If $0 \leq j \leq k$ then $\widetilde{k} = k-1$ since we have eaten the label of a $\widehat{\phi}$ operator, whereas if $(k+1) \leq j \leq n$ then $\widetilde{k} = k$ since we have eaten the label of a $\widehat{\pi}$ operator.  With this notation, the commutation relations are
\begin{align}
\nonumber \\
 [S_{0}^{(0)}, S_{n}^{(k)}(\vec{y}_1,...,\vec{y}_n)] &= 0 \\ \nonumber \\
[S_{1}^{(0)}(\vec{x}_1), S_{n}^{(k)}(\vec{y}_1,...,\vec{y}_n)] 
&= - i \sum_{j=1}^k \delta^{(d)}(\vec{x}_1-\vec{y}_j) \, S_{n-1}^{(k-1)}(\vec{y}_1,...,\vec{y}_k\,;\,\vec{y}_{k+1},...,\vec{y}_n\,;\,\vec{y}_j) \\ \nonumber \\
[S_{1}^{(1)}(\vec{x}_1), S_{n}^{(k)}(\vec{y}_1,...,\vec{y}_n)]
&= i \sum_{j=k+1}^n \delta^{(d)}(\vec{x}_1-\vec{y}_j) \, S_{n-1}^{(k)}(\vec{y}_1,...,\vec{y}_k\,;\,\vec{y}_{k+1},...,\vec{y}_n\,;\,\vec{y}_j)
\end{align}
\begin{align}
[S_{2}^{(0)}(\vec{x}_1,\vec{x}_2), S_{n}^{(k)}(\vec{y}_1,...,\vec{y}_n)]
&=  - i \sum_{j=1}^k \big(\delta^{(d)}(\vec{x}_1-\vec{y}_j) \, S_{n}^{(k-1)}(\vec{y}_1,...,\vec{y}_k\,;\,\vec{x}_2, \vec{y}_{k+1},...,\vec{y}_n\,;\,\vec{y}_j) \, \widehat{\pi}(\vec{x}_2) \nonumber \\ 
&\qquad \qquad + \delta^{(d)}(\vec{x}_2 - \vec{y}_j)\, \widehat{\pi}(\vec{x}_1)\,S_{n}^{(k-1)}(\vec{y}_1,...,\vec{y}_k\,;\,\vec{x}_1, \vec{y}_{k+1},...,\vec{y}_n\,;\,\vec{y}_j)\big) \label{unsimplified1} \\ \nonumber \\
[S_{2}^{(1)}(\vec{x}_1,\vec{x}_2), S_{n}^{(k)}(\vec{y}_1,...,\vec{y}_n)] 
&= 2 i \sum_{j=k+1}^n \delta^{(d)}(\vec{x}_1-\vec{y}_j) \, S_{n}^{(k)}(\vec{y}_1,...,\vec{y}_k\,;\,\vec{x}_2, \vec{y}_{k+1},...,\vec{y}_n\,;\,\vec{y}_j) \nonumber \\
&\quad + 2 i \sum_{j=1}^k \delta^{(d)}(\vec{x}_2-\vec{y}_j) \, S_{n}^{(k)}(\vec{x}_1, \vec{y}_1,...,\vec{y}_k\,;\, \vec{y}_{k+1},...,\vec{y}_n\,;\,\vec{y}_j)
 \\\nonumber \\
[S_{2}^{(2)}(\vec{x}_1,\vec{x}_2), S_{n}^{(k)}(\vec{y}_1,...,\vec{y}_n)]
&=  i \sum_{j=k+1}^n \big(\delta^{(d)}(\vec{x}_1-\vec{y}_j) \, S_{n}^{(k+1)}(\vec{x}_2,\vec{y}_1,...,\vec{y}_k\,;\,\vec{y}_{k+1},...,\vec{y}_n\,;\,\vec{y}_j) \, \widehat{\phi}(\vec{x}_2) \nonumber \\ & \qquad \qquad + \delta^{(d)}(\vec{x}_2 - \vec{y}_j)\, \widehat{\phi}(\vec{x}_1)\,S_{n}^{(k+1)}(\vec{x}_1,\vec{y}_1,...,\vec{y}_k\,;\,\vec{y}_{k+1},...,\vec{y}_n\,;\,\vec{y}_j)\big)
\label{unsimplified2}
\end{align}
where the reader should pay attention to the lower and upper bounds of the indices of summation.  We have not fully simplified the right-hand sides of Eqn.'s~\eqref{unsimplified1} and~\eqref{unsimplified2} to be solely in terms of $S_m^{(\ell)}$ operators because the explicit forms are cumbersome, but we still see that the right-hand sides of both equations are order $n$.  Indeed, all of the equations above have right-hand sides which are order $n$ or lower order.  It follows that commuting order $2$ and lower order operators with an arbitrary $S_{n}^{(k)}$ yields an operator which is order $n$ or lower order.  Taking linear combinations of the order $2$ and lower order $S_{n}^{(k)}$ operators, we find
$$[\text{order }2, \, \text{order }n] \leq \text{order }n$$
as in the $(0+1)$--dimensional case.

Repeating the same arguments as before, if $Q_2$ is an operator of order $2$ and $Q_n$ is an operator of order $n$ then $e^{-i Q_2} \, Q_n \, e^{i Q_2}$ has at most order $n$, and so
\begin{align}
\label{result2}
&\langle \Psi | e^{- i \sum_{m=0}^2 \sum_{\ell=0}^{m} \int d^d \vec{x}_1 \, d^d \vec{x}_2 \, c_{m}^{(\ell)}(\vec{x}_1,\vec{x}_2) \, S_{m}^{(\ell)}(\vec{x}_1,\vec{x}_2)} \, S_n^{(k)}(\vec{x}_1,...,\vec{x}_n) \,  e^{i \sum_{m=0}^2 \sum_{\ell=0}^{m} \int d^d \vec{x}_1 \, d^d \vec{x}_2 \, c_{m}^{(\ell)}(\vec{x}_1,\vec{x}_2) \, S_{m}^{(\ell)}(\vec{x}_1,\vec{x}_2)}|\Psi\rangle \nonumber \\
&= \sum_{m=0}^n \sum_{\ell = 0}^m \int d^d \vec{x}_1 \cdots d^d \vec{x}_n \, \widetilde{c}_{m,(\ell)}^{\,n,(k)}(\vec{x}_1,...,\vec{x}_n) \, \langle \Psi | S_{m}^{(\ell)}(\vec{x}_1,...,\vec{x}_n) |\Psi\rangle \,.
\end{align}
which is a sum of at most $\frac{1}{2}(n+2)(n+1)$ calculable terms.  Then we get the following partial answer to Question 2: \\ \\
\textbf{Partial answer to Question 2:} Given an initial state $|\Psi\rangle$ compute any equal-time correlation function $\langle \Psi| S_{n}^{(k)}(\vec{x}_1,...,\vec{x}_n) |\Psi\rangle$ for $n \geq 0$ and $0 \leq k \leq n$, for unitaries $U$ with Hermitian generators of at most order $2$, we can explicitly compute $\langle \Psi|U^\dagger\, S_{n}^{(k)}(\vec{x}_1,...,\vec{x}_n) \,U|\Psi\rangle$ for $n \geq 0$ and $0 \leq k \leq n$. \\ \\ \\

Similar to the $(0+1)$--dimensional case, for any Hermitian operator $Q_j$ with order $j > 2$, there is some other Hermitian operator $Q_n$ of order $n$ such that $[Q_j, Q_k]$ has order \textit{greater than }$n$.  As before, $e^{-i Q_j} Q_n e^{i Q_j}$ is typically an operator of infinite order, and so unitaries of the form $U = e^{i Q_j}$ fall into the possibilities 2.(a), 2.(b) and 2.(c) on page 8.

Although we have established that unitaries with Hermitian operators of at most order $2$ are analytically tractable within the contexts of Questions 1 and 2, we have not yet shown how to explicitly perform computations.  Furthermore, the class of unitaries with Hermitian operators of order $2$ or less is severely limiting, and so we would like to develop techniques to go beyond order $2$.  One attractive option is to perform perturbation theory in Hermitian generators which are greater than order $2$.  In the next section, we will formulate this type of ``quantum circuit perturbation theory,'' and show how to compute with it.

It would be interesting to understand which unitaries fall into possibility 2.(c) on page 8, namely when we can explicitly sum up correlation functions like $\langle \Psi|U^\dagger \, S_{n}^{(k)} \, U|\Psi\rangle$, even when $U^\dagger \, S_{n}^{(k)} \, U$ is of infinite order.  At this time, we do not have a complete understanding of when such an explicit infinite summation can be performed.

\subsection{Continuum circuits and perturbation theory}

We begin by working with $(0+1)$--dimensional field theory before generalizing to the $(d+1)$--dimensional case.  Let us again consider a single particle in one spatial dimension.  Given some initial state $|\Psi\rangle$ and unitary
\begin{equation}
U = \exp\left(-i \sum_{0 \leq p+q \leq N} c_{p,q} \, T_{p,q}\right)
\end{equation}
where $N$ is the maximum order of the Hermitian generator, we would like to compute the transformed state $U |\Psi\rangle$.  In the previous section, we explained that it is generally not possible to obtain explicit expressions for $U |\Psi\rangle$ if $U$ has a Hermitian generator greater than order $2$.  However, suppose that $U$ has the form
\begin{equation}
\label{Uepsilon1}
U = \exp\left(-i\left(\sum_{p+q \,\leq\, 2} c_{p,q} \, T_{p,q} + \epsilon \, \sum_{2 \,<\,r+s\,\leq\,N} c_{r,s} \, T_{r,s} \right) \right)
\end{equation}
where $\epsilon$ is a small parameter.  In words, we are considering unitaries with Hermitian generators which are order $2$ plus higher order perturbative corrections, where $\epsilon$ is the perturbation parameter.

For ease of notation, let us define
\begin{align}
Q_2 &:= \sum_{p+q \,\leq\, 2} c_{p,q} \, T_{p,q} \\ \nonumber \\
Q_{\text{higher}} &:= \sum_{0 \,\leq\,r+s\,\leq\,N} c_{r,s} \, T_{r,s}
\label{Qhigher1}
\end{align}
so that
\begin{equation}
\label{Q2plusQhigher}
U = \exp\left(-i \left(Q_2 + \epsilon \, Q_{\text{higher}}\right)\right)\,.
\end{equation}
Using the expansion
\begin{align}
\label{ABexpansion1}
e^{A + \epsilon  B} &= \left(1 + \epsilon \, B + \frac{1}{2!}[A,\epsilon \, B] + \frac{1}{3!} [A,[A,\epsilon\,B]] + \frac{1}{4!}[A,[A,[A,\epsilon\,B]]] + \cdots \right) \, e^A \,\,+\mathcal{O}(\epsilon^2) \\
\label{ABexpansion2}
&= \left(1 + \epsilon \, \frac{e^{\text{ad}_A}-1}{\text{ad}_A} \, B\right) \, e^{A} \,\,+\mathcal{O}(\epsilon^2)\,,
\end{align}
we can expand $U$ to first order in $\epsilon$ to obtain
\begin{align}
\label{QhigherExpand1}
U &= \left(1 + \epsilon \, \frac{e^{-i\,\text{ad}_{Q_2}}-1}{\text{ad}_{Q_2}} \, Q_{\text{higher}}\right) \, e^{-i\,Q_2} \,\,+\mathcal{O}(\epsilon^2)\,.
\end{align}
This expansion can be carried out to higher orders (for relevant techniques, see \cite{Replica1}), but we will only consider the first order expansion here.  Recalling from Eqn.~\eqref{Qhigher1} that $Q_{\text{higher}}$ is order $N$, note that in Eqn.~\eqref{QhigherExpand1} the term $\frac{e^{-i\,\text{ad}_{Q_2}}-1}{\text{ad}_{Q_2}} \, Q_{\text{higher}}$ is also at most order $N$.  The reason is that nested commutators $\text{ad}_{Q_2}^n Q_{\text{higher}} = [Q_2,[Q_2,[Q_2,...[Q_2,Q_{\text{higher}}]...]]]$ have at most the same order as $Q_{\text{higher}}$, since commutators with quadratic order operators are order non-increasing.  This feature leads to a well-controlled perturbative expansion for $U$ : to first order in $\epsilon$, we will only obtain nested commutators with quadratic operators and \textit{at most} one $Q_{\text{higher}}$.  Similarly, at $k$th order in $\epsilon$, we will only obtain nested commutators with quadratic operators and \textit{at most} $k$ copies $Q_{\text{higher}}$.  The general structure is:
\begin{equation}
\label{generalstructure1}
U = \bigg(1 + \epsilon \,[\text{order }N] + \epsilon^2 \, [\text{order }2(N-1)]+\cdots+\epsilon^k \, [\text{order }k(N-1)]+\cdots \bigg) \, e^{-i Q_2}\,.
\end{equation}
Indeed, the perturbation series is under control at any finite order in $\epsilon$.

If we specify some particular $Q_2$ and $Q_{\text{higher}}$, then we can explicitly compute terms like $\frac{e^{-i\,\text{ad}_{Q_2}}-1}{\text{ad}_{Q_2}} \, Q_{\text{higher}}$ in Eqn.~\eqref{QhigherExpand1}, and we will do this in later sections.  For a given state $|\Psi\rangle$, if we can calculate both $e^{-i Q_2} |\Psi\rangle$ and $\frac{e^{-i\,\text{ad}_{Q_2}}-1}{\text{ad}_{Q_2}} \, Q_{\text{higher}}\hspace{1mm}e^{-i Q_2} |\Psi\rangle$ explicitly, then we can compute $U|\Psi\rangle$ explicitly to first order in $\epsilon$.

All of the analysis above extends immediately to $(d+1)$--dimensional field theories.  If instead of Eqn.~\eqref{Uepsilon1} we have
\begin{align}
\label{Uepsilon2}
U = \exp\bigg(&-i \sum_{n=0}^2 \sum_{k=0}^n \int d^d \vec{x}_1 \, \cdots \, d^d \vec{x}_n \, c_n^{(k)}(\vec{x}_1,...,\vec{x}_n) \, S_{n}^{(k)}(\vec{x}_1,...,\vec{x}_n) \nonumber \\
& - i \epsilon\, \sum_{m=0}^{N} \sum_{\ell=0}^m \int d^d \vec{x}_1 \, \cdots \, d^d \vec{x}_m \, c_m^{(\ell)}(\vec{x}_1,...,\vec{x}_m) \, S_{m}^{(\ell)}(\vec{x}_1,...,\vec{x}_m)\bigg)\,,
\end{align}
then using the new definitions
\begin{align}
Q_2 &:=  \sum_{n=0}^2 \sum_{k=0}^n \int d^d \vec{x}_1 \, \cdots \, d^d \vec{x}_n \, c_n^{(k)}(\vec{x}_1,...,\vec{x}_n) \, S_{n}^{(k)}(\vec{x}_1,...,\vec{x}_n) \\ \nonumber \\
Q_{\text{higher}} &:= \sum_{m=0}^{N} \sum_{\ell=0}^m \int d^d \vec{x}_1 \, \cdots \, d^d \vec{x}_m \, c_m^{(\ell)}(\vec{x}_1,...,\vec{x}_m) \, S_{m}^{(\ell)}(\vec{x}_1,...,\vec{x}_m)
\end{align}
we similarly have
$$U = \exp\left(-i \left(Q_2 + \epsilon \, Q_{\text{higher}}\right)\right)$$
and
$$U = \left(1 + \epsilon \, \frac{e^{-i\,\text{ad}_{Q_2}}-1}{\text{ad}_{Q_2}} \, Q_{\text{higher}}\right) \, e^{-i\,Q_2} \,\,+\mathcal{O}(\epsilon^2)$$
as before.  The general structure of the perturbation theory given in Eqn.~\eqref{generalstructure1} likewise holds.

In the next sections, we explicitly demonstrate quantum circuit perturbation theory for continuum circuits in specific examples, and highlight general techniques. 

\subsection{Circuits between Gaussian states}

In order to construct circuits which map between Gaussian states and perturbations of Gaussian states (such as the ground states of weakly interacting field theories), we first need to construct circuits which map between Gaussian states and other Gaussian states.

\subsubsection{Single-particle quantum mechanics}

Again, we start with $(0+1)$--dimensional quantum mechanics for a single particle in one spatial dimension.  In this context, a Gaussian state $|\Psi_0\rangle$ can be expressed in the position basis as
\begin{equation}
\label{GaussianPos0}
\langle x | \Psi_0\rangle = \left( \frac{M_0}{\pi}\right)^{\frac{1}{4}} \, e^{- \frac{1}{2}\, M_0 \, (x-x_0)^2}\,.
\end{equation}
It is nicer to have a basis-independent characterization of a Gaussian state.  In terms of $\widehat{x}$ and $\widehat{p}$ operators, we have
\begin{equation}
\label{GaussianBasisIndep0}
\left( \sqrt{M_0} \,\, (\widehat{x}-x_0) + \frac{i}{\sqrt{M_0}}\,\widehat{p} \right) |\Psi_0\rangle = 0\,.
\end{equation}
In the position basis, the above equation corresponds to a first order ODE whose unique solution is given by Eqn.~\eqref{GaussianPos0}.  However, we emphasize again that as written, Eqn.~\eqref{GaussianBasisIndep0} is basis-\textit{independent}.

Suppose we have a second Gaussian state $|\Psi_1\rangle$, which we write as
\begin{equation}
\langle x | \Psi_1\rangle = \left( \frac{M_1}{\pi}\right)^{\frac{1}{4}} \, e^{- \frac{1}{2}\, M_1 \, (x-x_1)^2}
\end{equation}
or
\begin{equation}
\label{Psi1def1}
\left( \sqrt{M_1} \,\, (\widehat{x}-x_1) + \frac{i}{\sqrt{M_1}}\,\widehat{p} \right) |\Psi_1\rangle = 0\,.
\end{equation}
We would like to find a unitary $U$ such that $U|\Psi_0\rangle = |\Psi_1\rangle$.  First, we note that by inserting an identity operator $\textbf{1} = U^\dagger U$ in Eqn.~\eqref{GaussianBasisIndep0} and left-multiplying by a $U$, we obtain
\begin{equation}
U\left( \sqrt{M_0} \,\, (\widehat{x}-x_0) + \frac{i}{\sqrt{M_0}}\,\widehat{p} \right)U^\dagger U |\Psi_0\rangle = 0\,.
\end{equation}
If we have
\begin{equation}
\label{basisinvimp1}
U\left( \sqrt{M_0} \,\, (\widehat{x}-x_0) + \frac{i}{\sqrt{M_0}}\,\widehat{p} \right)U^\dagger = \sqrt{M_1} \,\, (\widehat{x}-x_1) + \frac{i}{\sqrt{M_1}}\,\widehat{p}\,,
\end{equation}
then
\begin{equation}
\left( \sqrt{M_1} \,\, (\widehat{x}-x_1) + \frac{i}{\sqrt{M_1}}\,\widehat{p} \right) \, U |\Psi_0\rangle = 0\,,
\end{equation}
and so $U|\Psi_0\rangle = |\Psi_1\rangle$ via Eqn.~\eqref{Psi1def1}.  Unitaries $U$ satisfying Eqn.~\eqref{basisinvimp1} can be constructed out of the displacement operator $e^{-i \alpha \widehat{p}}$ which satisfies
\begin{equation}
e^{-i \alpha \widehat{p}} \, \widehat{x} \,e^{i \alpha \widehat{p}} = \widehat{x} - \alpha\,,
\end{equation}
and the squeezing operator $e^{-i \beta (\widehat{x}\widehat{p} + \widehat{p} \widehat{x})}$ which satisfies
\begin{equation}
e^{-i \beta (\widehat{x}\widehat{p} + \widehat{p} \widehat{x})} \,\widehat{x}\, e^{i \beta (\widehat{x}\widehat{p} + \widehat{p} \widehat{x})} = e^{-2\beta} \widehat{x}\,,\qquad e^{-i \beta (\widehat{x}\widehat{p} + \widehat{p} \widehat{x})} \,\widehat{p}\, e^{i \beta (\widehat{x}\widehat{p} + \widehat{p} \widehat{x})} = e^{2\beta} \widehat{p}\,.
\end{equation}  
Then two viable $U$'s satisfying Eqn.~\eqref{basisinvimp1}, and hence mapping $U|\Psi_0\rangle = |\Psi_1\rangle$, are
\begin{equation}
\label{GaussianU1}
U = e^{-i (x_1-x_0) \widehat{p}} \, e^{-\frac{i}{4} \log\left( \frac{M_0}{M_1}\right)((\widehat{x}-x_0)\widehat{p} + \widehat{p} (\widehat{x}-x_0))} = e^{- i \big(\frac{1}{2} \frac{\log\left(\frac{M_0}{M_1}\right) \sqrt{M_1}}{\sqrt{M_0} - \sqrt{M_1}} \big) (x_1 - x_0) \widehat{p} -\frac{i}{4} \log\left( \frac{M_0}{M_1}\right)((\widehat{x}-x_0)\widehat{p} + \widehat{p} (\widehat{x}-x_0))}
\end{equation}
and
\begin{equation}
\label{GaussianU2}
U =  e^{-\frac{i}{4} \log\left( \frac{M_0}{M_1}\right)((\widehat{x}-x_1)\widehat{p} + \widehat{p} (\widehat{x}-x_1))}\,e^{-i (x_1-x_0) \widehat{p}} = e^{- i \big(\frac{1}{2} \frac{\log\left(\frac{M_0}{M_1}\right) \sqrt{M_0}}{\sqrt{M_0} - \sqrt{M_1}} \big) (x_1 - x_0) \widehat{p} -\frac{i}{4} \log\left( \frac{M_0}{M_1}\right)((\widehat{x}-x_1)\widehat{p} + \widehat{p} (\widehat{x}-x_1))}\,.
\end{equation}
In the two equations above, we have used the Baker-Campbell-Hausdorff formula to combine the product of a displacement operator and a squeezing operator into a single unitary.  Useful formulas for Baker-Campbell-Hausdorff manipulations can be found in \cite{Visser1}.  Further, we note that the unitaries in Eqn.'s~\eqref{GaussianU1} and~\eqref{GaussianU2} above have Hermitian generators of order $2$, which is why mapping between Gaussians is analytically tractable.

\subsubsection{Concrete example: harmonic oscillators with different masses}

Let us construct a unitary mapping the ground state of a quantum harmonic oscillator with mass $m_1$ to the ground state of another quantum harmonic oscillator with mass $m_2$.  Recall that the Hamiltonian for a quantum harmonic oscillator with mass $m$ is
\begin{equation}
H = \frac{1}{2m}\,\widehat{p}^2 + \frac{1}{2}\,m \omega^2 \widehat{x}^2
\end{equation} 
which has ground state $|\psi_0\rangle$ given by
\begin{equation}
\langle x |\psi_0\rangle = \left(\frac{m \omega}{\pi}\right)^{1/4} e^{- \frac{1}{2} \, m \omega x^2}\,.
\end{equation}
We set $\omega = 1$.  Then the ground states of the quantum harmonic oscillators with masses $m_1$ and $m_2$ are
\begin{align}
\langle x | \psi_0^{m_1}\rangle &= \left(\frac{m_1}{\pi}\right)^{1/4} e^{- \frac{1}{2} \, m_1 x^2} \\
\langle x | \psi_0^{m_2}\rangle &= \left(\frac{m_2}{\pi}\right)^{1/4} e^{- \frac{1}{2} \, m_2 x^2}\,.
\end{align}
We want to find a unitary $U$ such that $U |\psi_{0}^{m_1}\rangle = |\psi_0^{m_2}\rangle$.  Considering Eqn.'s~\eqref{GaussianU1} and~\eqref{GaussianU2}, we let $x_0 = x_1 = 0$, $M_0 = m_1$, and $M_1 = m_2$.  Then both Eqn.'s~\eqref{GaussianU1} and~\eqref{GaussianU2} simplify to
\begin{equation}
U = \exp\left(- \frac{i}{4} \, \log\left(\frac{m_1}{m_2}\right) (\widehat{x}\widehat{p} + \widehat{p} \widehat{x})\right)
\end{equation}
which satisfies $U |\psi_{0}^{m_1}\rangle = |\psi_0^{m_2}\rangle$.

\subsubsection{Scalar field theory}
We can perform a similar analysis for $(d+1)$--dimensional field theories.  We will work with scalar field theories in this paper, but a similar analysis holds for fermionic theories.  For a scalar field theory, recall $[\widehat{\phi}(\vec{x}), \widehat{\pi}(\vec{y})] = i\,\delta^{(d)}(\vec{x}-\vec{y})$, and that in the $\phi$-space representation we can write $\widehat{\pi}(\vec{y})$ as the functional derivative $\frac{1}{i} \frac{\delta}{\delta \phi(\vec{y})}$.  A Gaussian state has the wavefunctional representation
\begin{equation}
\label{wavefunctional0pos}
\langle \phi |\Psi_0\rangle = \text{det}^{\frac{1}{4}}\left(\frac{\Omega_0}{\pi}\right) \, \exp\left(-\frac{1}{2} \int d^d \vec{x} \, d^d \vec{y} \, (\phi(\vec{x}) - \phi_0(\vec{x})) \, \Omega_0(\vec{x},\vec{y}) \, (\phi(\vec{y}) - \phi_0(\vec{y})) \right)
\end{equation}
for some symmetric, invertible, positive-definite kernel $\Omega_0(\vec{x},\vec{y})$, and some function $\phi_0(\vec{x})$.  We further suppose that $\Omega_0(\vec{x},\vec{y}) = \Omega_0(|\vec{x}-\vec{y}|)$, i.e.\! that it is translation and rotation-invariant.  Then it is convenient to work in momentum space, instead of position space.  Letting
\begin{equation}
\widetilde{\Omega}_0(|\vec{k}|) = \int d^d \vec{x}\, e^{i\vec{k}\cdot\vec{x}} \, \Omega_0(|\vec{x}|)
\end{equation}
and
\begin{align}
\phi(\vec{k}) &:= \frac{1}{(2\pi)^{d/2}}\int d^d \vec{x} \, e^{i\vec{k} \cdot \vec{x}} \phi(\vec{x}) \\
\phi_0(\vec{k}) &:= \frac{1}{(2\pi)^{d/2}}\int d^d \vec{x} \, e^{i\vec{k} \cdot \vec{x}} \phi_0(\vec{x})\,,
\end{align}
we can rewrite Eqn.~\eqref{wavefunctional0pos} as
\begin{equation}
\label{wavefunctional0momen}
\langle \phi |\Psi_0\rangle = \text{det}^{\frac{1}{4}}\left(\frac{\widetilde{\Omega}_0}{\pi}\right) \, \exp\left(-\frac{1}{2} \int d^d \vec{k} \, (\phi(\vec{k}) - \phi_0(\vec{k})) \, \widetilde{\Omega}_0(k) \, (\phi(-\vec{k}) - \phi_0(-\vec{k})) \right)
\end{equation}
Above we use $|\vec{k}|$ and $k$ interchangeably.

Let us define the momentum space representations of the $\widehat{\phi}(\vec{x})$ and $\widehat{\pi}(\vec{y})$ operators by
\begin{align}
\widehat{\phi}(\vec{k})  &:= \frac{1}{(2\pi)^{d/2}}\int d^d \vec{x} \, e^{i\vec{k} \cdot \vec{x}} \widehat{\phi}(\vec{x}) \\
\widehat{\pi}(\vec{p})  &:= \frac{1}{(2\pi)^{d/2}}\int d^d \vec{x}\, e^{i\vec{p} \cdot \vec{y}} \widehat{\pi}(\vec{y})\,.
\end{align}
The definitions for $\widehat{\phi}(\vec{k})$ and $\widehat{\pi}(\vec{p})$ are chosen so that
\begin{equation}
[\widehat{\phi}(\vec{k}),\widehat{\pi}(\vec{p})] = i \delta^{(d)}(\vec{k}+\vec{p})\,.
\end{equation}
This convention is convenient, since for most of the rest of the paper we will work in the momentum space representation, and it is cumbersome to have factors of $(2\pi)^d$ floating around in the $[\widehat{\phi}(\vec{k}),\widehat{\pi}(\vec{p})]$ commutation relation.
Using $\widehat{\phi}(\vec{k})$ and $\widehat{\pi}(\vec{p})$, we can also characterize Eqn.~\eqref{wavefunctional0momen} by
\begin{equation}
\left(\sqrt{\widetilde{\Omega}_0(k)}\,(\widehat{\phi}(\vec{k}) - \phi_0(\vec{k})) + \frac{i}{\sqrt{\widetilde{\Omega}_0(k)}} \, \widehat{\pi}(\vec{k})\right) |\Psi_0\rangle = 0\,, \quad \text{for all } \vec{k} \in \mathbb{R}^d\,.
\end{equation}

Suppose we have a second Gaussian state $|\Psi_1\rangle$ given by
\begin{equation}
\langle \phi |\Psi_1\rangle = \text{det}^{\frac{1}{4}}\left(\frac{\widetilde{\Omega}_1}{\pi}\right) \, \exp\left(-\frac{1}{2} \int d^d \vec{k} \, (\phi(\vec{k}) - \phi_1(\vec{k})) \, \widetilde{\Omega}_1(k) \, (\phi(-\vec{k}) - \phi_1(-\vec{k})) \right)
\end{equation}
and also characterized by
\begin{equation}
\left(\sqrt{\widetilde{\Omega}_1(k)}\,(\widehat{\phi}(\vec{k}) - \phi_1(\vec{k})) + \frac{i}{\sqrt{\widetilde{\Omega}_1(k)}} \, \widehat{\pi}(\vec{k})\right) |\Psi_1\rangle = 0\,, \quad \text{for all } \vec{k} \in \mathbb{R}^d\,.
\end{equation}
We would like to find a unitary $U$ so that $U|\Psi_0\rangle = |\Psi_1\rangle$.  It will be useful here to use the field-theoretic generalization of the displacement operator, namely
$$\exp\left(- i \int d^d \vec{p} \, \alpha(\vec{p}) \, \widehat{\pi}(\vec{p}) \right)\,,$$
which satisfies
\begin{equation}
e^{- i \int d^d \vec{p} \, \alpha(\vec{p}) \, \widehat{\pi}(\vec{p}) } \, \widehat{\phi}(\vec{k}) \,e^{ i \int d^d \vec{p} \, \alpha(\vec{p}) \, \widehat{\pi}(\vec{p}) } = \widehat{\phi}(\vec{k}) - \alpha(-\vec{k})\,,
\end{equation}
and the field-theoretic generalization of the squeezing operator
$$\exp\left(-i \int d^d \vec{p} \, \beta(\vec{p}) \left(\widehat{\phi}(\vec{p}) \widehat{\pi}(-\vec{p}) + \widehat{\pi}(-\vec{p}) \widehat{\phi}(\vec{p}) \right) \right)\,,$$
which satisfies
\begin{align}
e^{-i \int d^d \vec{p} \, \beta(\vec{p}) \left(\widehat{\phi}(\vec{p}) \widehat{\pi}(-\vec{p}) + \widehat{\pi}(-\vec{p}) \widehat{\phi}(\vec{p}) \right) } \, \widehat{\phi}(\vec{k}) \, e^{i \int d^d \vec{p} \, \beta(\vec{p}) \left(\widehat{\phi}(\vec{p}) \widehat{\pi}(-\vec{p}) + \widehat{\pi}(-\vec{p}) \widehat{\phi}(\vec{p}) \right) } &= e^{-2 \beta(\vec{k})}\, \widehat{\phi}(\vec{k})  \\
e^{-i \int d^d \vec{p} \, \beta(\vec{p}) \left(\widehat{\phi}(\vec{p}) \widehat{\pi}(-\vec{p}) + \widehat{\pi}(-\vec{p}) \widehat{\phi}(\vec{p}) \right) } \, \widehat{\pi}(\vec{k}) \, e^{i \int d^d \vec{p} \, \beta(\vec{p}) \left(\widehat{\phi}(\vec{p}) \widehat{\pi}(-\vec{p}) + \widehat{\pi}(-\vec{p}) \widehat{\phi}(\vec{p}) \right) } &= e^{2 \beta(\vec{k})}\, \widehat{\pi}(\vec{k}) \,.
\end{align}
Similar to the $(0+1)$--dimensional case, two viable $U's$ that map $U|\Psi_0\rangle = |\Psi_1\rangle$ are
\begin{align}
\label{GaussianU3}
U &= e^{- i \int d^d \vec{p} \, (\phi_1(-\vec{p}) - \phi_0(-\vec{p})) \, \widehat{\pi}(\vec{p}) } \times e^{-\frac{i}{4} \int d^d \vec{p} \, \log\left(\frac{\widetilde{\Omega}_0(p)}{\widetilde{\Omega}_1(p)}\right)\,\left((\widehat{\phi}(\vec{p}) - \phi_0(\vec{p})) \widehat{\pi}(-\vec{p}) + \widehat{\pi}(-\vec{p})(\widehat{\phi}(\vec{p}) - \phi_0(\vec{p})) \right) } \nonumber \\
&= e^{- \frac{i}{2} \int d^d \vec{p} \,\frac{\log\left( \frac{\widetilde{\Omega}_0(p)}{\widetilde{\Omega}_1(p)}\right) \, \sqrt{\widetilde{\Omega}_1(p)}}{\sqrt{\widetilde{\Omega}_0(p)} - \sqrt{\widetilde{\Omega}_1(p)}}\, (\phi_1(-\vec{p}) - \phi_0(-\vec{p})) \, \widehat{\pi}(\vec{p}) -\frac{i}{4} \int d^d \vec{p} \, \log\left(\frac{\widetilde{\Omega}_0(p)}{\widetilde{\Omega}_1(p)}\right)\,\left((\widehat{\phi}(\vec{p}) - \phi_0(\vec{p})) \widehat{\pi}(-\vec{p}) + \widehat{\pi}(-\vec{p})(\widehat{\phi}(\vec{p}) - \phi_0(\vec{p})) \right)}
\end{align}
and
\begin{align}
\label{GaussianU4}
U &= e^{-\frac{i}{4} \int d^d \vec{p} \, \log\left(\frac{\widetilde{\Omega}_0(p)}{\widetilde{\Omega}_1(p)}\right)\,\left((\widehat{\phi}(\vec{p}) - \phi_1(\vec{p})) \widehat{\pi}(-\vec{p}) + \widehat{\pi}(-\vec{p})(\widehat{\phi}(\vec{p}) - \phi_1(\vec{p})) \right) } \times e^{- i \int d^d \vec{p} \, (\phi_1(-\vec{p}) - \phi_0(-\vec{p})) \, \widehat{\pi}(\vec{p}) } \nonumber \\
&= e^{- \frac{i}{2} \int d^d \vec{p} \,\frac{\log\left( \frac{\widetilde{\Omega}_0(p)}{\widetilde{\Omega}_1(p)}\right) \, \sqrt{\widetilde{\Omega}_0(p)}}{\sqrt{\widetilde{\Omega}_0(p)} - \sqrt{\widetilde{\Omega}_1(p)}}\, (\phi_1(-\vec{p}) - \phi_0(-\vec{p})) \, \widehat{\pi}(\vec{p}) -\frac{i}{4} \int d^d \vec{p} \, \log\left(\frac{\widetilde{\Omega}_0(p)}{\widetilde{\Omega}_1(p)}\right)\,\left((\widehat{\phi}(\vec{p}) - \phi_1(\vec{p})) \widehat{\pi}(-\vec{p}) + \widehat{\pi}(-\vec{p})(\widehat{\phi}(\vec{p}) - \phi_1(\vec{p})) \right)}\,.
\end{align}
Notice that Eqn.'s~\eqref{GaussianU3} and~\eqref{GaussianU4} are structurally similar to Eqn.'s~\eqref{GaussianU1} and~\eqref{GaussianU2} above.  The reason is that, for Gaussian wavefunctionals with translation and rotation-invariant kernels, the Gaussian decouples in momentum space:
\begin{equation}
\text{det}^{\frac{1}{4}}\left(\frac{\widetilde{\Omega}_0}{\pi}\right) \, e^{-\frac{1}{2} \int d^d \vec{k} \, (\phi(\vec{k}) - \phi_0(\vec{k})) \, \widetilde{\Omega}_0(k) \, (\phi(-\vec{k}) - \phi_0(-\vec{k})) } = \prod_{\vec{k}} \left(\frac{d^d \vec{k} \, \widetilde{\Omega}_0(\vec{k})}{\pi}\right)^{\frac{1}{4}} \, e^{-\frac{d^d \vec{k}}{2} \left((\phi(\vec{k}) - \phi_0(\vec{k})) \, \widetilde{\Omega}_0(k) \, (\phi(-\vec{k}) - \phi_0(-\vec{k}))  \right)}\,.
\end{equation}
Hence, the $(d+1)$--dimensional case essentially reduces to many independent copies of the $(0+1)$--dimensional case.  As before, we note that the unitaries in Eqn.'s~\eqref{GaussianU3} and~\eqref{GaussianU4} above have Hermitian generators of order $2$.

\subsubsection{Concrete example: scalar field theories with different masses}

Here we apply the techniques in the previous subsection, which parallel our analysis of harmonic oscillator with different masses above.  We construct a unitary which maps the ground state of a free scalar field theory with mass $m_1$ to the ground state of another free scalar field theory with mass $m_2$.  Recall that the Hamiltonian of a free scalar field of mass $m$ is
\begin{equation}
\label{freeHam1}
H = \frac{1}{2} \int d^d \vec{x} \, \left(\widehat{\pi}(x)^2 + \widehat{\phi}(x) \left(-\nabla^2 + m^2\right) \widehat{\phi}(x) \right)
\end{equation} 
and that its ground state wavefunctional is
\begin{equation}
\label{groundstate0}
\langle \phi |\Psi_0\rangle = \mathcal{N} \, \exp\left(-\frac{1}{2} \int d^d \vec{p} \,\, \phi(\vec{p}) \, \sqrt{\vec{p}^2 + m^2} \, \phi(-\vec{p}) \right)
\end{equation}
where $\mathcal{N}$ is the normalization.  Notice that the kernel in the exponent of Eqn.~\eqref{groundstate0} is related to the inverse equal-time Green's function, since
\begin{equation}
\langle \Psi_0| \,\widehat{\phi}(\vec{p}) \, \widehat{\phi}(\vec{k}) |\Psi_0\rangle = \frac{1}{2 \sqrt{\vec{p}^2 + m^2}} \, \delta^{(d)}(\vec{p}+\vec{k})\,.
\end{equation}
Considering two massive scalar field theories with masses $m_1$ and $m_2$, respectively, their ground states are
\begin{align*}
\langle \phi |\Psi_0^{m_1}\rangle &= \mathcal{N} \, \exp\left(-\frac{1}{2} \int d^d \vec{p} \,\, \phi(\vec{p}) \, \sqrt{\vec{p}^2 + m_1^2} \, \phi(-\vec{p}) \right) \\
\langle \phi |\Psi_0^{m_2}\rangle &= \mathcal{N} \, \exp\left(-\frac{1}{2} \int d^d \vec{p} \,\, \phi(\vec{p}) \, \sqrt{\vec{p}^2 + m_2^2} \, \phi(-\vec{p}) \right)\,.
\end{align*}
In the above equations, we have written $\mathcal{N}$ as a placeholder for the normalization of the wavefunctionals, although the different normalizations are not equal.  We will continue to use this convention for the remainder of the paper.

Now we find a $U$ such that $U|\Psi_0^{m_1}\rangle = |\Psi_0^{m_2}\rangle$.  Considering Eqn.'s~\eqref{GaussianU3} and~\eqref{GaussianU4} above, we let $\phi_0(\vec{p}) = \phi_1(\vec{p}) = 0$, $\widetilde{\Omega}_0(p) = \sqrt{\vec{p}^2 + m_1^2}$ and $\widetilde{\Omega}_1(\vec{p}) = \sqrt{\vec{p}^2 + m_2^2}$\,.  Then both Eqn.'s~\eqref{GaussianU3} and~\eqref{GaussianU4} simplify to the same unitary, namely
\begin{equation}
U = \exp\left(-\frac{i}{4} \int d^d \vec{p} \, \log\left(\frac{\sqrt{\vec{p}^2 + m_1^2}}{\sqrt{\vec{p}^2 + m_2^2}}\right) \, \left(\widehat{\phi}(\vec{p}) \, \widehat{\pi}(-\vec{p}) + \widehat{\pi}(-\vec{p}) \, \widehat{\phi}(\vec{p}) \right) \right)
\end{equation}
which indeed satisfies $U|\Psi_0^{m_1}\rangle = |\Psi_0^{m_2}\rangle$.

\subsection{From Gaussians to non-Gaussians}

In previous sections, we showed how to construct unitaries which map between Gaussian wavefunctions.  These unitaries had Hermitian generators of order $2$, which made various manipulations tractable.  If we want to map between a Gaussian state and a \textit{non}-Gaussian state (or even between two non-Gaussian states), then we generally need a unitary with a Hermitian generator greater than order $2$.  But suppose we want to build a circuit between a Gaussian state and another state which is \textit{mildly} non-Gaussian -- that is, Gaussian plus perturbative corrections.  Then we expect the corresponding circuit to be generated by an order $2$ Hermitian operator plus higher order corrections.  It is tractable to compute such a circuit using the perturbative techniques outlined in the previous sections.

By example, we will construct a unitary which perturbatively maps a Gaussian to the ground state of the quantum anharmonic oscillator.  We also consider the $(d+1)$--dimensional analog, namely constructing a unitary which perturbatively maps a Gaussian wavefunctional to the ground state wavefunctional of scalar $\varphi^4$ theory.

\subsubsection{Anharmonic oscillator}

Let us define the anharmonic oscillator Hamiltonian
\begin{equation}
\label{AnhOsc1}
H = \frac{1}{2m}\,\widehat{p}^2 + \frac{1}{2}\,m \widehat{x}^2 + \lambda \, \widehat{x}^4\,,
\end{equation}
(where we have set $\omega = 1$) with ground state $|\Psi\rangle$ given by
\begin{equation}
\label{AnhGS1}
\langle x | \psi \rangle = \left(\frac{m}{\pi}\right)^{1/4} \, \exp\left(- \frac{1}{2} \, m x^2 + \lambda \,\left(\frac{9}{16 m^2} - \frac{3}{4m} \, x^2 - \frac{1}{4} \, x^4 \right)  \right)\,\,+\mathcal{O}(\lambda^2)\,.
\end{equation}
We take $\lambda$ to be perturbatively small.  We also define a reference Gaussian state $|\psi_0 \rangle$, given by
\begin{equation}
\label{QMref1}
\langle x | \psi_0\rangle = \left(\frac{m_0}{\pi}\right)^{1/4} \, \exp\left(-\frac{1}{2} \, m_0 x^2 \right)\,.
\end{equation}
Our objective is to construct a unitary $U$ which maps
\begin{equation}
U |\psi_0\rangle = |\psi\rangle \,\,+\mathcal{O}(\lambda^2)\,.
\end{equation}
If we compare Eqn.'s~\eqref{AnhGS1} and~\eqref{QMref1}, we see that $U$ must have the form
\begin{equation}
\label{putativeUnitary1}
U = \exp\left(- \frac{i}{4} \, \log\left(\frac{m_0}{m}\right)(\widehat{x} \widehat{p} + \widehat{p} \widehat{x})- i \, \lambda \, [\text{order }4] \right)
\end{equation} 
which is the same form as $\exp(-i(Q_2 + \lambda\, Q_{\text{higher}}))$ in Eqn.~\eqref{Q2plusQhigher} above.  But what is a principled way of determining the $[\text{order }4]$ terms?

Let us denote the $[\text{order }4]$ terms in Eqn.~\eqref{putativeUnitary1} by $Q_4$.  Then $Q_4$ can be expressed in the Bender-Dunne basis as
\begin{equation}
Q_4 = \sum_{r+s \leq 4} c_{r,s} \, T_{r,s}\,.
\end{equation}
for some constants $c_{r,s}$.  Considering Eqn.~\eqref{QhigherExpand1} with $Q_2 = \frac{1}{2} \log(m_0/m) \, T_{1,1}$ and $\epsilon = \lambda$, we observe that $U$ in Eqn.~\eqref{putativeUnitary1} can be written as
\begin{equation}
\label{putativeUnitary2}
U = \left(1 + \lambda \, \frac{e^{-\frac{i}{2} \, \log(m_0/m)\, \text{ad}_{T_{1,1}}}-1}{\frac{1}{2} \log(m_0/m) \, \text{ad}_{T_{1,1}}} \,\, Q_4 \right) \, e^{- \frac{i}{2} \log(m_0/m) \, T_{1,1}} \,\, + \mathcal{O}(\lambda^2)\,.
\end{equation}
Recalling from Eqn.~\eqref{T11comm1} that
\begin{equation}
\text{ad}_{T_{1,1}}(T_{r,s}) = [T_{1,1}, T_{r,s}] = i (r-s) T_{r,s}\,,
\end{equation}
Eqn.~\eqref{putativeUnitary2} can be written as
\begin{equation}
\label{putativeUnitary3}
U = \left(1 - i\, \lambda \, \sum_{r+s\leq 4}\left(\frac{e^{\frac{1}{2} \, \log(m_0/m)\, (r-s)}-1}{\frac{1}{2} \log(m_0/m) \, (r-s)}\right) c_{r,s}\,T_{r,s} \right) \, e^{- \frac{i}{2} \log(m_0/m) \, T_{1,1}}\,\, + \mathcal{O}(\lambda^2)\,.
\end{equation}
For terms where $r=s$, coefficients $\frac{e^{\frac{1}{2} \, \log(m_0/m)\, (r-s)}-1}{\frac{1}{2} \log(m_0/m) \, (r-s)}$ become
\begin{equation}
\frac{e^{\frac{1}{2} \, \log(m_0/m)\, (r-s)}-1}{\frac{1}{2}\log(m_0/m) \, (r-s)} \,\, \xrightarrow{\,\, r = s \,\,}\,\, 1\,.
\end{equation}
Expressing $U$ in the form of Eqn.~\eqref{putativeUnitary3} is \textit{very} useful.  To see why, let us express $U |\psi_0\rangle$ in the position basis.  Since $T_{r,s}$ depends on $\widehat{p},\widehat{x}$, we write $T_{r,s}(\widehat{p},\widehat{x})$.  Then in the position basis, we have
\begin{align}
\label{posbasisexpand1}
&\langle x | U |\psi_0\rangle \nonumber \\
& = \left(1 - i\, \lambda \, \sum_{r+s\leq 4}\left(\frac{e^{\frac{1}{2} \, \log(m_0/m)\, (r-s)}-1}{\frac{1}{2} \log(m_0/m) \, (r-s)}\right) c_{r,s}\,T_{r,s}(\frac{1}{i} \frac{\partial}{\partial x},x) \right) \, e^{- \frac{i}{2} \log(m_0/m) \, T_{1,1}(\frac{1}{i} \frac{\partial}{\partial x},x)} \, \left(\frac{m_0}{\pi} \right)^{1/4} \, e^{-\frac{1}{2} \, m_0 x^2} \nonumber \\
& \qquad \qquad \qquad \qquad \qquad \qquad \qquad \qquad \qquad \qquad \qquad \qquad \qquad \qquad \qquad \qquad \qquad \qquad \qquad \qquad \,\, + \mathcal{O}(\lambda^2) \nonumber \\
& = \left(1 - i\, \lambda \, \sum_{r+s\leq 4}\left(\frac{e^{\frac{1}{2} \, \log(m_0/m)\, (r-s)}-1}{\frac{1}{2} \log(m_0/m) \, (r-s)}\right) c_{r,s}\,T_{r,s}(\frac{1}{i} \frac{\partial}{\partial x},x) \right) \, \left(\frac{m}{\pi} \right)^{1/4} \, e^{-\frac{1}{2} \, m x^2} \,\, + \mathcal{O}(\lambda^2)
\end{align}  
where we have used $e^{- \frac{i}{2} \log(m_0/m) \, T_{1,1}( \frac{1}{i} \frac{\partial}{\partial x},x)} \, \left(\frac{m_0}{\pi} \right)^{1/4} \, e^{-\frac{1}{2} \, m_0 x^2} = \left(\frac{m}{\pi} \right)^{1/4} \, e^{-\frac{1}{2} \, m x^2}$.  In words, we have accounted for the \textit{non-perturbative} part of the mapping from $|\psi_0\rangle$ to $|\psi\rangle$, in which we changed the mass appearing in the leading Gaussian  term.  To determine the $\mathcal{O}(\lambda)$ corrections, we set equal $\langle x | U |\psi_0\rangle + \mathcal{O}(\lambda^2)$ and $\langle x |\psi\rangle + \mathcal{O}(\lambda^2)$ to obtain
\begin{align}
\label{tomatch1}
&\left(1 - i\, \lambda \, \sum_{r+s\leq 4}\left(\frac{e^{\frac{1}{2} \, \log(m_0/m)\, (r-s)}-1}{\frac{1}{2} \log(m_0/m) \, (r-s)}\right) c_{r,s}\,T_{r,s}(\frac{1}{i} \frac{\partial}{\partial x},x) \right) \, \left(\frac{m}{\pi} \right)^{1/4} \, e^{-\frac{1}{2} \, m x^2} \,\,+\mathcal{O}(\lambda^2)\nonumber \\ \nonumber \\
=\,&\left(1 + \lambda \, \left(\frac{9}{16 m^2} - \frac{3}{4m} \, x^2 - \frac{1}{4} \, x^4 \right)  \right) \, \left(\frac{m}{\pi} \right)^{1/4} \, e^{-\frac{1}{2} \, m x^2} \,\,+ \mathcal{O}(\lambda^2)\,,
\end{align}
and determine the real coefficients $c_{r,s}$ by matching the two sides of the equation.

Before carrying out the matching, there are a few simplifications we can make.  Considering the right hand side of Eqn.~\eqref{tomatch1}, we notice that the $\mathcal{O}(\lambda)$ corrections to the Gaussian piece of $\langle x |\psi\rangle$ are strictly real.  However, the left-hand side of Eqn.~\eqref{tomatch1} contains terms which are \textit{not} strictly real: namely, any term with $T_{r,s}(\frac{1}{i} \frac{\partial}{\partial x},x)$ such that $r$ is even, and hence an even number of $\frac{1}{i} \frac{\partial}{\partial x}$\,'s.  To be explicit, since all of the derivatives in each $T_{r,s}(\frac{1}{i} \frac{\partial}{\partial x},x)$ fall on terms of the form $\textit{polynomial}(x)  \cdot e^{-\frac{1}{2} m x^2}$ which are purely real, each $T_{r,s}(\frac{1}{i} \frac{\partial}{\partial x},x)$ carries with it a factor of $(1/i)^r$ which is real when $r$ is even.  Since on the left-hand side of the equation there is an overall factor of $i$ in front of the $\lambda$, terms $T_{r,s}$ with $r$ even are overall imaginary.  Thus, we can throw away those $T_{r,s}$ for which $r$ is even, since their coefficients $c_{r,s}$ must necessarily be zero when we match to the right-hand side.  This leaves us with
$$T_{1,0},\,\, T_{1,1},\,\, T_{1,2},\,\, T_{1,3},\,\, T_{3,0},\,\, T_{3,1}\,.$$
We also notice that on the right-hand side of Eqn.~\eqref{tomatch1}, all of the $\mathcal{O}(\lambda)$ corrections to the Gaussian piece are even powers of $x$.  But considering the left-hand side, we see that the terms containing $T_{1,0}$, $T_{1,2}$ or $T_{3,0}$ will only generate odd powers of $x$ when acting on the Gaussian piece $e^{-\frac{1}{2} \, m x^2}$.  Then we can throw away the terms $T_{1,0}, T_{1,2}, T_{3,0}$ as well, leaving us with just
$$T_{1,1},\,\, T_{1,3},\,\, T_{3,1}\,.$$

With these simplifications in mind, we can rewrite Eqn.~\eqref{tomatch1} as
\begin{align}
\label{tomatch2}
&\left(1 - i\, \lambda \left(c_{1,1}\,T_{1,1} + \frac{(m/m_0)-1}{ \log(m/m_0)} \,c_{1,3}\,T_{1,3} + \frac{(m_0/m)-1}{\log(m_0/m)} \,c_{3,1}\,T_{3,1} \right)\right) \left(\frac{m}{\pi} \right)^{1/4} e^{-\frac{1}{2} \, m x^2} \,\, + \mathcal{O}(\lambda^2)  \nonumber \\
=\,&\left(1 + \lambda \, \left(\frac{9}{16 m^2} - \frac{3}{4m} \, x^2 - \frac{1}{4} \, x^4 \right)  \right) \, \left(\frac{m}{\pi} \right)^{1/4} \, e^{-\frac{1}{2} \, m x^2} \,\,+ \mathcal{O}(\lambda^2)\,.
\end{align}
Finally, recalling that
\begin{align*}
T_{1,3}(\widehat{p},\widehat{x}) &= \frac{1}{4}(\widehat{x}\,\widehat{x}\,\widehat{x}\,\widehat{p} + \widehat{x}\,\widehat{x}\,\widehat{p}\,\widehat{x} + \widehat{x}\,\widehat{p}\,\widehat{x}\,\widehat{x} + \widehat{p}\,\widehat{x}\,\widehat{x}\,\widehat{x}) \\
T_{3,1}(\widehat{p},\widehat{x}) &= \frac{1}{4}(\widehat{p}\,\widehat{p}\,\widehat{p}\,\widehat{x} + \widehat{p}\,\widehat{p}\,\widehat{x}\,\widehat{p} + \widehat{p}\,\widehat{x}\,\widehat{p}\,\widehat{p} + \widehat{x}\,\widehat{p}\,\widehat{p}\,\widehat{p})\,,
\end{align*}
we can expand the left-hand side of Eqn.~\eqref{tomatch2} and match it to the right-hand side to find the one-parameter family of solutions (parameterized by $z$)\,:
\begin{align}
c_{1,1}(z) &=  - \frac{9}{8 m^2} + \frac{3 m(m-m_0)}{m_0 \log(m_0/m)} \, z\\
c_{1,3}(z) &= \frac{\log(m_0/m)}{4(m-m_0)} + \frac{m^3}{m_0} \, z  \\
c_{3,1}(z) &= z \,.
\end{align}
Considering $z=0$, we have the particularly nice solutions
\begin{align}
c_{1,1}(z=0) &=  - \frac{9}{8 m^2} \\
c_{1,3}(z=0) &=  \frac{\log(m_0/m)}{4(m-m_0)} \\
c_{3,1}(z=0) &= 0 \,.
\end{align}
and so our final result is (choosing the $z=0$ solutions)\,:
\begin{equation}
U = \exp\left(-\frac{i}{2}\, \log\left(\frac{m_0}{m}\right) \, T_{1,1} - i \, \lambda \, \left(- \frac{9}{8 m^2} \, T_{1,1} + \frac{\log(m_0/m)}{4(m-m_0)} \, T_{1,3} \right) \right)\,\,+\mathcal{O}(\lambda^2)\,.
\end{equation}

Having gotten our desired answer, let us summarize the procedure:
\begin{enumerate}
\item Determine the order $2$ Hermitian generator $Q_2$ that satisfies $e^{- i Q_2} |\psi_0\rangle = |\psi\rangle + \mathcal{O}(\lambda)$.
\item For a general $Q_4 = \sum_{r+s\leq 4}c_{r,s} \, T_{r,s}$, expand out $e^{- i (Q_2 + \lambda \, Q_4)} |\psi_0\rangle$ to first order in $\lambda$ using Eqn.~\eqref{QhigherExpand1} to obtain
\begin{equation}
\langle x| U |\psi_0\rangle = \left(1 - i \, \lambda \sum_{r+s\leq 4} \widetilde{c}_{r,s} \, T_{r,s}\right) \, e^{-i Q_2}|\psi_0\rangle
\end{equation}
for modified coefficients $\widetilde{c}_{r,s}$ (cf. Eqn.~\eqref{posbasisexpand1}).
\item Match the $\mathcal{O}(\lambda)$ terms of $\langle x| U |\psi_0\rangle$ and $|\psi_1\rangle$ to determine the $\widetilde{c}_{r,s}$\,'s, and in turn the $c_{r,s}$'s.
\end{enumerate}
As we have shown, implementing the procedure is computationally straightforward.  We have chosen to go into many details in this example so that no step seems mysterious.  Next, we will solve an analogous problem for the ground state of scalar $\varphi^4$ theory.  The procedure will be conceptually the same, although the computations will be slightly more involved.
\subsubsection{Scalar $\varphi^4$ theory}
\label{sec:phi4circuit}
Now we upgrade to the field theory case, and consider the ground state wavefunctional of scalar $\varphi^4$ theory.  The Hamiltonian of $\varphi^4$ theory is
\begin{equation}
\label{phi4Ham1}
H = \int d^d \vec{x} \, \left[\frac{1}{2}\left(\widehat{\pi}(x)^2 + \widehat{\phi}(x) \left(-\nabla^2 + m^2\right) \widehat{\phi}(x)\right) + \frac{\lambda}{4!} \, \widehat{\phi}(x)^4 \right]
\end{equation} 
where we take $\lambda$ to be a perturbatively small parameter.  Note that Eqn.~\eqref{phi4Ham1} is the Hamiltonian of a free massive scalar theory up to perturbative corrections, and so its ground state wavefunctional will in turn be the ground state of a free massive scalar theory up to perturbative corrections.

First, we write Eqn.~\eqref{phi4Ham1} in momentum space, as
\begin{align}
H^\Lambda &= \frac{1}{2} \int^\Lambda d^d \vec{k} \, \left(\widehat{\pi}(\vec{k}) \widehat{\pi}(-\vec{k}) + \widehat{\phi}(\vec{k}) \left(\vec{k}^2 + m^2\right) \widehat{\phi}(-\vec{k})\right) \nonumber \\
& \qquad \qquad \qquad + \frac{\lambda}{4!}\frac{1}{(2\pi)^{d}}\int^\Lambda d^d \vec{k}_1 \, d^d \vec{k}_2 \, d^d \vec{k}_3 \, \widehat{\phi}(\vec{k}_1) \widehat{\phi}(\vec{k}_2) \widehat{\phi}(\vec{k}_3) \widehat{\phi}(-\vec{k}_1 - \vec{k}_2 - \vec{k}_3)
\end{align}
where we have imposed a UV cutoff at momentum scale $|\vec{k}| = \Lambda$.  Since it will be useful later, we renormalize the Hamiltonian to scale $\Lambda e^u$, where $-\infty < u \leq 0$.  After performing $1$-loop Wilsonian renormalization on the spatial momentum modes (see Appendix \ref{sec:AppA} for a review of Wilsonian renormalization on spatial momentum modes for scalar $\varphi^4$ theory), we obtain
\begin{align}
\label{phi4Ham1loop}
H_{1-\text{loop}}^{\Lambda e^u} &= \frac{1}{2} \int^\Lambda d^d \vec{k} \, \left(\widehat{\pi}(\vec{k}) \widehat{\pi}(-\vec{k}) + \widehat{\phi}(\vec{k}) \left(\vec{k}^2 + e^{-2u}\,\widetilde{m}^2\right) \widehat{\phi}(-\vec{k})\right) \nonumber \\
& \qquad \qquad + \frac{e^{(d-3)u}\lambda}{4!}\frac{1}{(2\pi)^{d}} \int^\Lambda d^d \vec{k}_1 \, d^d \vec{k}_2 \, d^d \vec{k}_3 \, \widehat{\phi}(\vec{k}_1) \widehat{\phi}(\vec{k}_2) \widehat{\phi}(\vec{k}_3) \widehat{\phi}(-\vec{k}_1 - \vec{k}_2 - \vec{k}_3)
\end{align}
where 
\begin{align}
\label{deltaM1}
\widetilde{m}^2 &= m^2 + \frac{\lambda}{2} \int_{\Lambda e^u}^\Lambda \frac{d^{d}\vec{k}}{(2\pi)^d}  \frac{1}{\vec{k}^2+ m^2}
\\&=: m^2 + \delta m^2
\end{align}
and the explicit form of $\delta m^2$ is given in Appendix \ref{sec:AppA} in equation \eqref{eq:deltam2}.

Using the Hamiltonian in Eqn.~\eqref{phi4Ham1loop}, we can calculate the ground state wavefunctional of $\varphi^4$ theory at scale $\Lambda e^u$ to $1$-loop \cite{Hatfield1}: It is
\begin{align}
\label{phi4gs1}
\langle \phi | \Psi(\Lambda e^u)\rangle &= \mathcal{N} \, \exp\left(- G[\phi] - e^{-2u} \delta m^2 \, R_1[\phi] - e^{(d-3)u} \lambda \, R_2[\phi] \right) \,\, + \mathcal{O}(\lambda^2)
\end{align}
where
\begin{align}
G[\phi] &= \frac{1}{2}\int^\Lambda d^d \vec{k} \, \phi(\vec{k}) \, \omega_k \, \phi(-\vec{k}) \\ \nonumber \\
R_1[\phi]&=\frac{1}{4}\int^\Lambda d^d \vec{k}\, \frac{1}{\omega_k}\,\phi(\vec{k}) \phi(-\vec{k}) \\ \nonumber \\
\begin{split}
R_2[\phi] &=\frac{1}{16}\int^\Lambda d^d \vec{k}\, \frac{1}{\omega_k}\,\left(\int \frac{d^d \vec{q}}{(2\pi)^{d}} \frac{1}{\omega_k + \omega_q} \right)\,\phi(\vec{k}) \phi(-\vec{k})
\\
&+ \frac{1}{24}\frac{1}{(2\pi)^d} \int^\Lambda \frac{d^d \vec{k}_1 \, d^d \vec{k}_2 \, d^d \vec{k}_3}{\omega_{k_1} + \omega_{k_2} + \omega_{k_3} + \omega_{-\vec{k}_1 - \vec{k}_2 - \vec{k}_3}} \, \phi(\vec{k}_1) \phi(\vec{k}_2) \phi(\vec{k}_3) \phi(-\vec{k}_1 - \vec{k}_2 - \vec{k}_3)
\end{split}
\end{align}
and where $\omega_k = \sqrt{\vec{k}^2 + e^{-2u} m^2}$.

Now we introduce a reference Gaussian state
\begin{equation}
\label{wavefunctional0momen1}
\langle \phi |\Psi_0\rangle = \text{det}^{\frac{1}{4}}\left(\frac{\widetilde{\Omega}_0}{\pi}\right) \, \exp\left(-\frac{1}{2} \int d^d \vec{k} \, \phi(\vec{k}) \, \widetilde{\Omega}_0(k) \, \phi(-\vec{k}) \right)
\end{equation}
which is the same as in Eqn.~\eqref{wavefunctional0momen} above, with $\widetilde{\Omega}_0(k) = \sqrt{\vec{k}^2 + m_0^2}$\,.  Similar to before, our goal will be to find a unitary $U$ so that
\begin{equation}
U|\Psi_0\rangle = |\Psi(\Lambda e^u)\rangle \,\,+ \mathcal{O}(\lambda^2)\,.
\end{equation}
Comparing Eqn.'s~\eqref{phi4gs1} and~\eqref{wavefunctional0momen1}, we see that $U$ must have the form
\begin{equation}
\label{putativeQFTunitary1}
U = \exp\left(-\frac{i}{4} \int d^d \vec{p} \, \log\left(\frac{\sqrt{\vec{p}^2 + m_0^2}}{\sqrt{\vec{p}^2 + e^{-2u} m^2}}\right) \left(\widehat{\phi}(\vec{p}) \, \widehat{\pi}(-\vec{p}) + \widehat{\pi}(-\vec{p}) \, \widehat{\phi}(\vec{p}) \right) - i\,\lambda \,[\text{order }4]\right)
\end{equation}
which is the same form as $\exp(-i(Q_2 + \lambda\, Q_{\text{higher}}))$ in Eqn.~\eqref{Q2plusQhigher} above.  Next, we apply the same general strategy we used for the anharmonic oscillator above.

Denoting the $[\text{order }4]$ terms in Eqn.~\eqref{putativeQFTunitary1} by $Q_4$, we most generally have
\begin{equation}
\label{Q4QFT1}
Q_4 = \sum_{r=0}^4 \sum_{s=0}^r \int d^d \vec{k}_1 \cdots d^d \vec{k}_r \, c_r^{(s)}(\vec{k}_1,...,\vec{k}_r) \, S_r^{(s)}(\vec{k}_1,...,\vec{k}_r) 
\end{equation}
where
\begin{equation}
S_r^{(s)}(\vec{k}_1,...,\vec{k}_r) = \widehat{\phi}(\vec{k}_1) \cdots \widehat{\phi}(\vec{k}_s) \widehat{\pi}(\vec{k}_{s+1}) \cdots \widehat{\pi}(\vec{k}_r) + \widehat{\pi}(\vec{k}_{s+1}) \cdots \widehat{\pi}(\vec{k}_r)\widehat{\phi}(\vec{k}_1) \cdots \widehat{\phi}(\vec{k}_s)
\end{equation}
is the momentum space analog of Eqn.~\eqref{fieldbasis1}.  Recall the notation for $S_r^{(s)}$\,: here, $r$ denotes the total number of operators per term, and $s$ denotes the number of $\widehat{\phi}$\,'s per term.  We will not need all of the terms in $Q_4$ in Eqn.~\eqref{Q4QFT1} -- since the ground state of scalar $\varphi^4$ theory to $\mathcal{O}(\lambda)$ is strictly real and an even functional of $\phi$, the only terms in $Q_4$ with non-zero $c_r^{(s)}$\,'s are
$$S_2^{(1)}(\vec{k}_1, \vec{k}_2)\,,\,\, S_4^{(1)}(\vec{k}_1,\vec{k}_2,\vec{k}_3,\vec{k}_4)\,,\,\,S_4^{(3)}(\vec{k}_1,\vec{k}_2,\vec{k}_3,\vec{k}_4)\,.$$
The argument is essentially the same as in the anharmonic oscillator case above, in which only $T_{1,1}, \, T_{1,3},\,T_{3,1}$ contributed to the unitary to first order in the anharmonic coupling.

Since we only have to work with $S_2^{(1)}(\vec{k}_1, \vec{k}_2),\, S_4^{(1)}(\vec{k}_1,\vec{k}_2,\vec{k}_3,\vec{k}_4),\,S_4^{(3)}(\vec{k}_1,\vec{k}_2,\vec{k}_3,\vec{k}_4)$, let us specialize our notation.  We define
\begin{align}
K_{2,0} &:= - \int d^d \vec{k}_1 \, d^d \vec{k}_2 \, \delta^{(d)}(\vec{k}_1 + \vec{k}_2)\,g_{2,0}(\vec{k}_1) \, S_2^{(1)}(\vec{k}_1, \vec{k}_2) \\
K_{2,1} &:= - \int d^d \vec{k}_1 \, d^d \vec{k}_2 \,\delta^{(d)}(\vec{k}_1 + \vec{k}_2)\, g_{2,1}(\vec{k}_1) \, S_2^{(1)}(\vec{k}_1, \vec{k}_2) \\
K_{4} &:= \int d^d \vec{k}_1 \, d^d \vec{k}_2 \, d^d \vec{k}_3 \, d^d \vec{k}_4 \, \delta^{(d)}(\vec{k}_1 + \vec{k}_2 + \vec{k}_3 + \vec{k}_4) \bigg( g_{4}^{(1)}(\vec{k}_1,\vec{k}_2,\vec{k}_3,\vec{k}_4) \, S_4^{(1)}(\vec{k}_1, \vec{k}_2, \vec{k}_3, \vec{k}_4) \nonumber \\
& \qquad \qquad \qquad \qquad \qquad \qquad \qquad \qquad \qquad \qquad \qquad +  g_{4}^{(3)}(\vec{k}_1,\vec{k}_2, \vec{k}_3, \vec{k}_4) \, S_4^{(3)}(\vec{k}_1, \vec{k}_2, \vec{k}_3, \vec{k}_4) \bigg)\,.
\end{align}
If we set
\begin{align}
g_{2,0}(\vec{k}) &= \frac{1}{4} \log \left(\frac{\widetilde{\Omega}_0(k)}{\sqrt{\vec{k}^2 + e^{-2u} m^2}}\right) \nonumber \\
&= \frac{1}{4} \log \left(\frac{\widetilde{\Omega}_0(k)}{\omega_k}\right)
\end{align}
then our unitary $U$ will take the form
\begin{equation}
\label{QFTUnitary1}
U = \exp\left(i \, K_{2,0} + i \, \lambda \left(K_{2,1} + K_4 \right) \right)\,.
\end{equation}
Our goal is to solve for $g_{2,1}$, $g_{4}^{(1)}$, and $g_4^{(3)}$ so that $U|\Psi_0\rangle = |\Psi(\Lambda e^u)\rangle + \mathcal{O}(\lambda^2)$.

We have the commutation relations
\begin{align}
\label{QFTcomm1}
[K_{2,0}, K_{2,1}] &= 0 \\
\label{QFTcomm2}
[K_{2,0}, K_4] &= i\int d^d \vec{k}_1 \, d^d \vec{k}_2 \, d^d \vec{k}_3 \, d^d \vec{k}_4 \, \delta^{(d)}(\vec{k}_1 + \vec{k}_2 + \vec{k}_3 + \vec{k}_4) \bigg(\mathcal{G}_1\, g_{4}^{(1)}\, S_4^{(1)} +  \mathcal{G}_3\, g_{4}^{(3)} \, S_4^{(3)} \bigg)\,,
\end{align}
where
\begin{align}
\mathcal{G}_1(\vec{k}_1,\vec{k}_2,\vec{k}_3,\vec{k}_4) &:= 2\big(g_{2,0}(\vec{k}_1) - g_{2,0}(\vec{k}_2) - g_{2,0}(\vec{k}_3) - g_{2,0}(\vec{k}_4)\big) \\
\mathcal{G}_3(\vec{k}_1,\vec{k}_2,\vec{k}_3,\vec{k}_4) &:= 2\big(g_{2,0}(\vec{k}_1) + g_{2,0}(\vec{k}_2) + g_{2,0}(\vec{k}_3) - g_{2,0}(\vec{k}_4)\big)\,.
\end{align}
Using the commutation relations in Eqn.'s~\eqref{QFTcomm1} and~\eqref{QFTcomm2} along with Eqn.~\eqref{QhigherExpand1}, we can simplify the unitary $U$ in Eqn.~\eqref{QFTUnitary1} as
\begin{equation}
\label{QFTUnitary2}
U = \left(1 + i \lambda \, K_{2,1} - i \lambda \int d^d \textbf{k} \, \delta^{(d)}(\textbf{k}) \left(\frac{e^{-\mathcal{G}_1}-1}{\mathcal{G}_1}\,g_4^{(1)} \, S_{4}^{(1)} + \frac{e^{-\mathcal{G}_3}-1}{\mathcal{G}_3} \, g_4^{(3)} \, S_{4}^{(3)}\right) \right) \, e^{i K_{2,0}}\,\,+\mathcal{O}(\lambda^2)\,.
\end{equation}
where $d^d \textbf{k} := \prod_{i=1}^d d^d \vec{k}_i$ and $\delta^{(d)}(\textbf{k}) := \delta^{(d)}(\vec{k}_1+\vec{k}_2+\vec{k}_3+\vec{k}_4)$.  We notice that Eqn.~\eqref{QFTUnitary2} above has the same form as Eqn.~\eqref{putativeUnitary3} in the anharmonic oscillator case, which is indeed no coincidence.  It is convenient to define
\begin{align}
\widetilde{g}_4^{(j)}(\vec{k}_1, \vec{k}_2, \vec{k}_3, \vec{k}_4) &:= \frac{e^{- \mathcal{G}_j(\vec{k}_1,\vec{k}_2,\vec{k}_3,\vec{k}_4)}-1}{\mathcal{G}_j(\vec{k}_1,\vec{k}_2,\vec{k}_3,\vec{k}_4)} \, g_4^{(j)}(\vec{k}_1, \vec{k}_2, \vec{k}_3, \vec{k}_4)
\end{align}
for $j=1,3$ so that Eqn.~\eqref{QFTUnitary2} is simply
\begin{align}
\label{QFTUnitary3}
U &= \bigg(1 - i \lambda \int d^d \vec{k}_1 \, d^d \vec{k}_2 \, \delta^{(d)}(\vec{k}_1 + \vec{k}_2) \, g_{2,1}(\vec{k}_1) \, S_{2}^{(1)}(\vec{k}_1,\vec{k}_2) \nonumber \\
&\qquad \qquad \qquad \qquad \qquad - i \lambda \int d^d \textbf{k} \, \delta^{(d)}(\textbf{k}) \left(\widetilde{g}_4^{(1)} \, S_{4}^{(1)} + \widetilde{g}_4^{(3)} \, S_{4}^{(3)}\right) \bigg) \, e^{i K_{2,0}}\,\,+\mathcal{O}(\lambda^2)\,.
\end{align}
Applying $U$ to the Gaussian state $|\Psi_0\rangle$ and expressing the result in the $\phi$--basis, we obtain
\begin{align}\label{eqq:UVstate}
&\langle \phi| U |\Psi_0\rangle \nonumber \\
&=
\Bigg\{1-\lambda\bigg(\int^{\Lambda} d^d\vec{k}_1\,
g_{2,1}(\vec{k}_1)-\int^{\Lambda} d^d\vec{k}_1\,d^d\vec{k}_2\,\omega_{k_{2}}\hspace{1mm} \nonumber
\\&
\hspace{4.7cm}\times\left[\tilde{g}^{(1)}_4(\vec{k}_1,\vec{k}_1,\vec{k}_2,\vec{k}_2)+\tilde{g}^{(1)}_4(\vec{k}_1,\vec{k}_2,\vec{k}_1,\vec{k}_2)+\tilde{g}^{(1)}_4(\vec{k}_1,\vec{k}_2,\vec{k}_2,\vec{k}_1)\right]\bigg)\delta^{(d)}(0)
\nonumber \\ \nonumber \\
&
\hspace{1cm}+\lambda\bigg(\int^{\Lambda} d^d\vec{k}_1\,
2\omega_{k_1}\, g_{2,1}(\vec{k}_1)\phi(\vec{k}_1) \phi(-\vec{k}_1)
\nonumber 
\\ &
\hspace{1.7cm}-\int^{\Lambda} d^d\vec{k}_1\,d^d\vec{k}_2\, 
\bigg[
2\omega_{k_1}\omega_{k_2}\left[\tilde{g}^{(1)}_4(\vec{k}_1,\vec{k}_1,\vec{k}_2,\vec{k}_2)+\tilde{g}^{(1)}_4(\vec{k}_1,\vec{k}_2,\vec{k}_1,\vec{k}_2)+\tilde{g}^{(1)}_4(\vec{k}_1,\vec{k}_2,\vec{k}_2,\vec{k}_1)\right]
\nonumber
\\ & 
\hspace{4.4cm}+\omega_{k_1}^2 \left[\tilde{g}^{(1)}_4(\vec{k}_2,\vec{k}_2,\vec{k}_1,\vec{k}_1)+\tilde{g}^{(1)}_4(\vec{k}_2,\vec{k}_1,\vec{k}_2,\vec{k}_1)+\tilde{g}^{(1)}_4(\vec{k}_2,\vec{k}_1,\vec{k}_1,\vec{k}_2)\right]
\nonumber
\\ & 
\hspace{4.4cm}-\left[\tilde{g}^{(3)}_4(\vec{k}_1,\vec{k}_1,\vec{k}_2,\vec{k}_2)+\tilde{g}^{(3)}_4(\vec{k}_1,\vec{k}_2,\vec{k}_1,\vec{k}_2)+\tilde{g}^{(3)}_4(\vec{k}_2,\vec{k}_1,\vec{k}_1,\vec{k}_2)\right]
\bigg]\phi(\vec{k}_1) \phi(-\vec{k}_1)
\bigg)
\nonumber \\ \nonumber
\\&
\hspace{1cm}+ 2\,\lambda \int^\Lambda d^d \textbf{k} \, \delta^{(d)}(\textbf{k}) \big(\omega_{k_2}\omega_{k_3}\omega_{k_4}\, \widetilde{g}_{4}^{(1)}(\vec{k}_1,\vec{k}_2,\vec{k}_3,\vec{k}_4)  \nonumber \\
& \qquad \,\,\, \qquad \qquad \qquad \qquad \qquad - \omega_{k_4}\,\widetilde{g}_{4}^{(3)}(\vec{k}_1,\vec{k}_2,\vec{k}_3,\vec{k}_4)\big) \, \phi(\vec{k}_1)\phi(\vec{k}_2)\phi(\vec{k}_3)\phi(\vec{k}_4)
\Bigg\}
\nonumber \\
& \qquad \qquad \qquad \qquad \qquad \qquad \qquad \qquad \qquad \qquad \qquad \qquad\qquad\times \mathcal{N} \, e^{-\frac{1}{2}\int^\Lambda d^d \vec{k}_1 \, \phi(\vec{k}_1) \, \omega_{k_1} \, \phi(-\vec{k}_1)} + \mathcal{O}(\lambda^2)\,, \nonumber \\ \nonumber \\
\end{align}
The ground state wavefunctional in Eqn.~\eqref{phi4gs1} can be written in the $\phi$--basis as
\begin{align}
\label{phi4phiBasis1}
&\langle \phi|\Psi(\Lambda e^u)\rangle \nonumber \\
&= \bigg\{1 + \lambda \left(\frac{e^{-2u} (\delta m^2/\lambda)}{8}\int^\Lambda \frac{d^d \vec{k}_1}{\omega_{k_1}^2} + \frac{e^{(d-3)u}}{32}\frac{1}{(2\pi)^{d}} \int^\Lambda d^d \vec{k}_1 \, d^d \vec{k}_2\, \left(\frac{1}{\omega_{k_1}^2 \omega_{k_2}} - \frac{1}{2} \frac{1}{\omega_{k_1} \omega_{k_2}(\omega_{k_1} + \omega_{k_2})}  \right) \right) \delta^{(d)}(0) \nonumber \\ \nonumber \\
& \qquad \,\,\, - \lambda\bigg(\frac{e^{-2u}(\delta m^2/\lambda)}{4}\int^\Lambda \frac{d^d \vec{k}_1}{\omega_{k_1}} \cdot \phi(\vec{k}_1) \phi(-\vec{k}_1) + \frac{e^{(d-3)u}}{16}\frac{1}{(2\pi)^{d}}\int^\Lambda \frac{d^d \vec{k}_1 \, d^d \vec{k}_2}{\omega_{k_1}(\omega_{k_1} + \omega_{k_2})}\cdot \phi(\vec{k}_1)\phi(-\vec{k}_1) \bigg) \nonumber \\ \nonumber \\
& \qquad \,\,\, - \frac{\lambda}{24}\frac{e^{(d-3)u}}{(2\pi)^{d}} \int^\Lambda d^d \textbf{k} \, \delta^{(d)}(\textbf{k}) \, \frac{1}{\omega_{k_1} + \omega_{k_2} + \omega_{k_3} + \omega_{k_4}} \cdot \phi(\vec{k}_1)\phi(\vec{k}_2)\phi(\vec{k}_3)\phi(\vec{k}_4) \bigg\} \nonumber \\ \nonumber \\
& \qquad \qquad \qquad \qquad \qquad \qquad \qquad \qquad \qquad \qquad \qquad \qquad \times \mathcal{N} \, e^{-\frac{1}{2}\int^\Lambda d^d \vec{k}_1 \, \phi(\vec{k}_1) \, \omega_{k_1} \, \phi(-\vec{k}_1)} + \mathcal{O}(\lambda^2)\,, \nonumber \\ \nonumber \\
\end{align}
which is convenient to rewrite as
\begin{align}
\label{phi4phiBasis2}
&\langle \phi|\Psi(\Lambda e^u)\rangle \nonumber \\
&= \bigg\{1 + \lambda \bigg(\frac{e^{-2u} (\delta m^2/\lambda)}{8}\int^\Lambda d^d \vec{k}_1 \,\frac{1}{\omega_{k_1}^2} + \nonumber \\
& \qquad \qquad \qquad  +  \frac{e^{(d-3)u}}{32} \frac{1}{(2\pi)^{d}}\int^\Lambda d^d \vec{k}_1 \, d^d \vec{k}_2\, \left(\frac{1}{\omega_{k_1}^2 (\omega_{k_1}+\omega_{k_2})} + \frac{1}{2} \frac{1}{\omega_{k_1} \omega_{k_2}(\omega_{k_1} + \omega_{k_2})}  \right) \bigg) \delta^{(d)}(0) \nonumber \\ \nonumber \\
& \qquad \,\,\, - \lambda \bigg(\frac{e^{-2u}(\delta m^2/\lambda)}{4}\int^\Lambda d^d \vec{k}_1 \, \frac{1}{\omega_{k_1}} \cdot \phi(\vec{k}_1) \phi(-\vec{k}_1) \nonumber \\
& \qquad \qquad \quad \,\,\,\, + \frac{e^{(d-3)u}}{16}\frac{1}{(2\pi)^{d}}\int^\Lambda d^d \vec{k}_1 \, d^d \vec{k}_2 \, \frac{1}{\omega_{k_1}(\omega_{k_1} + \omega_{k_2})}\cdot \phi(\vec{k}_1)\phi(-\vec{k}_1) \bigg) \nonumber \\ \nonumber \\
& \qquad \,\,\, - \frac{\lambda}{24} \frac{e^{(d-3)u}}{(2\pi)^{d}}\int^\Lambda d^d \textbf{k} \, \delta^{(d)}(\textbf{k}) \, \frac{1}{\omega_{k_1} + \omega_{k_2} + \omega_{k_3} + \omega_{k_4}} \cdot \phi(\vec{k}_1)\phi(\vec{k}_2)\phi(\vec{k}_3)\phi(\vec{k}_4) \bigg\} \nonumber \\ \nonumber \\
& \qquad \qquad \qquad \qquad \qquad \qquad \qquad \qquad \qquad \qquad \qquad \qquad \times \mathcal{N} \, e^{-\frac{1}{2}\int^\Lambda d^d \vec{k}_1 \, \phi(\vec{k}_1) \, \omega_{k_1} \, \phi(-\vec{k}_1)} + \mathcal{O}(\lambda^2)\,, \nonumber \\ \nonumber \\
\end{align}
where we have rearranged the two terms in the $\int^\Lambda d^d \vec{k}_1 \, d^d \vec{k}_2$ integral which multiplies the $\delta^{(d)}(0)$ term.  Since $\delta m^2$ is proportional to the perturbative coupling $\lambda$ by Eqn.~\eqref{deltaM1}, we find that Eqn.~\eqref{phi4phiBasis2} above has the form
\begin{align}
&\langle \phi|\Psi(\Lambda e^u)\rangle \nonumber \\
&= \bigg\{1 + \lambda\, \bigg((\cdots) \, \delta^{(d)}(0) + \int^\Lambda d^d \vec{k}_1 \, (\cdots) \, \phi(\vec{k}_1) \phi(-\vec{k}_1)\nonumber \\
& \qquad \qquad \qquad + \int^\Lambda d^d \textbf{k} \, \delta^{(d)}(\textbf{k}) \, (\cdots) \, \phi(\vec{k}_1)\phi(\vec{k}_2)\phi(\vec{k}_3)\phi(\vec{k}_4) \bigg)\bigg\} \, \times \mathcal{N} \, e^{-\frac{1}{2}\int^\Lambda d^d \vec{k}_1 \, \phi(\vec{k}_1) \, \omega_{k_1} \, \phi(-\vec{k}_1)}+ \mathcal{O}(\lambda^2)
\end{align}
which is exactly the form of Eqn.~\eqref{eqq:UVstate}.  Matching the $\delta^{(d)}(0)$ term, the $\phi\,\phi$ term, and the $\phi\,\phi\,\phi\,\phi$ term between Eqn.~\eqref{eqq:UVstate} and Eqn.~\eqref{phi4phiBasis2}, we obtain three ``matching'' equations:
\begin{align}\label{new.one}
-\int^\Lambda d^d \vec{k}_1 \,g_{2,1}(\vec{k}_1)+\int^\Lambda d^d \vec{k}_1 \, d^d \vec{k}_2 \,\,\omega_{k_2}\left(\tilde{g}^{(1)}_4(\vec{k}_1,\vec{k}_1,\vec{k}_2,\vec{k}_2)+\tilde{g}^{(1)}_4(\vec{k}_1,\vec{k}_2,\vec{k}_1,\vec{k}_2)+\tilde{g}^{(1)}_4(\vec{k}_1,\vec{k}_2,\vec{k}_2,\vec{k}_1)\right)
\nonumber \\
&
\hspace{-17cm}=\frac{e^{-2u} (\delta m^2/\lambda)}{8}\int^\Lambda d^d \vec{k}_1 \,\frac{1}{\omega_{k_1}^2} +  \frac{e^{(d-3)u} \lambda}{32}\frac{1}{(2\pi)^{d}} \int^\Lambda d^d \vec{k}_1 \, d^d \vec{k}_2\, \left(\frac{1}{\omega_{k_1}^2 (\omega_{k_1}+\omega_{k_2})} + \frac{1}{2} \frac{1}{\omega_{k_1} \omega_{k_2}(\omega_{k_1} + \omega_{k_2})}  \right)
\nonumber \\ \nonumber\\
&
\hspace{-17cm}2\omega_{k_1}\, g_{2,1}(\vec{k}_1)-\int^{\Lambda} d^d\vec{k}_2\, 
\bigg(
2\omega_{k_1}\omega_{k_2}\left[\tilde{g}^{(1)}_4(\vec{k}_1,\vec{k}_1,\vec{k}_2,\vec{k}_2)+\tilde{g}^{(1)}_4(\vec{k}_1,\vec{k}_2,\vec{k}_1,\vec{k}_2)+\tilde{g}^{(1)}_4(\vec{k}_1,\vec{k}_2,\vec{k}_2,\vec{k}_1)\right]
\nonumber \\ & \hspace{-12.5cm} 
+\omega_{k_1}^2 \left[\tilde{g}^{(1)}_4(\vec{k}_2,\vec{k}_2,\vec{k}_1,\vec{k}_1)+\tilde{g}^{(1)}_4(\vec{k}_2,\vec{k}_1,\vec{k}_2,\vec{k}_1)+\tilde{g}^{(1)}_4(\vec{k}_2,\vec{k}_1,\vec{k}_1,\vec{k}_2)\right]
\nonumber \\ & \hspace{-12.5cm} 
-\left[\tilde{g}^{(3)}_4(\vec{k}_1,\vec{k}_1,\vec{k}_2,\vec{k}_2)+\tilde{g}^{(3)}_4(\vec{k}_1,\vec{k}_2,\vec{k}_1,\vec{k}_2)+\tilde{g}^{(3)}_4(\vec{k}_2,\vec{k}_1,\vec{k}_1,\vec{k}_2)\right]
\bigg)
\nonumber \\ & \hspace{-12cm} 
=-\frac{e^{-2u}\left(\delta m^2/\lambda\right)}{4}\frac{1}{\omega_{k_1}}
-\frac{e^{(d-3)u}}{16}\frac{1}{(2\pi)^d}\int^{\Lambda}  \frac{d^d\vec{k}_2}{\omega_{k_1}(\omega_{k_1}+\omega_{k_2})} 
\nonumber\\ \nonumber\\&
\hspace{-17cm} 
2\left(\omega_{k_2}\omega_{k_3}\omega_{k_4}\tilde{g}^{(1)}_4(\vec{k}_1,\vec{k}_2,\vec{k}_3,\vec{k}_4)-\omega_{k_4}\hspace{.5mm}\tilde{g}^{(3)}_4(\vec{k}_1,\vec{k}_2,\vec{k}_3,\vec{k}_4)\right)
=
-\frac{e^{(d-3)u}}{24}\frac{1}{(2\pi)^d}\frac{1}{\omega_{k_1}+\omega_{k_2}+\omega_{k_3}+\omega_{k_4}}.\nonumber\\\nonumber\\
\end{align}
Since $S_4^{(1)}(\vec{k}_1,\vec{k}_2,\vec{k}_3,\vec{k}_4)$ is symmetric in $\vec{k}_2,\vec{k}_3,\vec{k}_4$, we ansatz that $g_{4}^{(1)}(\vec{k}_1,\vec{k}_2,\vec{k}_3,\vec{k}_4)$ is also symmetric in $\vec{k}_2,\vec{k}_3,\vec{k}_4$.  Since $\mathcal{G}_1$ is also symmetric in $\vec{k}_2,\vec{k}_3,\vec{k}_4$, our ansatz implies that $\widetilde{g}_{4}^{(1)}(\vec{k}_1,\vec{k}_2,\vec{k}_3,\vec{k}_4)$ is likewise symmetric in $\vec{k}_2,\vec{k}_3,\vec{k}_4$.  Similarly, $S_4^{(3)}(\vec{k}_1,\vec{k}_2,\vec{k}_3,\vec{k}_4)$ is symmetric in $\vec{k}_1,\vec{k}_2,\vec{k}_3$, and so we ansatz that $g_{4}^{(3)}(\vec{k}_1,\vec{k}_2,\vec{k}_3,\vec{k}_4)$ is likewise symmetric in $\vec{k}_1,\vec{k}_2,\vec{k}_3$.  Analogous to the previous case, $\mathcal{G}_3$ is also symmetric in $\vec{k}_1,\vec{k}_2,\vec{k}_3$, and so our ansatz implies that $\widetilde{g}_{4}^{(3)}(\vec{k}_1,\vec{k}_2,\vec{k}_3,\vec{k}_4)$ is also symmetric in $\vec{k}_1,\vec{k}_2,\vec{k}_3$.

We also assume that $g_4^{(1)}$ and $g_4^{(3)}$ are even functions with respect to each of their arguments (i.e., they are invariant under $\vec{k}_i \to - \vec{k}_i$ for any fixed $i=1,2,3,4$).  Since $\mathcal{G}_1$ and $\mathcal{G}_3$ are even in each of their arguments, we have that $\widetilde{g}_4^{(1)}$ and $\widetilde{g}_4^{(3)}$ are also even in each of their arguments.  These assumptions will make the next computations easier to follow.  With these considerations, Eqn.'s~\eqref{eqq:UVstate} and~\eqref{new.one} reduce to
\begin{align}
\label{phi4genericCircuit1}
&\langle \phi| U |\Psi_0\rangle \nonumber \\
&= \bigg\{1 - \lambda \left(\int^\Lambda d^d \vec{k}_1 \, g_{2,1}(\vec{k}_1) - 3 \int^\Lambda d^d \vec{k}_1 \, d^d \vec{k}_2 \, \omega_{k_2}\,\widetilde{g}_{4}^{(1)}(\vec{k}_1,\vec{k}_1,\vec{k}_2,\vec{k}_2) \right) \, \delta^{(d)}(0) \nonumber \\ \nonumber \\
& \qquad \,\,\, + \lambda \bigg(2\int^\Lambda d^d \vec{k}_1 \, \omega_{k_1}\,g_{2,1}(\vec{k}_1) \, \phi(\vec{k}_1) \phi(-\vec{k}_1) \nonumber \\
& \qquad \qquad \quad \,\,\,\, + 3\int^\Lambda d^d \vec{k}_1 \, d^d \vec{k}_2 \big(- 2 \,\omega_{k_1} \omega_{k_2}\, \widetilde{g}_4^{(1)}(\vec{k}_1,\vec{k}_1,\vec{k}_2,\vec{k}_2) - \,\omega_{k_1}^2\, \widetilde{g}_4^{(1)}(\vec{k}_2,\vec{k}_2,\vec{k}_1,\vec{k}_1) \nonumber \\
& \qquad \qquad \qquad \qquad \qquad \qquad \qquad \qquad \qquad \qquad \qquad \qquad \qquad \quad + \,\widetilde{g}_4^{(3)}(\vec{k}_1,\vec{k}_1,\vec{k}_2,\vec{k}_2)\big) \, \phi(\vec{k}_1)\phi(-\vec{k}_1) \bigg) \nonumber \\
& \qquad \,\,\, + 2\,\lambda \int^\Lambda d^d \textbf{k} \, \delta^{(d)}(\textbf{k}) \big(\omega_{k_2}\omega_{k_3}\omega_{k_4}\, \widetilde{g}_{4}^{(1)}(\vec{k}_1,\vec{k}_2,\vec{k}_3,\vec{k}_4)  \nonumber \\
& \qquad \,\,\, \qquad \qquad \qquad \qquad \qquad - \omega_{k_4}\,\widetilde{g}_{4}^{(3)}(\vec{k}_1,\vec{k}_2,\vec{k}_3,\vec{k}_4)\big) \, \phi(\vec{k}_1)\phi(\vec{k}_2)\phi(\vec{k}_3)\phi(\vec{k}_4) \bigg\} \nonumber \\
& \qquad \qquad \qquad \qquad \qquad \qquad \qquad \qquad \qquad \qquad \qquad \qquad \times \mathcal{N} \, e^{-\frac{1}{2}\int^\Lambda d^d \vec{k}_1 \, \phi(\vec{k}_1) \, \omega_{k_1} \, \phi(-\vec{k}_1)} + \mathcal{O}(\lambda^2)\,. \nonumber \\ \nonumber \\
\end{align}
and
\begin{align}\label{final.eq}
&-\int^\Lambda d^d \vec{k}_1 \, g_{2,1}(\vec{k}_1) + 3 \int^\Lambda d^d \vec{k}_1 \, d^d \vec{k}_2 \, \omega_{k_2}\,\widetilde{g}_{4}^{(1)}(\vec{k}_1,\vec{k}_1,\vec{k}_2,\vec{k}_2) \nonumber \\
&= \frac{e^{-2u} (\delta m^2/\lambda)}{8}\int^\Lambda d^d \vec{k}_1 \,\frac{1}{\omega_{k_1}^2} +  \frac{e^{(d-3)u} \lambda}{32}\frac{1}{(2\pi)^{d}} \int^\Lambda d^d \vec{k}_1 \, d^d \vec{k}_2\, \left(\frac{1}{\omega_{k_1}^2 (\omega_{k_1}+\omega_{k_2})} + \frac{1}{2} \frac{1}{\omega_{k_1} \omega_{k_2}(\omega_{k_1} + \omega_{k_2})}  \right) \nonumber \\ \nonumber \\
&
2\omega_{k_1}\, g_{2,1}(\vec{k}_1)-3\int^{\Lambda} d^d\vec{k}_2\, 
\bigg(
2\omega_{k_1}\omega_{k_2}\hspace{.5mm}\tilde{g}^{(1)}_4(\vec{k}_1,\vec{k}_1,\vec{k}_2,\vec{k}_2)
+\omega_{k_1}^2 \tilde{g}^{(1)}_4(\vec{k}_2,\vec{k}_2,\vec{k}_1,\vec{k}_1)
-\tilde{g}^{(3)}_4(\vec{k}_1,\vec{k}_1,\vec{k}_2,\vec{k}_2)
\bigg)
\nonumber \\ & = -\frac{e^{-2u}(\delta m^2/\lambda)}{4} \, \frac{1}{\omega_{k_1}} - \frac{e^{(d-3)u} \lambda}{16}\frac{1}{(2\pi)^{d}}\int^\Lambda d^d \vec{k}_2 \, \frac{1}{\omega_{k_1}(\omega_{k_1} + \omega_{k_2})}
\nonumber \\ \nonumber \\
&
2 \, \delta^{(d)}(\textbf{k}) \big(\omega_{k_2}\omega_{k_3}\omega_{k_4}\, g_{4}^{(1)}(\vec{k}_1,\vec{k}_2,\vec{k}_3,\vec{k}_4) - 
\omega_{k_4}\,\widetilde{g}_{4}^{(3)}(\vec{k}_1,\vec{k}_2,\vec{k}_3,\vec{k}_4)\big) \nonumber\\
&= - \frac{1}{24} \frac{e^{(d-3)u}}{(2\pi)^{d}}\, \delta^{(d)}(\textbf{k}) \, \frac{1}{\omega_{k_1} + \omega_{k_2} + \omega_{k_3} + \omega_{k_4}}\,. 
\end{align}
The equations (\ref{final.eq}) are solved by
\begin{align}\label{eq:gs}
g_{2,1}(\vec{k}_1) &= - \frac{1}{\omega_{k_1}^2} \left(\frac{e^{-2u} (\delta m^2/\lambda)}{8} + \frac{e^{(d-3)u}}{32} \frac{1}{(2\pi)^{d}}\int d^d \vec{k}_2 \, \frac{1}{\omega_{k_1} + \omega_{k_2}} \right) \\  \nonumber \\
\widetilde{g}_{4}^{(1)}(\vec{k}_1,\vec{k}_2,\vec{k}_3,\vec{k}_4) &= \frac{1}{96}\frac{e^{(d-3)u}}{(2\pi)^{d}}\,\frac{1}{\omega_{k_2} \omega_{k_3} \omega_{k_4}(\omega_{k_1} + \omega_{k_2} + \omega_{k_3} + \omega_{k_4})} \\ \nonumber \\
\widetilde{g}_{4}^{(3)}(\vec{k}_1,\vec{k}_2,\vec{k}_3,\vec{k}_4) &= \frac{1}{32}\frac{e^{(d-3)u}}{(2\pi)^{d}}\,\frac{1}{\omega_{k_4}(\omega_{k_1} + \omega_{k_2} + \omega_{k_3} + \omega_{k_4})}\,.
\end{align}
We note that these solutions are not unique, which we remark on below.

We have succeeded in constructing a unitary which maps from an arbitrary translation and rotation-invariant Gaussian wavefunctional to the ground state of $\varphi^4$ theory to 1-loop in perturbation theory.  Our result is summarized below. \\ \\
\noindent \textbf{Summary of $1$-loop circuit from Gaussian to the ground state of $\varphi^4$ theory:} \\ \\
The translation and rotation-invariant Gaussian wavefunctional is denoted by $|\Psi_0\rangle$ and the ground state wavefunctional of $\varphi^4$ theory to 1-loop (and Wilsonian renormalized to spatial momentum scale $\Lambda e^u$) is denoted by $|\Psi(\Lambda e^u)\rangle$.  Then $|\Psi_0\rangle$ and $|\Psi(\Lambda e^u)\rangle$ are given in the $\phi$-basis by
\begin{equation}
\langle \phi |\Psi_0\rangle = \text{det}^{\frac{1}{4}}\left(\frac{\widetilde{\Omega}_0}{\pi}\right) \, \exp\left(-\frac{1}{2} \int d^d \vec{k} \, \phi(\vec{k}) \, \widetilde{\Omega}_0(k) \, \phi(-\vec{k}) \right)
\end{equation}
\begin{align}
\langle \phi | \Psi(\Lambda e^u)\rangle &= \mathcal{N} \, \exp\left(- G[\phi] - e^{-2u} \delta m^2 \, R_1[\phi] - e^{(d-3)u} \lambda \, R_2[\phi] \right) \,\, + \mathcal{O}(\lambda^2)
\end{align}
where
\begin{align}
G[\phi] &= \frac{1}{2}\int^\Lambda d^d \vec{k} \, \phi(\vec{k}) \, \omega_k \, \phi(-\vec{k}) \\ \nonumber \\
R_1[\phi]&=\frac{1}{4}\int^\Lambda d^d \vec{k}\, \frac{1}{\omega_k}\,\phi(\vec{k}) \phi(-\vec{k}) \\ \nonumber \\
\begin{split}
R_2[\phi] &=\frac{1}{16}\int^\Lambda d^d \vec{k}\, \frac{1}{\omega_k}\,\left(\int \frac{d^d \vec{q}}{(2\pi)^{d}} \frac{1}{\omega_k + \omega_q} \right)\,\phi(\vec{k}) \phi(-\vec{k})
\\
&+ \frac{1}{24}\frac{1}{(2\pi)^d} \int^\Lambda \frac{d^d \vec{k}_1 \, d^d \vec{k}_2 \, d^d \vec{k}_3}{\omega_{k_1} + \omega_{k_2} + \omega_{k_3} + \omega_{-\vec{k}_1 - \vec{k}_2 - \vec{k}_3}} \, \phi(\vec{k}_1) \phi(\vec{k}_2) \phi(\vec{k}_3) \phi(-\vec{k}_1 - \vec{k}_2 - \vec{k}_3)\,.
\end{split}
\end{align}
Consider the unitary $U$ given by
\begin{align}
U &= \exp\left( i \, K_{2,0} + i \lambda \, (K_{2,1} + K_4)\right)
\end{align}
with
\begin{align}
K_{2,0} &= - \int d^d \vec{k}_1 \, d^d \vec{k}_2 \, \delta^{(d)}(\vec{k}_1 + \vec{k}_2)\,g_{2,0}(\vec{k}_1) \, S_2^{(1)}(\vec{k}_1, \vec{k}_2) \\
K_{2,1} &= - \int d^d \vec{k}_1 \, d^d \vec{k}_2 \,\delta^{(d)}(\vec{k}_1 + \vec{k}_2)\, g_{2,1}(\vec{k}_1) \, S_2^{(1)}(\vec{k}_1, \vec{k}_2) \\
K_{4} &= \int d^d \vec{k}_1 \, d^d \vec{k}_2 \, d^d \vec{k}_3 \, d^d \vec{k}_4 \, \delta^{(d)}(\vec{k}_1 + \vec{k}_2 + \vec{k}_3 + \vec{k}_4) \bigg( g_{4}^{(1)}(\vec{k}_1,\vec{k}_2,\vec{k}_3,\vec{k}_4) \, S_4^{(1)}(\vec{k}_1, \vec{k}_2, \vec{k}_3, \vec{k}_4) \nonumber \\
& \qquad \qquad \qquad \qquad \qquad \qquad \qquad \qquad \qquad \qquad \qquad +  g_{4}^{(3)}(\vec{k}_1,\vec{k}_2, \vec{k}_3, \vec{k}_4) \, S_4^{(3)}(\vec{k}_1, \vec{k}_2, \vec{k}_3, \vec{k}_4) \bigg)\,.
\end{align}
Letting
\begin{align}
\mathcal{G}_1(\vec{k}_1,\vec{k}_2,\vec{k}_3,\vec{k}_4) &:= 2\big(g_{2,0}(\vec{k}_1) - g_{2,0}(\vec{k}_2) - g_{2,0}(\vec{k}_3) - g_{2,0}(\vec{k}_4)\big) \\
\mathcal{G}_3(\vec{k}_1,\vec{k}_2,\vec{k}_3,\vec{k}_4) &:= 2\big(g_{2,0}(\vec{k}_1) + g_{2,0}(\vec{k}_2) + g_{2,0}(\vec{k}_3) - g_{2,0}(\vec{k}_4)\big)
\end{align}
and defining
\begin{align}
\widetilde{g}_4^{(j)}(\vec{k}_1, \vec{k}_2, \vec{k}_3, \vec{k}_4) &:= \frac{e^{- \mathcal{G}_j(\vec{k}_1,\vec{k}_2,\vec{k}_3,\vec{k}_4)}-1}{\mathcal{G}_j(\vec{k}_1,\vec{k}_2,\vec{k}_3,\vec{k}_4)} \, g_4^{(j)}(\vec{k}_1, \vec{k}_2, \vec{k}_3, \vec{k}_4)
\end{align}
with $j=1,3$, the functions in $U$ above are given by
\begin{align}
\label{sol000}
g_{2,0}(\vec{k}) &= \frac{1}{4} \log\left(\frac{\widetilde{\Omega}_0(k)}{\omega_k}\right) \\ \nonumber \\
\label{sol01}
g_{2,1}(\vec{k}_1) &= - \frac{1}{\omega_{k_1}^2} \left(\frac{e^{-2u} (\delta m^2/\lambda)}{8} + \frac{e^{(d-3)u}}{32} \frac{1}{(2\pi)^{d}}\int d^d \vec{k}_2 \, \frac{1}{\omega_{k_1} + \omega_{k_2}} \right) \\  \nonumber \\
\label{sol02}
\widetilde{g}_{4}^{(1)}(\vec{k}_1,\vec{k}_2,\vec{k}_3,\vec{k}_4) &= \frac{1}{96}\frac{e^{(d-3)u}}{(2\pi)^{d}}\,\frac{1}{\omega_{k_2} \omega_{k_3} \omega_{k_4}(\omega_{k_1} + \omega_{k_2} + \omega_{k_3} + \omega_{k_4})} \\ \nonumber \\
\label{sol03}
\widetilde{g}_{4}^{(3)}(\vec{k}_1,\vec{k}_2,\vec{k}_3,\vec{k}_4) &= \frac{1}{32}\frac{e^{(d-3)u}}{(2\pi)^{d}}\,\frac{1}{\omega_{k_4}(\omega_{k_1} + \omega_{k_2} + \omega_{k_3} + \omega_{k_4})}\,.
\end{align}
The unitary $U$ satisfies
\begin{equation}
U|\Psi_0\rangle = |\Psi(\Lambda e^u)\rangle \,\,+ \mathcal{O}(\lambda^2)\,,
\end{equation}
and so it maps $|\Psi_0\rangle$ to $|\Psi(\Lambda e^u)\rangle$ up to $\mathcal{O}(\lambda^2)$ corrections.
\newpage
\noindent \textbf{Non-uniqueness of $1$-loop circuit from Gaussian to the ground state of $\varphi^4$ theory} \\ \\
As commented above, the solution we found for the $1$-loop circuit from Gaussian to the ground state of $\varphi^4$ theory is not unique.  This is perhaps not surprising -- given an initial and final state, there are typically many unitaries which map between them, even if the form of the unitaries are somewhat constrained.  Indeed, we found a one-parameter family of solutions for the anharmonic oscillator example above.

For instance, if we relax the conditions on the permutation symmetry of the arguments of $g_{2,1}$, $g_4^{(1)}$, $g_4^{(3)}$, from (\ref{new.one}) we find the more general solutions
\begin{align}
g_{2,1}(\vec{k}) &= - \frac{1}{\omega_{k_1}^2} \left(\frac{e^{-2u} (\delta m^2/\lambda)}{8} + \frac{e^{(d-3)u}}{32} \frac{1}{(2\pi)^{d}}\int d^d \vec{k}_2 \, \frac{1}{\omega_{k_1} + \omega_{k_2}} \right) \nonumber \\ \nonumber \\
& \quad \, \hspace{-.5cm}+ \frac{1}{\omega_{k_1}}\int d^d \vec{k}_2 \, \frac{1}{\omega_{k_2}}\left(F(\vec{k}_1,-\vec{k}_1,\vec{k}_2, -\vec{k}_2) + F(\vec{k}_1,\vec{k}_2,-\vec{k}_1, -\vec{k}_2) + F(\vec{k}_1,\vec{k}_2,-\vec{k}_2, -\vec{k}_1) \right) \nonumber \\ \nonumber \\
\widetilde{g}_{4}^{(1)}(\vec{k}_1,\vec{k}_2,\vec{k}_3,\vec{k}_4) &= \frac{1}{96}\frac{e^{(d-3)u}}{(2\pi)^{d}}\,\frac{1}{\omega_{k_2} \omega_{k_3} \omega_{k_4}(\omega_{k_1} + \omega_{k_2} + \omega_{k_3} + \omega_{k_4})} + \frac{F(\vec{k}_1,\vec{k}_2,\vec{k}_3, \vec{k}_4)}{\omega_{k_2} \omega_{k_3} \omega_{k_4}} \\ \nonumber \\
\widetilde{g}_{4}^{(3)}(\vec{k}_1,\vec{k}_2,\vec{k}_3,\vec{k}_4) &= \frac{1}{32}\frac{e^{(d-3)u}}{(2\pi)^{d}}\,\frac{1}{\omega_{k_4}(\omega_{k_1} + \omega_{k_2} + \omega_{k_3} + \omega_{k_4})}+ \frac{F(\vec{k}_1,\vec{k}_2,\vec{k}_3, \vec{k}_4)}{\omega_{k_4}}\,,
\end{align}
where $F(\vec{k}_1,\vec{k}_2,\vec{k}_3, \vec{k}_4)$ is a function satisfying the constraint 
\begin{align}
&\int d^d \vec{k}_2 \, \frac{1}{\omega_{k_2}}\bigg(-F(\vec{k}_2,-\vec{k}_2,\vec{k}_1, -\vec{k}_1) - F(\vec{k}_{2},\vec{k}_{1},-\vec{k}_2, -\vec{k}_1) \nonumber \\
& \qquad \qquad \qquad \qquad \qquad \qquad \quad + F(\vec{k}_1,\vec{k}_2,-\vec{k}_1, -\vec{k}_2) + F(\vec{k}_1,-\vec{k}_1,\vec{k}_2, -\vec{k}_2) \bigg) = 0\,.
\end{align}
For the remainder of this paper, will will specialize to the case that $F(\vec{k}_1,\vec{k}_2,\vec{k}_3, \vec{k}_4)=0$, which agree with Eqn.'s~\eqref{sol01},~\eqref{sol02}, and~\eqref{sol03} above.

\subsection{Comments on circuit non-uniqueness}

Suppose that we have a unitary $U$ which maps $|\Psi\rangle$ to $|\Phi\rangle$, in particular $U |\Psi\rangle = |\Phi\rangle$.  This unitary is not the unique unitary mapping $|\Psi\rangle$ to $|\Phi\rangle$, since for any unitaries $U_1,U_2$ satisfying $U_1 |\Psi\rangle = |\Psi\rangle$ and $U_2 |\Phi\rangle = |\Phi\rangle$ we have
\begin{equation}
U_2 U U_1 |\Psi\rangle = |\Phi \rangle\,.
\end{equation} 
Clearly $U_2 U U_1$ is also a unitary mapping $|\Psi\rangle$ to $|\Phi\rangle$.  Geometrically, a unitary maps a basis $\{|\Psi_1\rangle, |\Psi_2\rangle,...,|\Psi_N\rangle\}$ to another basis  $\{U|\Psi_1\rangle, U|\Psi_2\rangle,...,U|\Psi_N\rangle\}$.  If we specify that $U$ must map $|\Psi_1\rangle$ to $|\Phi_1\rangle$, then we still have the freedom to choose how $U$ maps the remaining basis vectors $\{|\Psi_2\rangle,...,|\Psi_N\rangle\}$.

In the framework of quantum circuit perturbation theory, consider a unitary $U = e^{-i (Q_2 + \epsilon \, Q_k)}$ where $Q_k$ is order $k$ and
\begin{equation}
e^{- i (Q_2 + \epsilon \, Q_k)}|\Psi\rangle = |\Phi\rangle\, + \mathcal{O}(\epsilon^2)\,.
\end{equation}
Suppose we want construct the most general unitary of the form $e^{-i(\widetilde{Q}_2 + \epsilon \, \widetilde{Q}_k)}$ that maps $|\Psi\rangle$ to $|\Phi\rangle$ up to $\mathcal{O}(\epsilon^2)$ corrections.  Then we find the parametric family of operators $Q_2^{(1)}, Q_k^{(1)},Q_2^{(2)}, Q_k^{(2)}$ which satisfy
\begin{equation}
(Q_2^{(1)} + \epsilon \, Q_k^{(1)}) |\Psi\rangle = \mathcal{O}(\epsilon^2)\,\,, \qquad (Q_2^{(2)} + \epsilon \, Q_k^{(2)})  |\Phi\rangle = \mathcal{O}(\epsilon^2)\,.
\end{equation}
The $Q_2^{(1)}, Q_k^{(1)},Q_2^{(2)}, Q_k^{(2)}$ satisfying the above equations may have many free parameters (and in the context of field theory, free functions).  Then we can find the most general unitary of the form $e^{-i(\widetilde{Q}_2 + \epsilon \, \widetilde{Q}_k)}$ that maps $|\Psi\rangle$ to $|\Phi\rangle$ up to $\mathcal{O}(\epsilon^2)$ corrections by solving
\begin{equation}
e^{-i(\widetilde{Q}_2 + \epsilon \, \widetilde{Q}_k)} = e^{-i (Q_2^{(2)} + \epsilon \, Q_k^{(2)})} \, e^{-i (Q_2 + \epsilon \, Q_k)} \, e^{-i (Q_2^{(1)} + \epsilon \, Q_k^{(1)})} \,\, + \mathcal{O}(\epsilon^2)
\end{equation} 
for $\widetilde{Q}_2, \widetilde{Q}_k$ which in turn may depend on many free parameters (or free functions, in the context of field theory).  There is an analogous procedure if we work to higher order in perturbation theory.  In this paper, we do not attempt to find the most generic circuits mapping between two states of interest -- rather, we find particular solutions that suit our purposes.

\subsection{More general perturbative examples}

So far, we have given various constructions of unitaries which map between Gaussian states and perturbatively non-Gaussian states.  Here we detail a more general class of unitary mappings which goes beyond our chosen examples.  Suppose that we have two states $|\Psi_1\rangle$ and $|\Psi_2\rangle$ which are Gaussian up to perturbative corrections in some smaller parameter $\epsilon$.  Then we can write
\begin{align}
\label{moregeneralexpand1}
|\Psi_1\rangle &= \left(1 - i\,\epsilon \, Q_{\text{higher},1}\right)  |\Psi_1^G\rangle \,\,+\mathcal{O}(\epsilon^2) = e^{- i\, \epsilon  \, Q_{\text{higher},1}} |\Psi_1^G\rangle \,\,+\mathcal{O}(\epsilon^2) \\
\label{moregeneralexpand2}
|\Psi_2\rangle &= \left(1 - i\, \epsilon \, Q_{\text{higher},2}\right) |\Psi_2^G\rangle \,\,+\mathcal{O}(\epsilon^2) = e^{- i\, \epsilon\, Q_{\text{higher},2}}|\Psi_2^G\rangle\,\,+\mathcal{O}(\epsilon^2)\,,
\end{align}
where $Q_{\text{higher},1}, Q_{\text{higher},2}$ each have order greater than $2$, and $|\Psi_1^{\text{G}}\rangle, |\Psi_2^{\text{G}}\rangle$ are Gaussian states.  As explained in previous sections, given any two Gaussian states we can explicitly construct a unitary $e^{- i Q_2}$ with a quadratic generator $Q_2$ which maps between them.  So suppose that
\begin{equation}
|\Psi_2^G\rangle = e^{- i Q_2} \, |\Psi_1^G\rangle\,.
\end{equation}
Then using the above equation along with Eqn.'s~\eqref{moregeneralexpand1} and~\eqref{moregeneralexpand2}, we can write
\begin{align}
\label{moregeneralexpand3}
|\Psi_2\rangle &=  e^{-i \,\epsilon\, Q_{\text{higher},2}} |\Psi_2^G\rangle \,\,+\mathcal{O}(\epsilon^2) \nonumber \\
&= e^{-i \,\epsilon\, Q_{\text{higher},2}} \, e^{-i Q_2} |\Psi_1^G\rangle \,\,+\mathcal{O}(\epsilon^2) \nonumber \\
&= e^{-i \,\epsilon \, Q_{\text{higher},2}} \, e^{-i Q_2} \, e^{i\,\epsilon \, Q_{\text{higher},1}}|\Psi_1\rangle \,\,+\mathcal{O}(\epsilon^2)
\end{align}
Generalizing Eqn.'s~\eqref{ABexpansion1} and~\eqref{ABexpansion2}, we have
\begin{align}
e^{A + \epsilon \,B + \epsilon \, C} = \left(1 + \epsilon \, \frac{e^{\text{ad}_A} - 1}{\text{ad}_A} \, B\right) \, e^A \, \left(1 + \epsilon \, \frac{1 - e^{-\text{ad}_A}}{\text{ad}_A} \, C\right) \,\,+\mathcal{O}(\epsilon^2)\,,
\end{align}
and so
\begin{align}
\label{ABCeq1}
\exp\left(A + \epsilon \,\frac{\text{ad}_A}{e^{\text{ad}_A} - 1}\,B + \epsilon \,\frac{\text{ad}_A}{1 - e^{-\text{ad}_A}}\, C\right) &= \left(1 + \epsilon \, B\right) \, e^A \, \left(1 + \epsilon\, C\right) \,\,+\mathcal{O}(\epsilon^2) \nonumber \\
&= e^{\epsilon \, B} \, e^A \, e^{\epsilon \, C}\,\,+\mathcal{O}(\epsilon^2)\,.
\end{align}
Applying Eqn.'s~\eqref{ABCeq1} to Eqn.~\eqref{moregeneralexpand3}, we find that
\begin{equation}
|\Psi_2\rangle = \exp\left(- i \left(Q_2 + \epsilon \,\frac{i\,\text{ad}_{Q_2}}{1 - e^{- i \,\text{ad}_{Q_2}}} \, Q_{\text{higher},2} + \epsilon \,\frac{i\,\text{ad}_{Q_2}}{1 - e^{i \,\text{ad}_{Q_2}}}\, Q_{\text{higher,1}} \right) \right) |\Psi_1\rangle \,\,+\mathcal{O}(\epsilon^2)\,,
\end{equation}
and so the unitary
\begin{equation}
U =  \exp\left(- i \left(Q_2 + \epsilon \,\frac{i\,\text{ad}_{Q_2}}{1 - e^{- i \,\text{ad}_{Q_2}}} \, Q_{\text{higher},2} + \epsilon \,\frac{i\,\text{ad}_{Q_2}}{1 - e^{i \,\text{ad}_{Q_2}}}\, Q_{\text{higher,1}} \right) \right)\,\,+\mathcal{O}(\epsilon^2)
\end{equation}
maps $|\Psi_1\rangle$ to $|\Psi_2\rangle$ up to $\mathcal{O}(\epsilon^2)$ corrections.

We have thus given a systematic procedure to construct a unitary which maps between two states which are each Gaussian up to first order corrections.  An analogous, more elaborate procedure is possible for constructing unitary operators which map between states which are each Gaussian up to higher than first order corrections.

\section{cMERA for interacting fields}

\subsection{Overview of MERA}

The multiscale entanglement renormalization ansatz, or MERA, is a class of quantum states which capture quantum correlations across a hierarchy of distance scales \cite{ER1, ER2, ER3, Hauru1}.  MERA has been used with great success in numerical studies as a robust variational approximation to low-energy states of local lattice theories, especially in one spatial dimension \cite{nuMERAcle1}--\cite{nuMERAcle8}.  We briefly review the lattice MERA formalism before considering its generalization to the continuum.

Suppose we have a translation-invariant lattice system in one spatial dimension comprised of $2^n$ sites, where each site has local dimension $d$.  Then the Hilbert space $\mathcal{H}_{2^n}$ of the system is
\begin{equation}
\mathcal{H}_{2^n} = \underbrace{\mathbb{C}^d \otimes \cdots \otimes \mathbb{C}^d }_{2^n\text{ times}} \simeq \mathbb{C}^{d^{2^n}}\,.
\end{equation}
Since each site is a $d$-level system, we will refer to each site as a qudit.\footnote{Recall that a $2$-level system is called a qubit.}  Choosing some integer $m<n$, we begin with the smaller Hilbert space $\mathcal{H}_{2^m}$ given by
\begin{equation}
\mathcal{H}_{2^m} = \underbrace{\mathbb{C}^d \otimes \cdots \otimes \mathbb{C}^d }_{2^m\text{ times}} \simeq \mathbb{C}^{d^{2^m}}\,.
\end{equation}
and choose the initial state $|\Omega_{2^m}\rangle \in \mathcal{H}_{2^m}$ given by
\begin{equation}
|\Omega_{2^m}\rangle = \underbrace{|0\rangle \otimes |0\rangle \otimes \cdots \otimes |0\rangle}_{2^m\text{ times}}\,,
\end{equation}
which factorizes with respect to each site.  We imagine that the $2^m$ sites are organized in a line, and that adjacent tensor factors in the decomposition above correspond to adjacent sites in space.  Letting $U_{2^m}$ be a spatially local circuit, we write
\begin{equation}
U_{2^m} |\Omega_{2^m}\rangle = \sum_{i_1,...,i_{2^m} = 1}^d c_{i_1,i_2,...,i_{2^m}} |i_1\rangle \otimes |i_2\rangle \otimes \cdots \otimes |i_{2^m}\rangle\,.
\end{equation}
Next, we will embed $U_{2^m} |\Omega_{2^m}\rangle \in \mathcal{H}_{2^m}$ into the larger Hilbert space $\mathcal{H}_{2^{m+1}}$ comprised of $2^{m+1}$ qudits, i.e.\! twice as many as before.  We use the isometry $V_{2^m} : \mathcal{H}_{2^{m}} \hookrightarrow \mathcal{H}_{2^{m+1}}$, satisfying $V_{2^m}^\dagger V_{2^m} = \textbf{1}_{\mathcal{H}_{2^m}}$\,, which acts by
\begin{equation}
V_{2^m} U_{2^m} |\Omega_{2^m}\rangle = \sum_{i_1,...,i_{2^m} = 1}^d c_{i_1,i_2,...,i_{2^m}} |0\rangle \otimes |i_1\rangle \otimes |0\rangle \otimes |i_2\rangle \otimes |0\rangle \otimes \cdots \otimes |0\rangle \otimes |i_{2^m}\rangle \,,
\end{equation}
namely, interlacing the existing state with qudits in the $|0\rangle$ state.  Similar to above, we imagine that the $2^{m+1}$ sites are organized in a line, and that adjacent tensor factors in the decomposition above correspond to adjacent sites in space.  Thus we can think of the isometry as putting in a new site in the ``space between'' each pair of existing sites.

We then further transform $V_{2^m} U_{2^m} |\Omega_{2^m}\rangle$ by another local circuit $U_{2^{m+1}}$ which acts on all $2^{m+1}$ sites, and then embed the resulting state into the even larger Hilbert space $\mathcal{H}_{2^{m+2}}$ using the isometry $V_{2^{m+1}}$.  We repeat this procedure until we have a state in the full Hilbert space $\mathcal{H}_{2^n}$ of the form
\begin{equation}
\label{MERAansatz1}
|\Psi_{\text{MERA}}\rangle := U_{2^n} V_{2^{n-1}} U_{2^{n-1}} \cdots V_{2^{m+1}} U_{2^{m+1}} V_{2^m} U_{2^m}|\Omega_{2^m}\rangle\,,
\end{equation}
which is a MERA state.  The MERA state depends on our choice of the local unitaries $U_{2^m},U_{2^{m+1}},...,U_{2^n}$, as well as our choice of $m$.  The local unitaries $U_{2^m},U_{2^{m+1}},...,U_{2^n}$ are called \textit{entangling unitaries} since they entangle existing lattice sites with newly introduced lattice sites.  Their inverses are appropriately called \textit{disentangling unitaries}.

Typically, when performing variational calculations, one fixes $m$ and parametrizes the local unitaries $U_{2^m},U_{2^{m+1}},...,U_{2^n}$.  Then the variational optimization is performed with respect to the parameters of the unitaries.  For instance, given some Hamiltonian $H$ acting on $\mathcal{H}_{2^n}$, we would minimize
\begin{equation}
E_{\text{min}} = \min_{\{a_i\}} \, \langle \Psi_{\text{MERA}}(\{a_i\})| H |\Psi_{\text{MERA}}(\{a_i\}) \rangle
\end{equation}
as per the Rayleigh-Ritz variational principle, where $\{a_i\}$ are the parameters which define the local unitaries $U_{2^m},U_{2^{m+1}},...,U_{2^n}$ and $E_{\text{min}}$ is the variational approximation to the ground state energy of $H$.  One of the key features of the MERA ansatz is that it is not exclusively sensitive to short-distance correlations.  Due to the hierarchical structure of the MERA ansatz, variational optimization more equitably energetically favors the correctness of long-distance and intermediate-distance correlations as well as short-distance correlations.

The state $|\Psi_{\text{MERA}}\rangle$ has several important features.  First, we notice that the initial $2^m$ sites are now $(2^{n-m} - 1)$ sites apart in the final state $|\Psi_{\text{MERA}}\rangle$.  Thus the unitary $U_{2^m}$ which acted \textit{locally} on the initial $2^m$ sites is responsible for \textit{long-range} correlations (i.e., across distances of around $2^{n-m}$ sites) in the final state.  Similarly, each unitary $U_{2^k}$ for $m \leq k \leq n$ is responsible for correlations across a characteristic scale of $2^{n-k}$ sites in the final state $|\Psi_{\text{MERA}}\rangle$.

In words, local correlations introduced by each $U_{2^k}$ on $2^k$ sites are stretched to larger distance scales by the isometric embeddings into larger lattices.  Furthermore, the form of Eqn.~\eqref{MERAansatz1} shows that the correlations are introduced \textit{hierarchically}: (what will become the) long-range correlations are introduced first, and then new degrees of freedom are introduced in succession so that progressively shorter-range correlations can be built on top.

The MERA state $|\Psi_{\text{MERA}}\rangle$ is closely tied with the renormalization group.  Suppose that we have a state $|\Phi\rangle$ of a lattice theory of $2^n$ qudits in one spatial dimension that can be written in the form of Eqn.~\eqref{MERAansatz1}.  If we wanted to strip away short-distance correlations from the state, we could strip away the final $U_{2^n}$ unitary and then zoom in on the state by omitting the $V_{2^{n-1}}$ isometry.  Then we would be left with
\begin{equation}
\label{MERAansatz2}
U_{2^{n-1}} \cdots V_{2^{m+1}} U_{2^{m+1}} V_{2^m} U_{2^m}|\Omega_{2^m}\rangle\,.
\end{equation}
From this structure, we see that the mapping 
\begin{equation}
U_{2^n} V_{2^{n-1}} U_{2^{n-1}} \cdots V_{2^{m+1}} U_{2^{m+1}} V_{2^m} U_{2^m}|\Omega_{2^m}\rangle \longrightarrow U_{2^{n-1}} \cdots V_{2^{m+1}} U_{2^{m+1}} V_{2^m} U_{2^m}|\Omega_{2^m}\rangle
\end{equation}
is essentially a step of Kadanoff block renormalization,\footnote{Kadanoff block renormalization is typically applied to a Hamiltonian, although here we are working directly with a state.} in which we first eliminate certain short-range interactions and then coarse grain the system by mapping it to fewer sites (in this case, half of the number of sites).  So if a state $|\Phi\rangle$ of $2^n$ qudits can be expressed in the form of a MERA as in Eqn.~\eqref{MERAansatz1}, then the state \textit{renormalized to scale $2^{n-k}$} is
\begin{equation}
\label{MERArenormalizedtoscale}
|\Phi(2^{n-k})\rangle := U_{2^{n-k}} V_{2^{n-k-1}} U_{2^{n-k-1}} \cdots V_{2^{m+1}} U_{2^{m+1}} V_{2^m} U_{2^m} |\Omega_{2^m}\rangle\,,
\end{equation}
which is the same as Eqn.~\eqref{MERAansatz1} but with the appropriate unitaries and isometries stripped off the front.  We call $|\Phi(2^{n-m})\rangle = |\Omega_{2^m}\rangle$ the ``IR state'' and $|\Phi(2^n)\rangle = |\Phi\rangle$ the ``UV state.''

The basic MERA construction can be generalized to lattice systems in higher dimensions.  So far, we have chosen isometries which embed a state into a Hilbert space with double the number of sites, although we could have chosen other embeddings (e.g., into a Hilbert space with triple the number of sites, into a Hilbert space with 1.2 times the number of sites, etc.).

Before turning to the generalization of MERA to the continuum, there is an alternative perspective of the MERA construction which is useful to keep in mind.  In the construction of MERA states given above, we have started with a smaller number of sites than the full lattice, and added sites after applying each entangling unitary $U_{2^k}$.  Instead, we could have initially started with a state $|\Omega_{2^n}\rangle \in \mathcal{H}_{2^n}$ given by
\begin{equation}
\label{productstate1}
|\Omega_{2^n}\rangle = \underbrace{|0\rangle \otimes |0\rangle \otimes \cdots \otimes |0\rangle}_{2^n\text{ times}}\,.
\end{equation}
Forgoing the isometries $V$, we would let $U_{2^m}$ act on sites positioned at multiples of $2^{m}$.  Then $U_{2^{m-1}}$ would only act on sites positioned at multiples of $2^{m-1}$, and more generally $U_{2^{m-k}}$ would only act on sites positioned at multiples of $2^{m-k}$.  In words, we do not need the isometries $V$ since we just start with a state in $\mathcal{H}_{2^n}$ and ``skip over'' sites if we do not need them.  In this setup, after each application of a $U_{2^k}$ there are still many sites in the state $|0\rangle$ which are like a reservoir of unused UV degrees of freedom.

One virtue of this alternative approach is that the MERA state can be written as a unitary transformation of $|\Omega_{2^n}\rangle$, namely
\begin{equation}
\label{MERAansatz3}
|\Psi_{\text{MERA}}\rangle := \widetilde{U}_{2^n} \widetilde{U}_{2^{n-1}} \cdots \widetilde{U}_{2^{m+1}} \widetilde{U}_{2^m}|\Omega_{2^m}\rangle\,,
\end{equation}
where we have put tildes over the unitaries to denote that they each act on an appropriate subset of the sites in $\mathcal{H}_{2^n}$ (and are thus similar to but distinct from the corresponding unitaries $U_{2^k}$ which act on $\mathcal{H}_{2^k}$).  Similarly, if a state $|\Phi\rangle$ of $2^n$ qudits can be written in the form of Eqn.~\eqref{MERAansatz3}, then the state \textit{renormalized to scale $2^{n-k}$} is
\begin{equation}
\label{MERAansatz4}
|\Phi(2^{n-k})\rangle := \widetilde{U}_{2^{n-k}} \widetilde{U}_{2^{n-k-1}} \cdots \widetilde{U}_{2^{m+1}} \widetilde{U}_{2^m} |\Omega_{2^m}\rangle\,.
\end{equation}
We will use this alternative, \textit{fully unitary} approach in our presentation of the continuum generalization of MERA below.

\subsection{Overview of cMERA}

While the MERA ansatz has had great success for numerical simulations of low-dimensional local lattice systems, there are some drawbacks:
\begin{enumerate}
\item MERA is a lattice ansatz, and does not work directly in the continuum.
\item MERA does not work directly in the infinite-volume limit since one has to specify the number of sites of the lattice.
\item Although it is possible generalize MERA to more than one spatial dimension, such generalizations can be both analytically cumbersome and numerically difficult.
\end{enumerate}
A continuum version of MERA called cMERA was developed to address these three issues above \cite{cMERA1, cMERA2, cMERA3}.

The cMERA ansatz is structurally similar to the form of the MERA ansatz in Eqn.~\eqref{MERAansatz4} above.  Suppose the Hilbert space of our continuum field theory is $\mathcal{H}$.  We require a simple IR state $|\Omega\rangle \in \mathcal{H}$, upon which we will build correlations at progressively finer distance scales.  We would like $|\Omega\rangle$ to be scale-invariant\footnote{We mean scale-invariant with respect to \textit{spatial} dilatations.}, so that it has the same correlations across \textit{any} distance scale.  If $L$ is a generator of spatial dilatations for our continuum theory, then we would like $e^{-i L u} |\Omega\rangle = |\Omega\rangle$ for all $u$, or equivalently
\begin{equation}
L |\Omega\rangle = 0\,.
\end{equation}
The scale-invariance of $|\Omega\rangle$ is inspired by the fact that theories tend to become scale-invariant as they are flowed by the renormalization group into the IR.  In other words, we would like to start with a scale-invariant state $|\Omega\rangle$ in the IR, and then build up correlations into the UV.

In analogy with Eqn.~\eqref{MERAansatz4}, we need to apply a sequence of unitaries to $|\Omega\rangle$ to build up correlations at progressively shorter distance scales.  To do this, we construct a path-ordered exponential and apply it to $|\Omega\rangle$, namely
\begin{equation}
\label{pathordered1}
|\Psi_{\text{cMERA}}\rangle = \mathcal{P}_s \, \exp\left(-i \int_{u_{\text{IR}}}^{u_{\text{UV}}} ds \, (K(s) + L)\right) |\Omega\rangle
\end{equation}
where $L$ is the spatial dilatation operator, and $K(s)$ is a one-parameter family of Hermitian operators called the \textit{entangler}.  Note that $K(s)$ is called the \textit{disentangler} if we run the path-ordered exponential in reverse -- so $K(s)$ is synonymously the entangler and disentangler.  The unitary
\begin{equation}
\label{entanglingunitary1}
U(u_1,u_2) = \mathcal{P}_s \, \exp\left(-i \int_{u_1}^{u_2} ds \, (K(s) + L)\right)
\end{equation}
is called the \textit{entangling unitary}.

The path-ordered exponential acting on $|\Omega\rangle$ as per Eqn.~\eqref{pathordered1} has a simple interpretation.  It can be expanded as
\begin{equation}
\label{pathorderexpanded1}
e^{- i (K(u_{\text{UV}}) + L) \, \Delta s} \,\, e^{- i (K(u_{\text{UV}} - \Delta s) + L) \, \Delta s} \cdots e^{- i (K(u_{\text{IR}} + \Delta s) + L) \, \Delta s} \, e^{- i (K(u_{\text{IR}}) + L) \, \Delta s} |\Omega\rangle
\end{equation}
in the limit of $\Delta s \to 0$.  Then Eqn.~\eqref{pathorderexpanded1} shows us that the path-ordered exponential creates correlations at the $u_{\text{IR}}$ scale via the entangler $K(u_{\text{IR}})$ and then dilates the state by $e^{\Delta s}$ via $L$ to ``zoom in.''  Then correlations are created at the new zoomed-in scale via $K(u_{\text{IR}}+ \Delta s)$, and the state is further zoomed in by another factor of $e^{\Delta s}$ (so that the net zoom is $e^{2 \Delta s}$) via $L$.  The procedure is repeated until we reach the UV scale corresponding to a net zoom of $e^{(u_{\text{UV}} - u_{\text{IR}})}$.  For concreteness, we let \cite{cMERA2}
\begin{align}
u_{\text{IR}} &= -\infty \\
u_{\text{UV}} &= 0
\end{align}
and follow these conventions for the rest of the paper.  Then we can interpret $K(u_{\text{IR}}) = K(-\infty)$ as creating correlations at a distance scale which goes as $\sim e^{-u_{\text{IR}}} \sim \infty$, and $K(u_{\text{UV}}) = K(0)$ as creating correlations at a distance scale which goes as $\sim e^{-u_{\text{UV}}} = 1$.  Thus,
\begin{equation}
\label{Kscalecreate1}
K(s)\text{ creates correlations at a distance scale }\sim \, \exp(-s)\,, \quad -\infty < s \leq 0\,.
\end{equation}
Since it is often useful to work in momentum space, we note that
\begin{equation}
\label{Kscalecreate2}
\,\,\, K(s)\text{ creates correlations at a momentum scale }\sim \, \exp(s)\,, \quad -\infty < s \leq 0\,,
\end{equation}
which is complementary to Eqn.~\eqref{Kscalecreate1}.

It follows that if we want to capture the correlations of $|\Psi_{\text{cMERA}}\rangle$ in Eqn.~\eqref{pathordered1} at an intermediate length scale $\sim e^{-u}$ for $-\infty \leq u \leq 0$, then we would write
\begin{equation}
\label{pathordered2}
|\Psi_{\text{cMERA}}(\ell = e^{-u})\rangle = \mathcal{P}_s \, \exp\left(-i \int_{u_{\text{IR}}}^{u} ds \, (K(s) + L)\right) |\Omega\rangle
\end{equation}
which is directly analogous to Eqn.~\eqref{MERArenormalizedtoscale} above.  Thus we say that $|\Psi_{\text{cMERA}}\rangle$, renormalized to distance scale $e^{-u}$, is $|\Psi_{\text{cMERA}}(\ell = e^{-u})\rangle$.

So far, we have not specified the form of the entangler $K(s)$.  Up until now, it has only been understood how to construct $K(s)$ for Gaussian wavefunctionals, such as the ground states of free bosonic or fermionic theories \cite{cMERA1, cMERA2, cMERA3, cMERA4}.  This drawback has rendered cMERA to have limited utility.  Although mean field theory has been applied to cMERA \cite{CMM1}, this approach still approximates the ground states of interacting theories by Gaussian wavefunctionals.

One reason why it is so difficult to determine $K(s)$ for non-Gaussian states, such as the ground state of an interacting theory, is that the RG structure of a non-Gaussian state is often non-trivial.  Since $K(s)$ needs to encode correlations of a state at all intermediate distance scales, knowing information about the RG flow of a non-Gaussian state is tantamount to knowing $K(s)$.  Conversely, since the RG flow of a Gaussian state \textit{is} trivial, its corresponding entangler $K(s)$ is easy to construct.

Using the tools developed in the first part of this paper, we can make progress on the problem of constructing $K(s)$ for non-Gaussian states.  In particular, we will show how to construct $K(s)$ perturbatively for a state which is a Gaussian wavefunctional with perturbative non-Gaussian corrections.  This will require understanding the perturbative RG structure of such states, which we will use as input data to construct $K(s)$.  Even though our $K(s)$'s will be constructed perturbatively, their forms are illuminating, and suggest how to construct ansatzes for non-perturbative variational calculations of the ground states of interacting field theories.


\subsection{Outline of cMERA procedure}
\label{sec:OutlineProcedure}
Before proceeding with calculations, let us outline our strategy for perturbatively computing $K(s)$ for states of a quantum field theory.  Suppose we have a state $|\Psi\rangle$ in a theory with a UV \textit{spatial momentum} cutoff $|\vec{p}| \leq \Lambda$.  The strategy for the computation is:
\begin{enumerate}
\item Using some RG scheme, calculate $|\Psi(\Lambda e^u)\rangle$, namely the wavefunctional renormalized to momentum scale $|\vec{p}| = \Lambda e^u$, for $-\infty < u \leq 0$.
\item Given an IR state $|\Omega\rangle$, find a unitary $U(u)$ such that $\langle \phi(\vec{p})|\Psi(\Lambda e^u)\rangle = \langle \phi(\vec{p})| U(u) |\Omega\rangle$ for $\phi(\vec{p})$ for which $\phi(\vec{p}) = 0$ when $|\vec{p}| \geq \Lambda$.  In other words, $|\Psi(\Lambda e^u)\rangle$ equals $U(u) |\Omega\rangle$ for modes with $|\vec{p}| \leq \Lambda$.
\item Determine $K(s)$ such that
$$\mathcal{P}_s \, e^{-i \int_{-\infty}^u ds \, (K(s) + L)}|\Omega\rangle = U(u) |\Omega\rangle\,.$$
\end{enumerate}
We will consider the case that $|\Psi\rangle$ is a \textit{non}-Gaussian state, which is Gaussian up to perturbative corrections in a parameter $\lambda$.  While we cannot implement the above strategy exactly, we can implement each step perturbatively to a fixed order in the perturbative coupling $\lambda$.

Note that our approach is slightly different than that of \cite{cMERA1, cMERA2, cMERA3, cMERA4}.  These papers only desire to find a $K(s)$ such that $\mathcal{P}_s \, \exp(-i \int_{-\infty}^0 ds \, (K(s) + L))|\Omega\rangle$ equals the \textit{UV} state $|\Psi(\Lambda)\rangle$ (for spatial momentum modes with $|\vec{p}| \leq \Lambda$), and \textit{not} that $\mathcal{P}_s \, \exp(-i \int_{-\infty}^u ds \, (K(s) + L))|\Omega\rangle$ agree with $|\Psi(\Lambda e^u)\rangle$ for generic $u$ corresponding to intermediate RG scales.  This is reasonable if one only wants to understand the structure of $|\Psi\rangle$ in the UV.  However, the cMERA ansatz is powerful in part because it seems able to capture the RG structure of quantum states.  Therefore, it seems prudent to understand precisely how cMERA is able to capture RG in the continuum -- thus, we attack the harder question of finding an entangler $K(s)$ such that $\mathcal{P}_s \, \exp(-i \int_{-\infty}^u ds \, (K(s) + L))|\Omega\rangle$ equals $|\Psi(\Lambda e^u)\rangle$ (for spatial momentum modes with $|\vec{p}| \leq \Lambda e^u$) for \textit{all} admissible values of $u$, \textit{including} the UV value $u=0$.

As a warmup, we first calculate $K(s)$ for the ground state of a free scalar field theory using the steps of the above strategy.  We will subsequently consider cMERA for perturbatively non-Gaussian states.

\subsection{cMERA for free scalar fields revisited}
\label{sec:cMERAfreerevisited}

Consider again the Hamiltonian of a free scalar field theory with mass $m$ in ($d+1$) dimensions.
Recall that the Hamiltonian of a free scalar field of mass $m$ is
\begin{equation}
\label{freeHam2}
H^\Lambda = \frac{1}{2} \int^\Lambda d^d \vec{k} \, \left(\widehat{\pi}(\vec{k}) \widehat{\pi}(-\vec{k}) + \widehat{\phi}(\vec{k}) \left(\vec{k}^{\,2} + m^2\right) \widehat{\phi}(-\vec{k}) \right)
\end{equation} 
with ground state wavefunctional
\begin{equation}
\label{groundstate1}
\langle \phi |\Psi_0(\Lambda)\rangle = \mathcal{N} \, \exp\left(-\frac{1}{2} \int^\Lambda d^d \vec{k} \,\, \phi(\vec{k}) \, \sqrt{\vec{k}^{\,2} + m^2} \, \phi(-\vec{k}) \right)
\end{equation}
where $\mathcal{N}$ is the normalization and the $\vec{k}$--integral is up to scale $|\vec{k}| = \Lambda$.

The first step in our strategy above is to compute $|\Psi_0(\Lambda e^u)\rangle$, namely, the ground state wavefunction renormalized to scale $|\vec{k}| = \Lambda \, e^u$.  What properties do we want such a renormalized state $|\Psi_0(\Lambda e^u)\rangle$ to have?  Suppose we have an equal-time correlation function of the original UV state $|\Psi_0\rangle = |\Psi_0(\Lambda)\rangle$, of the form
\begin{equation}
\langle \Psi_0(\Lambda)| \mathcal{O}_1(\vec{p}_1,0) \cdots \mathcal{O}_n(\vec{p}_n,0)|\Psi_0(\Lambda)\rangle
\end{equation}
where $|\vec{p}_i|\leq \Lambda$ for $i=1,...,n$.  Note that each $\mathcal{O}_i(\vec{p}_i,0)$ can be written in terms of $\phi(\vec{p}_i,0)$'s and $\pi(\vec{p}_i,0)$'s.  For instance, we might have $\mathcal{O}_1(\vec{p}_1,0) = \pi(\vec{p}_1,0) (\partial_{\vec{p}_1} \phi(\vec{p}_1,0) \cdot \partial_{\vec{p}_1} \phi(\vec{p}_1,0))^4 + \pi(\vec{p}_i,0)^{5}$.  The state $|\Psi(\Lambda)\rangle$ lives in a Hilbert space\footnote{There are many subtleties in defining the Hilbert spaces of QFT's.  We do not attempt a rigorous approach, but instead provide an intuitive perspective which glosses over presently unnecessary details.}
\begin{equation}
\mathcal{H}^\Lambda = \bigotimes_{|\vec{p}| \leq \Lambda} \text{span}\{\,|\xi\rangle \, | \, \xi \in \mathbb{R}\}\,.
\end{equation}
Note that a state $|\phi(\vec{k})\rangle \in \mathcal{H}^\Lambda$, corresponding to a fixed field configuration $\phi(\vec{k})$ taking $\mathbb{R}^d \to \mathbb{R}$, can be written as
\begin{equation}
|\phi(\vec{k})\rangle = \bigotimes_{|\vec{p}| \leq \Lambda} |\!\left.\phi(\vec{k})\right|_{\vec{k} = \vec{p}}\,\rangle
\end{equation}
where $|\left.\phi(\vec{k})\right|_{\vec{k} = \vec{p}}\,\rangle$ denotes a state in $\text{span}\{\,|\xi\rangle \, | \, \xi \in \mathbb{R}\}$.  We emphasize that $\phi(\vec{k})$ is a function, whereas $\left.\phi(\vec{k})\right|_{\vec{k} = \vec{p}}$ is a scalar value.  

Suppose that instead of considering correlations functions with $|\vec{p}_i|\leq \Lambda$, we \textit{only} considered correlation functions for which $|\vec{p}_i|\leq \Lambda e^u$ for $-\infty < u < 0$. Then we might ask if there is a state $|\widetilde{\Psi}_0(\Lambda e^u)\rangle$ in the smaller Hilbert space
\begin{equation}
\label{smallerHilbert1}
\mathcal{H}^{\Lambda e^u} = \bigotimes_{|\vec{p}| \leq \Lambda e^u} \text{span}\{\,|\!\left.\phi(\vec{k})\right|_{\vec{k} = \vec{p}}\,\rangle \, | \, \left.\phi(\vec{k})\right|_{\vec{k} = \vec{p}} \in \mathbb{R}\}
\end{equation}
such that
\begin{equation}
\label{RGcorrelatormatching1}
\langle \widetilde{\Psi}_0(\Lambda e^u)| \mathcal{O}_1(\vec{p}_1,0) \cdots \mathcal{O}_n(\vec{p}_n,0)|\widetilde{\Psi}_0(\Lambda e^u)\rangle = \langle \Psi_0(\Lambda)| \mathcal{O}_1(\vec{p}_1,0) \cdots \mathcal{O}_n(\vec{p}_n,0)|\Psi_0(\Lambda)\rangle
\end{equation}
so long as $|\vec{p}_i|\leq \Lambda e^u$ for all $i$.  Such a renormalized state $|\widetilde{\Psi}_0(\Lambda e^u)\rangle$ need only capture correlations \textit{below} the momentum scale $|\vec{p}| \leq \Lambda e^u$.  Since we can write
\begin{align}
\mathcal{H}^\Lambda &= \mathcal{H}^{\Lambda e^u} \otimes \mathcal{H}^{\Lambda e^u < |\vec{p}| \leq \Lambda} \nonumber \\
&= \mathcal{H}^{\Lambda e^u} \otimes \bigotimes_{\Lambda e^u < |\vec{p}| \leq \Lambda} \text{span}\{\,|\!\left.\phi(\vec{k})\right|_{\vec{k} = \vec{p}}\,\rangle \, | \, \left.\phi(\vec{k})\right|_{\vec{k} = \vec{p}} \in \mathbb{R}\}
\end{align}
and an operator $\mathcal{O}_i(\vec{p}_i,0)$ for $|\vec{p}_i| \leq \Lambda e^u$ acts on $\mathcal{H}^\Lambda = \mathcal{H}^{\Lambda e^u} \otimes \mathcal{H}^{\Lambda e^u < |\vec{p}| \leq \Lambda}$ as the identity on the second tensor factor (i.e., as $\mathcal{O} \otimes \textbf{1}$), the restriction of $\mathcal{O}_i(\vec{p}_i,0)$ to $\mathcal{H}^{\Lambda e^u}$ is well-defined (i.e., by restricting $\mathcal{O} \otimes \textbf{1} \to \mathcal{O}$) and hence Eqn.~\eqref{RGcorrelatormatching1} is sensible.

From this perspective, our coarse graining of $|\Psi_0(\Lambda)\rangle \mapsto |\widetilde{\Psi}_0(\Lambda e^u)\rangle$ can be thought of as a mapping of states from $\mathcal{H}^\Lambda \to \mathcal{H}^{\Lambda e^u}$ that preserve the appropriate correlation functions below a certain spatial momentum scale.  However, the standard practice of RG is to spatially ``zoom out'' on a state after coarse graining it, and then rescale various operators (called \textit{renormalization}) so that correlation functions have a standard form (this procedure will be elucidated below).  As a result of the spatially zooming out, the cutoff scale $\Lambda e^u$ is restored to $\Lambda$, although no new correlations in the state are introduced since we are just spatially shrinking the entire state by zooming out.  Hence we would like a mapping $\mathcal{H}^{\Lambda  e^u} \to \mathcal{H}^{\Lambda}$ corresponding to this spatially zooming out and the ``renormalization'' of operators.  We denote the resulting state by $|\Psi_0(\Lambda e^u)\rangle \in \mathcal{H}^\Lambda$, this time without a tilde over the $\Psi$ to indicate that this state belongs to the same Hilbert space as $|\Psi_0(\Lambda)\rangle$.  We desire for our cMERA to capture $|\Psi_0(\Lambda e^u)\rangle$ for $-\infty < u \leq 0$, and not $|\widetilde{\Psi}_0(\Lambda e^u)\rangle$.  The state $|\Psi_0(\Lambda e^u)\rangle$ will satisfy
\begin{equation}
\label{RGcorrelatormatching2}
\langle \Psi_0(\Lambda e^u)| \mathcal{O}_1'(\vec{p}_1 \, e^{-u},0) \cdots \mathcal{O}_n'(\vec{p}_n \, e^{-u},0)|\Psi_0(\Lambda e^u)\rangle = \langle \Psi_0(\Lambda)| \mathcal{O}_1(\vec{p}_1,0) \cdots \mathcal{O}_n(\vec{p}_n,0)|\Psi_0(\Lambda)\rangle
\end{equation}
for $|\vec{p}_i| \leq \Lambda e^u$, which is analogous to Eqn.~\eqref{RGcorrelatormatching1} above.  The $\{\mathcal{O}_i'\}$ are related to the $\{\mathcal{O}_i\}$ by appropriately multiplicatively rescaling (e.g., renormalizing) the $\widehat{\phi}$'s and $\widehat{\pi}$'s which comprise the original un-primed operators (this will be detailed below).

We will use \textit{spatial Wilsonian RG} to find $|\Psi_0(\Lambda e^u)\rangle$.  Note that \textit{spatial} Wilsonian RG is distinct from the more standard form of \textit{spacetime} Wilsonian RG which is applied to Euclidean field theories.  Spacetime (Euclidean) Wilsonian RG is not suitable for our present purposes -- in particular, the truncation of space\textit{time} modes (i.e., $\vec{p}$ \textit{as well as} the frequency which is Fourier-conjugate to time $t$) renders a renormalized state and Hamiltonian as ill-defined.  

For completeness, we detail here the procedure for spatial Wilsonian renormalization of the free massive scalar field.  We invite readers who are already familiar with this or related schemes to skip to the bullet points which immediately follow.

\subsubsection*{Spatial Wilsonian RG for free massive scalar field theory}

The ground state of the free scalar field theory $|\Psi_0(\Lambda)\rangle$ at scales $|\vec{p}| \leq \Lambda$ is the lowest-energy eigenstate of the Hamiltonian $H^\Lambda$ in Eqn.~\eqref{freeHam2} (which has a momentum cutoff at $\Lambda$).  It is clear that the parameters of the Hamiltonian $H^\Lambda$ determine the parameters of the ground state $|\Psi_0(\Lambda)\rangle$.  In particular, since the only parameter of $H^\Lambda$ is the mass $m$, this is the only parameter that shows up in $|\Psi_0(\Lambda)\rangle$ (see Eqn.~\eqref{groundstate1}).

The ground state is also specified by all of its equal-time correlation functions.  Recall that equal-time correlation functions of the ground state can be computed via the partition function path integral
\begin{align}
\label{partition1}
&Z^\Lambda[\{J_i(\vec{p})\}] = \nonumber \\
& \qquad \lim_{T \to \infty(1-i \epsilon)}\int \prod_{|\vec{p}| \leq \Lambda} \mathcal{D}\phi(\vec{p},t) \, \mathcal{D}\pi(\vec{p},t) \, e^{i \int_{-T}^T dt \left(\int^\Lambda d^d \vec{p} \, \pi(\vec{p},t) \dot{\phi}(\vec{p},t) - \mathscr{H}^\Lambda\big(\phi(\vec{p},t), \pi(\vec{p},t)\big) \right)} \, e^{- i \int^\Lambda d^d \vec{p} \, \sum_i J_i(\vec{p}) \mathcal{O}_i(\vec{p},0) }
\end{align}
where $\epsilon$ is a positive, infinitesimal parameter, and $\mathscr{H}^\Lambda\big(\phi(\vec{p},t), \pi(\vec{p},t)\big)$ is the Hamiltonian density
\begin{equation}
\label{HamDensity1}
\mathscr{H}^\Lambda\big(\phi(\vec{p},t), \pi(\vec{p},t)\big) = \frac{1}{2}\int^\Lambda d^d \vec{p} \, \left( \pi(\vec{p},t) \pi(-\vec{p},t) + \phi(\vec{p},t) (\vec{p}\hspace{.5mm}^2 + m^2) \phi(-\vec{p},t)\right)\,. 
\end{equation}
Equal-time correlation functions of the ground state can be computed by the relation
\begin{equation}
\label{generatingfunction1}
\langle \Psi_0(\Lambda)| \mathcal{O}_{1}(\vec{p}_1,0) \cdots \mathcal{O}_{n}(\vec{p}_n,0) |\Psi_0(\Lambda)\rangle = \frac{1}{i^n} \, \frac{\delta}{\delta J_{1}(\vec{p}_1)} \cdots \frac{\delta}{\delta J_{n}(\vec{p}_n)} \, \log Z^\Lambda[\{J_i(\vec{p})\}]\bigg{|}_{J=0}\,
\end{equation}
It is no surprise that the Hamiltonian density $\mathscr{H}^\Lambda$ makes an appearance in the partition function, since the data of the ground state is of course encoded in the Hamiltonian and its parameters.

If we only wanted to compute equal-time correlation functions of operators $\mathcal{O}_i(\vec{p},0)$ with $|\vec{p}| \leq \Lambda e^u$ for $-\infty < u < 0$, then we can simplify the partition function in Eqn.~\eqref{partition1} by performing spatial Wilsonian RG.  Suppose we write
\begin{align}
\phi(\vec{p},t) &= \begin{cases} 
   \phi_{<}(\vec{p},t) & \text{if } |\vec{p}| \leq \Lambda e^u \\
   \phi_{>}(\vec{p},t) & \text{if } \Lambda e^u < |\vec{p}| \leq \Lambda
  \end{cases} \\ \nonumber \\
\pi(\vec{p},t) &= \begin{cases} 
   \pi_{<}(\vec{p},t) & \text{if } |\vec{p}| \leq \Lambda e^u \\
   \pi_{>}(\vec{p},t) & \text{if } \Lambda e^u < |\vec{p}| \leq \Lambda
  \end{cases}
\end{align}
Since we only want to consider correlation functions with $|\vec{p}| \leq \Lambda e^u$, we can assume that our $\mathcal{O}_i(\vec{p},0)$'s can be written solely in terms of $\phi_{<}(\vec{p},0)$'s and $\pi_{<}(\vec{p},0)$'s.  Then we can write $Z^\Lambda[\{J_i(\vec{p})\}]$ as
\begin{align}
\label{partitionintegrated0}
& \int\prod_{|\vec{p}| \leq \Lambda e^u}\mathcal{D}\phi_{<}(\vec{p},t) \, \mathcal{D}\pi_{<}(\vec{p},t) \prod_{\Lambda e^u < |\vec{p}| \leq \Lambda} \mathcal{D}\phi_{>}(\vec{p},t) \, \mathcal{D}\pi_{>}(\vec{p},t) \nonumber \\
& \,\, \times \, \exp\left(i \int_{-T}^T dt \int^{\Lambda e^u} d^d \vec{p} \, \left(\pi_{<}(\vec{p},t) \dot{\phi}_{<}(\vec{p},t) - \frac{1}{2}\pi_{<}(\vec{p},t)\pi_{<}(-\vec{p},t) - \frac{1}{2}\phi_{<}(\vec{p},t) (\vec{p}\hspace{.5mm}^2 + m^2) \phi_{<}(-\vec{p},t)\right)
\right) \nonumber \\
& \,\, \times \, \exp\left(i \int_{-T}^T dt \int_{\Lambda e^u}^{\Lambda} d^d \vec{p} \, \left(\pi_{>}(\vec{p},t) \dot{\phi}_{>}(\vec{p},t) - \frac{1}{2}\pi_{>}(\vec{p},t)\pi_{>}(-\vec{p},t) - \frac{1}{2}\phi_{>}(\vec{p},t) (\vec{p}\hspace{.5mm}^2 + m^2) \phi_{>}(-\vec{p},t)\right)
\right) \nonumber \\
& \,\, \times \, \exp\left(i \int^{\Lambda e^u} d^d \vec{p} \, \sum_i J_i(\vec{p}) \mathcal{O}_i(\vec{p},0) \right) \\ \nonumber \\
\label{partitionintegrated1}
=\, & \, C \int \prod_{|\vec{p}| \leq \Lambda e^u} \mathcal{D}\phi_{<}(\vec{p},t) \, \mathcal{D}\pi_{<}(\vec{p},t) \, e^{i \int_{-T}^T dt \int^{\Lambda e^u} d^d \vec{p} \, \left(\pi_{<}(\vec{p},t) \dot{\phi}_{<}(\vec{p},t) - \frac{1}{2}\pi_{<}(\vec{p},t)\pi_{<}(-\vec{p},t) - \frac{1}{2}\phi_{<}(\vec{p},t) (\vec{p}\hspace{.5mm}^2 + m^2) \phi_{<}(-\vec{p},t)\right)} \nonumber \\
& \qquad \qquad \qquad \qquad \qquad \qquad \qquad \qquad \qquad \qquad \qquad \qquad \times e^{i \int^{\Lambda e^u} d^d \vec{p} \, \sum_i J_i(\vec{p}) \mathcal{O}_i(\vec{p},0)}
\end{align}
where $C$ is some multiplicative constant.  This constant will not matter to us since we are always considering derivatives of the logarithm of $Z^\Lambda[\{J_i(\vec{p})\}]$ to compute correlation functions as per Eqn.~\eqref{generatingfunction1}, in which case the $C$ will cancel out.

Comparing Eqn.~\eqref{partitionintegrated1} with Eqn.~\eqref{partition1}, we see that the modified partition function has the exact same form as the original one, except that the momentum integrals in the exponentials are integrated up to $\Lambda e^u$ instead of $\Lambda$.  This happens because the integrals over $\int \mathcal{D}\phi_{>}(\vec{p},t) \, \mathcal{D}\pi_{>}(\vec{p},t)$ in Eqn.~\eqref{partitionintegrated0} decouple from the integrals over $\int \mathcal{D}\phi_{<}(\vec{p},t) \, \mathcal{D}\pi_{<}(\vec{p},t)$.  To be clear, this only happens for the free massive scalar field theory since it is quadratic in $\phi$'s and $\pi$'s, and such a decoupling does not occur for interacting theories with non-quadratic interaction terms (such as a $\lambda \, \phi^4$ interaction).  

We can read off from Eqn.~\eqref{partitionintegrated1} that the effective Hamiltonian $\widetilde{H}^{\Lambda e^u}$ on the Hilbert space $\mathcal{H}^{\Lambda e^u}$ defined in Eqn.~\eqref{smallerHilbert1} is simply
\begin{equation}
\widetilde{H}^{\Lambda e^u} = \frac{1}{2} \int^{\Lambda e^u} d^d \vec{k} \, \left(\widehat{\pi}(\vec{k}) \widehat{\pi}(-\vec{k}) + \widehat{\phi}(\vec{k}) \left(\vec{k}^{\,2} + m^2\right) \widehat{\phi}(-\vec{k}) \right)
\end{equation} 
and its ground state is
\begin{equation}
\langle \phi |\widetilde{\Psi}_0(\Lambda e^u)\rangle = \mathcal{N} \, \exp\left(-\frac{1}{2} \int^{\Lambda e^u} d^d \vec{k} \,\, \phi(\vec{k}) \, \sqrt{\vec{k}^{\,2} + m^2} \, \phi(-\vec{k}) \right)\,.
\end{equation}
However, we instead desire a spatially zoomed-out and ``renormalized'' state $|\Psi_0(\Lambda e^u)\rangle$ defined on $\mathcal{H}^\Lambda$.  We will find this state by completing the renormalization group procedure.

Let us turn our attention back to our result for the partition function in Eqn.~\eqref{partitionintegrated1} above.  If we wanted to ``zoom in'' on the remaining modes $|\vec{p}| \leq \Lambda e^u$, it is natural to rescale $\vec{p} \to e^{-u}\,\vec{p}$ so that the cutoff becomes $\Lambda$ again.  This corresponds to ``zooming out'' in position space by a factor of $e^{u}$ (i.e., $\vec{x} \to e^{u} \vec{x}$), and equivalently ``zooming in'' in momentum space by a factor of $e^{-u}$ (i.e., $\vec{p} \to e^{-u} \vec{p}$).  If we ``zoom out'' in position space, it is also natural to simultaneously ``zoom out'' in time by the same factor, namely $t \to e^{u} t$, so that we stretch the timescales under consideration.  This corresponds to rescaling the Fourier-conjugate to $t$, say $\omega$, as $\omega \to e^{-u} \omega$.  To summarize, we want $(\vec{x},t) \to (e^{u} \vec{x}, e^{u} t)$ or equivalently $(\vec{p},\omega) \to (e^{-u} \vec{p}, e^{-u} \omega)$.  Since we are working with $\vec{p}$ and $t$, we will take $(\vec{p},t) \to (e^{-u} \vec{p}, e^{u} t)$.

Changing variables $\vec{p} \to e^{-u}\,\vec{p}$ and $t \to e^{u} t$, Eqn.~\eqref{partitionintegrated1} becomes
\begin{align}
\label{partitionintegrated2}
& C \int \prod_{|\vec{p}| \leq \Lambda} \mathcal{D}\phi_{<}(e^u\vec{p},e^{-u}t) \, \mathcal{D}\pi_{<}(e^u\vec{p},e^{-u}t) \nonumber \\
& \qquad \times \, \exp\bigg(i \int_{-T e^{u}}^{T e^{u}} dt \int^{\Lambda} d^d \vec{p} \,e^{(d-1)u} \big(\pi_{<}(e^u \vec{p},e^{-u}t) \dot{\phi}_{<}(e^u \vec{p},e^{-u}t) \nonumber \\
&\qquad \qquad \qquad \qquad \qquad - \frac{1}{2}\pi_{<}(e^u \vec{p},e^{-u}t)\pi_{<}(-e^u \vec{p},e^{-u}t) - \frac{1}{2}\phi_{<}(e^u \vec{p},e^{-u}t) (e^{2u}\vec{p}\hspace{.5mm}^2 + m^2) \phi_{<}(-e^u\vec{p},e^{-u}t)\big)\bigg) \nonumber \\
& \qquad \times \, \exp\left(i \int^{\Lambda} d^d \vec{p} \,e^{du}\, \sum_i J_i(\vec{p}) \mathcal{O}_i(e^u\,\vec{p},0) \right)\,. \nonumber \\
\end{align}
We see that the integral over time in the action has been changed by $\int_{-T}^T dt \to \int_{- T e^u}^{T e^u} dt \, e^{-u}$.  This change of the limits of integration of time will not end up mattering, since in Eqn.~\eqref{partition1} the partition function is defined by taking the $T \to \infty(1-i\epsilon)$ limit.

Notice that in Eqn.~\eqref{partitionintegrated2}, the fluctuations of the $\phi_{<}$ field are rescaled relative to the original scalar field, since our kinetic term is
\begin{equation}
e^{(d+1)u}\,\phi_{<}(e^u \vec{p},e^{-u}t)\,\vec{p}^2\,\phi_{<}(-e^u \vec{p},e^{-u}t)\,,
\end{equation}
which differs from the standard kinetic term $\phi \, \vec{p}^2 \, \phi$ by an extra factor of $e^{(d+1)u}$ out front.  To put the fluctuations induced by the kinetic term into the same form as the original theory, we redefine the fields by
\begin{align}
\label{renormalize1}
e^{\frac{d+1}{2}  \cdot u} \, \phi_{<}(e^u \vec{p},e^{-u}t) \quad &\longrightarrow \quad \phi(\vec{p},t) \\
\label{renormalize2}
e^{\frac{d-1}{2}  \cdot u} \, \pi_{<}(e^u \vec{p}, e^{-u} t) \quad &\longrightarrow \quad \pi(\vec{p},t)\,.
\end{align}
Notice that we are renormalizing $\phi_{<}$ and $\pi_{<}$ differently: $\phi$ carries a factor of $e^{\frac{d+1}{2} \cdot u}$ relative to $\phi_{<}$, whereas $\pi$ carries a factor of  $e^{\frac{d-1}{2} \cdot u}$ relative to $\pi_{<}$\,.  This difference is to enforce the standard, equal-time canonical commutation relation for the corresponding operators $\hat{\phi}$ and $\widehat{\pi}$, so that $[\widehat{\phi}(\vec{p},0), \widehat{\pi}(\vec{k},0)] = i \delta^d(\vec{p} + \vec{k})$.  Explicitly,
\begin{align}
[\widehat{\phi}(\vec{p},0), \widehat{\pi}(\vec{k},0)] &= [e^{\frac{d+1}{2} \cdot u}\phi_{<}(e^u \vec{p}), e^{\frac{d-1}{2}\cdot u} \widehat{\pi}_{<}(e^{u}\vec{k},0)] \nonumber \\
&= i \, e^{du} \, \delta^d(e^{u} \vec{p} + e^{u} \vec{k}) \nonumber \\
&= i \, \delta^d(\vec{p} + \vec{k})\,. \nonumber
\end{align}
With the above rescalings, the partition function in Eqn.~\eqref{partitionintegrated2} is 
\begin{align}
\label{partitionintegrated3}
Z^{\Lambda e^u}[\{J_i(\vec{p})\}] :=& \,\,\lim_{T \to \infty(1-i \epsilon)} C \! \int \prod_{|\vec{p}| \leq \Lambda} \mathcal{D}\phi(\vec{p},t) \, \mathcal{D}\pi(\vec{p},t) \nonumber \\
& \qquad \qquad \times \, \exp\bigg(i \int_{-T}^T dt \int^{\Lambda} d^d \vec{p} \,\big(\pi(\vec{p},t) \dot{\phi}(\vec{p},t) \nonumber \\
&\qquad \qquad \qquad \qquad \qquad \qquad \qquad - \frac{1}{2}\pi(\vec{p},t)\pi(-\vec{p},t) - \frac{1}{2}\phi(\vec{p},t) (\vec{p}\hspace{.5mm}^2 + e^{-2u} m^2) \phi(-\vec{p},t)\big)\bigg) \nonumber \\
& \qquad \qquad \times \, \exp\left(i \int^{\Lambda} d^d \vec{p} \,e^{du}\, \sum_i J_i(\vec{p}) \mathcal{O}_i^{\,'}(\vec{p},0) \right)\,,
\end{align}
where the primed $\mathcal{O}_i^{\,'}$ operators are the same as the unprimed $\mathcal{O}_i$ operators, but written in terms of $\phi,\pi$ instead of $\phi_<\,, \pi_<$ as per Eqn.'s~\eqref{renormalize1} and~\eqref{renormalize2}.  In particular, writing $\mathcal{O}_i\big(\phi_<(\vec{p},0), \pi_<(\vec{p},0)\big)$ to denote the operator's dependence on $\phi_<(\vec{p},0)$ and $\pi_<(\vec{p},0)$, we have
\begin{equation}
\mathcal{O}_i^{\,'}\big(\phi(\vec{p},0), \pi(\vec{p},0)\big)\hspace{1mm} := \hspace{1mm} \mathcal{O}_i\big(e^{\frac{d+2}{2} \cdot u}\phi_<(e^u\vec{p},0), e^{-\frac{d+2}{2} \cdot u} \pi_<(e^u\vec{p},0)\big)\,.
\end{equation}
From Eqn.~\eqref{partitionintegrated3}, we read off that the renormalized Hamiltonian density at scale $\Lambda e^u$, namely $\mathscr{H}^{\Lambda e^u}$, is
\begin{equation}
\label{HamDensity2}
\mathscr{H}^{\Lambda e^u}\big(\phi(\vec{p},t), \pi(\vec{p},t)\big) = \frac{1}{2}\int^\Lambda d^d \vec{p} \, \left( \pi(\vec{p},t) \pi(-\vec{p},t) + \phi(\vec{p},t) (\vec{p}^2 + e^{-2u}m^2) \phi(-\vec{p},t)\right)\,,
\end{equation}
which differs from $\mathscr{H}^\Lambda$ in Eqn.~\eqref{HamDensity1} since the mass is rescaled by a factor of $e^{-2u} > 1$.  This represents the flow of the mass under the renormalization group.  Then the renormalized Hamiltonian at scale $\Lambda e^u$ is
\begin{equation}
\label{renormfreeHam1}
H^{\Lambda e^u} = \frac{1}{2} \int^{\Lambda} d^d \vec{k} \, \left(\widehat{\pi}(\vec{k}) \widehat{\pi}(-\vec{k}) + \widehat{\phi}(\vec{k}) \left(\vec{k}^2 + e^{-2u} m^2\right) \widehat{\phi}(-\vec{k}) \right)\,.
\end{equation} 
Finally, we see that since the Hamiltonian above has the same form as the original un-renormalized Hamiltonian $H^\Lambda$, except with a rescaled mass $m^2 \to e^{-2u} m^2$.  The renormalized ground state wavefunctional $|\Psi_0(\Lambda e^u)\rangle \in \mathcal{H}^\Lambda$ is\footnote{As mentioned before, we have used $\mathcal{N}$ as a placeholder for a normalization constant, although it is different from the $\mathcal{N}$ in Eqn.~\eqref{groundstate1} and other equations.}
\begin{equation}
\langle \phi |\Psi_0(\Lambda e^u)\rangle = \mathcal{N} \, \exp\left(-\frac{1}{2} \int^\Lambda d^d \vec{k} \,\, \phi(\vec{k}) \, \sqrt{\vec{k}^{\,2} + e^{-2u} m^2} \, \phi(-\vec{k}) \right)\,,
\end{equation}
which is the appropriate renormalized state we desire for our cMERA procedure.

To recap, we have found $|\Psi_0(\Lambda e^u)\rangle$ by performing spatial Wilsonian RG on the partition function $Z^\Lambda$ which yields the renormalized partition function $Z^{\Lambda e^u}$.  Just as $Z^\Lambda$ generates correlation functions of the state $|\Psi_0(\Lambda)\rangle$, the renormalized partition function $Z^{\Lambda e^u}$ generates correlation functions of the state $|\Psi_0(\Lambda e^u)\rangle$.  The only difference between $Z^{\Lambda}$ and $Z^{\Lambda e^u}$ is the mass of the scalar field, and so in turn the only difference between $|\Psi_0(\Lambda)\rangle$ and $|\Psi_0(\Lambda e^u)\rangle$ must be the rescaled mass.  One can also read off from $Z^{\Lambda e^u}$ the renormalized Hamiltonian density $\mathscr{H}^{\Lambda e^u}$ which allows us to determine the renormalized Hamiltonian $H^{\Lambda e^u}$.  Indeed, the ground state wavefunctional of $H^{\Lambda e^u}$ is $|\Psi_0(\Lambda e^u)\rangle$.

We emphasize that our ability to infer the renormalized state and Hamiltonian is due to the fact that we did \textit{not} integrate out any temporal modes.  As mentioned before, in standard Wilsonian RG, one usually considers a Euclidean field theory and then truncates the space\textit{time} modes (i.e., $\vec{p}$ \textit{as well as} the frequency which is Fourier-conjugate to time $t$), in which case a renormalized state and Hamiltonian are ill-defined.  This is why we have emphasized that we are performing \textit{spatial} Wilsonian RG, since we are not touching the temporal modes.

The spatial Wilsonian RG procedure is recapped in the bullet points below: \\ \\

\noindent \textbf{Summary of spatial Wilsonian RG for free massive scalar field theory}
\begin{itemize}
\item We compute the ground state wavefunctional of the free scalar field at scale $\Lambda$, namely $|\Psi_0(\Lambda)\rangle$, and observe that it only depends on the mass $m^2$.
\item We consider the partition function $Z^\Lambda$ which generates ground state correlation functions of $|\Psi_0(\Lambda)\rangle$.
\item We perform \textit{spatial} Wilsonian RG on $Z^\Lambda$ to obtain $Z^{\Lambda e^u}$.  By \textit{spatial} Wilsonian RG, we mean that we do not integrate out temporal degrees of freedom (i.e., the frequency Fourier-conjugate to time), and \textit{only} integrate out momentum modes.  This way, a renormalized ground state wavefunctional and renormalized Hamiltonian are well-defined.
\item We observe that $Z^\Lambda$ differs from $Z^{\Lambda e^u}$ only by a rescaling of the mass $m^2 \to e^{-2u} m^2$, and hence $|\Psi_0(\Lambda e^u)\rangle$ has the same form as $|\Psi_0(\Lambda)\rangle$ but with a rescaled mass:
\begin{equation}
\label{renormwf1}
\langle \phi |\Psi_0(\Lambda e^u)\rangle = \mathcal{N} \, \exp\left(-\frac{1}{2} \int^\Lambda d^d \vec{k} \,\, \phi(\vec{k}) \, \sqrt{\vec{k}^{\,2} + e^{-2u} m^2} \, \phi(-\vec{k}) \right)\,,
\end{equation}
\item The renormalized Hamiltonian $H^{\Lambda e^u}$ can be read off from $Z^{\Lambda e^u}$, and has a rescaled mass relative to $H^\Lambda$.  The ground state wavefunctional of $H^{\Lambda e^u}$ is $|\Psi_0(\Lambda e^u)\rangle$.
\end{itemize}
$$$$
\subsubsection*{Returning to cMERA for free massive scalar field theory}

Now we return to our construction of cMERA for $|\Psi_0(\Lambda e^u)\rangle$.  Our goal is to obtain $|\Psi_0(\Lambda e^u)\rangle$ by entangling momentum modes for $|\vec{k}| \leq \Lambda$ which comprise a simple, scale-invariant state $|\Omega\rangle$.  First, we discuss how to construct simple, scale-invariant states.

Following the conventions of \cite{cMERA1, cMERA2}, we let $L$ be the non-relativistic scale transformation
\begin{equation}
L = - \frac{1}{2} \int d^d \vec{x} \, \left(\widehat{\pi}(x) \left(\vec{x} \cdot \vec{\nabla}\widehat{\phi}(\vec{x})\right) + \left(\vec{x} \cdot \vec{\nabla}\widehat{\phi}(\vec{x})\right) \widehat{\pi}(\vec{x}) + \frac{d}{2} \, \widehat{\phi}(\vec{x}) \widehat{\pi}(\vec{x}) + \frac{d}{2} \, \widehat{\pi}(\vec{x}) \widehat{\phi}(\vec{x})  \right)
\end{equation}
so that
\begin{align}
e^{-i u L} \, \widehat{\phi}(x) \, e^{i u L} &= e^{\frac{d}{2} \, u}\, \widehat{\phi}(e^u x)\,, \qquad \quad e^{-i u L} \, \widehat{\phi}(\vec{k}) \, e^{i u L} = e^{-\frac{d}{2} \, u}\, \widehat{\phi}(e^{-u} \vec{k})\,, \nonumber \\ \nonumber \\
e^{-i u L} \, \widehat{\pi}(x) \, e^{i u L} &= e^{\frac{d}{2} \, u}\, \widehat{\pi}(e^u x)\,, \qquad \quad e^{-i u L} \, \widehat{\pi}(\vec{k}) \, e^{i u L} = e^{-\frac{d}{2} \, u}\, \widehat{\pi}(e^{-u} \vec{k})\,.
\end{align}
Then a scale-invariant state $|\Omega\rangle$ satisfies $L |\Omega\rangle = 0$, or equivalently $e^{-i u L}|\Omega\rangle = |\Omega\rangle$ for all $u$.

To construct a useful class of scale-invariant states, consider operators $\mathcal{O}(\vec{k})$ which are \textit{only} built out of $\widehat{\phi}$'s and satisfy
\begin{equation}
\label{Ocondition1}
e^{-i u L} \, \mathcal{O}(\vec{k}) \, e^{i u L} = e^{-\frac{d}{2} \,u} \, \mathcal{O}(e^{-u} \vec{k})\,.
\end{equation}
Then for some particular $\mathcal{O}(\vec{k})$ satisfying the above equation, we can define a scale-invariant state $|\Omega\rangle$ by
\begin{equation}
\label{scaleinvcond1}
\left(\mathcal{O}(\vec{k}) + i \, \widehat{\pi}(\vec{k}) \right) |\Omega\rangle = 0\, \quad \text{for all} \quad \vec{k} \in \mathbb{R}^d\,.
\end{equation}
Such a $|\Omega\rangle$ state is unique up to an overall constant, since it is defined by the separable first-order differential equations
\begin{equation}
\left(\mathcal{O}(\vec{k}) + \frac{\delta}{\delta \phi(-\vec{k})} \right) \Psi_{\Omega}[\phi] = 0 \quad \text{for all} \quad \vec{k} \in \mathbb{R}^d\,,
\end{equation}
where $\Psi_{\Omega}[\phi] = \langle \phi | \Omega\rangle$.  Choosing $|\Omega\rangle$ to be normalized (assuming that it is normalizable), $|\Omega\rangle$ is uniquely specified by Eqn.~\eqref{scaleinvcond1}.  For a good discussion of functional differential equations, see \cite{Hatfield1}.

Furthermore, such a state $|\Omega\rangle$ is scale-invariant since $\left(\mathcal{O}(\vec{k}) + i \, \pi(\vec{k})\right) |\Omega\rangle = 0$ is equivalent to $e^{-i u L}\left(\mathcal{O}(\vec{k}) + i \, \pi(\vec{k})\right) e^{i u L} \cdot e^{-i u L}|\Omega\rangle = 0$ and thus
\begin{equation}
e^{-\frac{d}{2}\,u}\left(\mathcal{O}(e^{-u}\vec{k}) + i \, \widehat{\pi}(e^{-u}\vec{k}) \right) e^{-i u L} |\Omega\rangle = 0\, \quad \text{for all} \quad \vec{k} \in \mathbb{R}^d
\end{equation}
which implies
\begin{equation}
\label{scaleinvcond2}
\left(\mathcal{O}(\vec{k}) + i \, \widehat{\pi}(\vec{k})\right) e^{- i u L}|\Omega\rangle = 0\, \quad \text{for all} \quad \vec{k} \in \mathbb{R}^d\,.
\end{equation}
Since $|\Omega\rangle$ is uniquely specified by Eqn.~\eqref{scaleinvcond1} along with the normalization condition, and since Eqn.~\eqref{scaleinvcond2} has the same form as Eqn.~\eqref{scaleinvcond1}, it follows that $|\Omega\rangle = e^{- i u L} |\Omega\rangle$ and hence $|\Omega\rangle$ is scale-invariant.

We will consider several examples of $|\Omega\rangle$'s, but the simplest is the one satisfying
\begin{equation}
\label{scaleinvcond3}
\left(\sqrt{M} \, \widehat{\phi}(\vec{k}) + \frac{i}{\sqrt{M}}\,\hat{\pi}(\vec{k})\right) |\Omega\rangle = 0\, \quad \text{for all} \quad \vec{k} \in \mathbb{R}^d
\end{equation} 
or equivalently
\begin{equation}
\label{scaleinvcond4}
\left(\sqrt{M} \, \widehat{\phi}(\vec{x}) + \frac{i}{\sqrt{M}}\,\widehat{\pi}(\vec{x})\right) |\Omega\rangle = 0\, \quad \text{for all} \quad \vec{x} \in \mathbb{R}^d
\end{equation} 
where $M$ is a constant.  Comparing Eqn.~\eqref{scaleinvcond3} to Eqn.~\eqref{scaleinvcond1}, we have $\mathcal{O}(\vec{k}) = M \widehat{\phi}(\vec{k})$ and we have also have rescaled the whole equation by an overall factor of $1/\sqrt{M}$.  The state $|\Omega\rangle$ can be expressed as a wavefunctional:
\begin{equation}
\label{IRstate1}
\langle \phi | \Omega\rangle = \mathcal{N} \, \exp\left(- \frac{1}{2} \int d^d \vec{k} \, \phi(\vec{k}) \, M \, \phi(-\vec{k}) \right) \,,
\end{equation}
or equivalently
\begin{equation}
\langle \phi | \Omega\rangle = \mathcal{N} \, \exp\left(- \frac{1}{2} \int d^d \vec{x} \, \phi(\vec{x}) \, M \, \phi(\vec{x}) \right) = \mathcal{N} \, \prod_{\vec{x}} \, \exp\left(- \frac{1}{2} \,d^d \vec{x} \, \phi(\vec{x}) \, M \, \phi(\vec{x}) \right)\,.
\end{equation}
From the above equation we see that $|\Omega\rangle$ is a \textit{product state} is position space, and thus is \textit{unentangled}.  This is directly analogous to the product state $|0\rangle \otimes |0\rangle \otimes \cdots \otimes |0\rangle$ in Eqn.~\eqref{productstate1} above.  Henceforth, we will refer to $|\Omega\rangle$ as the ``IR state.''  For a free scalar field, we expect that the ground state wavefunctional is not entangled in the IR, and has the form of $|\Omega\rangle$.  In the IR, the renormalized mass will diverge, causing two-point correlation functions in the ground state to spatially factorize.  One can also say that the heavy mass is spatially localizing entanglement in the ground state so that as the mass diverges entanglement goes away entirely.

Our goal is to find an entangler $K(s)$ so that $\mathcal{P}_s \, e^{-i \int_{-\infty}^u ds \, (K(s) + L)}|\Omega\rangle$ equals $|\Psi_0(\Lambda e^u)\rangle$ (given in Eqn.~\eqref{renormwf1}) for modes with $|\vec{k}| \leq \Lambda$.  Explicitly, we want
\begin{align}
\label{whatwewant1}
\langle \phi| \mathcal{P}_s \, e^{-i \int_{-\infty}^u ds \, (K(s) + L)}|\Omega\rangle = \mathcal{N} \, \exp\left\{-\frac{1}{2} \int^\Lambda d^d \vec{k} \, \phi(\vec{k}) \, \sqrt{\vec{k}^2 + e^{-2u} m^2} \, \phi(\vec{-k}) - \underbrace{\frac{1}{2} \int_{\Lambda}^\infty d^d \vec{k} \, \phi(\vec{k}) \, M \, \phi(-\vec{k})}_{\text{leftover, unused modes}} \right\}\,. \nonumber \\
\end{align}
Notice that the leftover, unused modes designated above are all above the cutoff scale $\Lambda$.  At the UV scale (i.e., when $u=0$), we see that there is a discontinuity in the wavefunctional unless $M = \sqrt{\Lambda^2 + m^2}$, which forces the ``entangled'' modes below $\Lambda$ to agree with the modes of IR ansatz above $\Lambda$, exactly at the cutoff scale $\Lambda$.  This continuity consideration is aesthetic instead of essential.  Henceforth we define
\begin{equation}
M := \sqrt{\Lambda^2 + m^2}
\end{equation}
and write Eqn.~\eqref{whatwewant1} more compactly as
\begin{align}
\label{whatwewant2}
\langle \phi| \mathcal{P}_s \,e^{-i \int_{-\infty}^u ds \, (K(s) + L)}|\Omega\rangle = \mathcal{N} \, \exp\left\{-\frac{1}{2} \int d^d \vec{k} \, \theta(1 - |\vec{k}|/\Lambda) \, \phi(\vec{k}) \, \sqrt{\vec{k}^2 + e^{-2u} m^2} \, \phi(\vec{-k}) \right\}\,,
\end{align}
where $\theta(z)$ is the Heaviside step function.

Instead of using the Heaviside step function which is discontinuous, we may alternatively use a smooth cutoff function, such as a sigmoid (which is infinitely differentiable).  This corresponds to choosing a ``softer'' UV cutoff for our theory.  The discontinuity of the Heaviside step function can lead to various pathologies, which are ameliorated when the discontinuity is smoothed over.  The effects of the discontinuity are rendered as non-physical.  Therefore, we will treat $\theta(z)$ as a smooth cutoff function when appropriate, and will point out when we do so.

Letting $U(u) = \mathcal{P}_s \,  e^{-i \int_{-\infty}^u ds \, (K(s) + L)}$, using Eqn.'s~\eqref{GaussianU3} and~\eqref{GaussianU4} we know that our desired state is given by
\begin{align}
\label{Ueq1}
U(u) \, |\Omega\rangle &= \exp\left\{ i \int^\Lambda d^d \vec{k} \, \frac{1}{8} \log\left(\frac{\vec{k}^2 + e^{-2u} m^2}{M^2}\right) \, \left(\phi(\vec{k})\pi(-\vec{k}) + \pi(\vec{k})\phi(-\vec{k}) \right)\right\} \, |\Omega\rangle \nonumber \\ \nonumber \\
&= \exp\left\{ i \int^\Lambda d^d \vec{k} \, \frac{1}{8} \log\left(\frac{\vec{k}^2 + e^{-2u} m^2}{\Lambda^2 + m^2}\right) \, \left(\phi(\vec{k})\pi(-\vec{k}) + \pi(\vec{k})\phi(-\vec{k}) \right)\right\} \, |\Omega\rangle
\end{align}
and we want to match this to $\mathcal{P}_s \, e^{-i \int_{-\infty}^u ds \, (K(s) + L)}|\Omega\rangle$.  Rewriting
\begin{equation}
\label{rewrite1}
\mathcal{P}_s \, e^{-i \int_{-\infty}^u ds \, (K(s) + L)}|\Omega\rangle = e^{-i u L}\, \mathcal{P}_s \, e^{-i \int_{-\infty}^u ds \, \widetilde{K}(s)}|\Omega\rangle
\end{equation}
where
\begin{equation}
\label{rewrite2}
\widetilde{K}(s) = e^{i s L} \, K(s) \, e^{- i s L}\,,
\end{equation}
since $|\Omega\rangle = e^{i u L} |\Omega\rangle$ we can write
\begin{equation}
\label{rewrite3}
U(u) \, |\Omega\rangle = e^{-iuL} \left(\mathcal{P}_s e^{-i \int_{-\infty}^u ds \, \widetilde{K}(s)}\right) \, e^{iuL} \, |\Omega\rangle\,.
\end{equation}
Using Eqn.~\eqref{Ueq1} we find that $\left(\mathcal{P}_s e^{-i \int_{-\infty}^u ds \, \widetilde{K}(s)}\right) \, |\Omega\rangle$ equals
\begin{equation}
\exp\left\{ i \int d^d \vec{k} \, \frac{1}{8} \log\left(e^{-2u}\frac{\vec{k}^2 + m^2}{\Lambda^2 + m^2}\right) \, \theta(1-|\vec{k}|/\Lambda e^u) \, \left(\phi(\vec{k})\pi(-\vec{k}) + \pi(\vec{k})\phi(-\vec{k}) \right)\right\} \, |\Omega\rangle\,.
\end{equation}
Since  $[\phi(\vec{k})\pi(-\vec{k}) + \pi(\vec{k})\phi(-\vec{k}), \phi(\vec{p})\pi(-\vec{p}) + \pi(\vec{p})\phi(-\vec{p})] = 0$ for any $\vec{k}, \vec{p}$, we do not need to worry about the path ordering.  Suppose that
\begin{equation}
K(s) = \int d^d\vec{k} \, f_{2,0}(\vec{k},s) \, \left(\phi(\vec{k})\pi(-\vec{k}) + \pi(\vec{k})\phi(-\vec{k}) \right)
\end{equation}
and thus
\begin{equation}
\widetilde{K}(s) = \int d^d\vec{k} \, f_{2,0}(\vec{k} e^{-s},s) \, \left(\phi(\vec{k})\pi(-\vec{k}) + \pi(\vec{k})\phi(-\vec{k}) \right)\,.
\end{equation}
Now we would like to find a function $f_{2,0}(\vec{k} e^{-s},s)$ satisfying
\begin{equation}
\label{satisfythis1}
\int_{-\infty}^u f_{2,0}(\vec{k} e^{-s},s) \, ds = - \frac{1}{8} \log\left(e^{-2u}\frac{\vec{k}^2 + m^2}{\Lambda^2 + m^2}\right) \, \theta(1-|\vec{k}|/\Lambda e^u)\,.
\end{equation}
To check that this equation is sensible, we would like to see if the right-hand side can be written as
\begin{equation}
f^{\text{(-1)}}(\vec{k},u) - f^{\text{(-1)}}(\vec{k},-\infty)
\end{equation}
where $f^{\text{(-1)}}(\vec{k},s)$ is an antiderivative of $f_{2,0}(\vec{k} e^{-s},s)$ with respect to $s$.  Indeed, letting
\begin{equation}
f^{\text{(-1)}}(\vec{k},s) = - \frac{1}{8} \log\left(e^{-2s}\frac{\vec{k}^2 + m^2}{\Lambda^2 + m^2}\right) \, \theta(1-|\vec{k}|/\Lambda e^s)
\end{equation}
we see that $\lim_{s \to -\infty} f^{\text{(-1)}}(\vec{k},s)$ goes to zero as a distribution in $\vec{k}$, since $f^{\text{(-1)}}(\vec{k},s)$ has a width in $\vec{k}$-space that goes as $\sim e^{-u}$ (due to the $\theta(1-|\vec{k}|/\Lambda e^s)$ term) but a height that only goes as $\sim u$ (from the logarithmic term).  Thus, Eqn.~\eqref{satisfythis1} is sensible, and so we can differentiate both sides of the equation with respect to $u$ to recover $f_{2,0}(\vec{k} e^{-s},s)$.

Differentiating both sides of Eqn.~\eqref{satisfythis1} with respect to $u$, we find
\begin{align}
f_{2,0}(\vec{k} e^{-u},u) &= \frac{1}{4} \theta(1-|\vec{k}|/\Lambda e^u) - \frac{1}{8} \log\left(e^{-2u}\frac{\vec{k}^2 + m^2}{\Lambda^2 + m^2}\right) \frac{|\vec{k}|}{\Lambda e^u} \, \theta'(1-|\vec{k}|/\Lambda e^u)\,.
\end{align}
Therefore,
\begin{align}
\widetilde{K}(s) = \int d^d \vec{k} \left[\frac{1}{4} \theta(1-|\vec{k}|/\Lambda e^s) - \frac{1}{8} \log\left(e^{-2s}\frac{\vec{k}^2 + m^2}{\Lambda^2 + m^2}\right) \frac{|\vec{k}|}{\Lambda e^s} \, \theta'(1-|\vec{k}|/\Lambda e^s)\right]\! \left[ \phi(\vec{k})\pi(-\vec{k}) + \pi(\vec{k})\phi(-\vec{k})\right]
\end{align}
and so
\begin{equation}
\label{entanglerMassive1}
K(s) = \int d^d \vec{k} \,  \left[\frac{1}{4} \theta(1-|\vec{k}|/\Lambda) - \frac{1}{8} \log\left(\frac{\vec{k}^2 + e^{-2s}m^2}{\Lambda^2 + m^2}\right) \frac{|\vec{k}|}{\Lambda} \, \theta'(1-|\vec{k}|/\Lambda)\right] \, \left[ \phi(\vec{k})\pi(-\vec{k}) + \pi(\vec{k})\phi(-\vec{k})\right]
\end{equation}
which is our result for the entangler of free massive scalar field theory.

Several comments are in order.  First, we emphasize that our result for the entangler is different from previous results \cite{cMERA1, cMERA2, cMERA3, cMERA4} which only generate the correct UV wavefunction.  By contrast, our entangler generates the correction wavefunction at all intermediate RG scales, as well as the UV.

There are several interesting features of our entangler in Eqn.~\eqref{entanglerMassive1} for non-zero mass $m$, in $d+1$ dimensions.  Here we take $\theta(z)$ to be a smooth version of the Heaviside step function.  Recall that $\phi(\vec{k})\pi(-\vec{k}) + \pi(\vec{k})\phi(-\vec{k})$ is the squeezing operator with respect to the $\vec{k}$ momentum modes.  Then the first term in Eqn.~\eqref{entanglerMassive1} term, namely
\begin{align*}
\int^\Lambda d^d \vec{k} \,  \frac{1}{4} \theta(1-|\vec{k}|/\Lambda) \, \left[ \phi(\vec{k})\pi(-\vec{k}) + \pi(\vec{k})\phi(-\vec{k})\right]\,,
\end{align*}
implements squeezing on all modes equitably for $|\vec{k}| \leq \Lambda$, and is negligible for $|\vec{k}| \geq \Lambda$.  In position space, this corresponds to
\begin{align*}
\int d^d \vec{x} \, d^d \vec{y} \,\,  B(|\vec{x}-\vec{y}|) \, \left[ \phi(\vec{x})\pi(\vec{y}) + \pi(\vec{y})\phi(\vec{x})\right]\,,
\end{align*}
where $B(z)$ is essentially a bump function with width $\sim 1/\Lambda$ and rapid decay for $z \gtrsim 1/\Lambda$.  So in position space, we are creating $2$-point correlations which are smeared over a characteristic length scale of $\sim 1/\Lambda$.

The second term in Eqn.~\eqref{entanglerMassive1} is
\begin{align*}
- \frac{1}{8} \int^\Lambda d^d \vec{k} \, \log\left(\frac{\vec{k}^2 + e^{-2u}m^2}{\Lambda^2 + m^2} \right) \frac{|\vec{k}|}{\Lambda} \, \theta'(1-|\vec{k}|/\Lambda) \, \left[ \phi(\vec{k})\pi(-\vec{k}) + \pi(\vec{k})\phi(-\vec{k})\right]\,.
\end{align*}
Since $\theta'(1-|k|/\Lambda)$ is peaked around $|\vec{k}| = \Lambda$, the $|\vec{k}|/\Lambda$ term can be approximately set to $1$. The presence of the $\theta'(1-|\vec{k}|/\Lambda)$ means that the above term is only ``activated'' for $|\vec{k}| \sim \Lambda$, namely at the cutoff scale.  The above term is easier to interpret in position space:
\begin{align*}
\int d^d \vec{x} \, d^d \vec{y} \,\,  C(|\vec{x}-\vec{y}|) \, D(|\vec{x}-\vec{y}|)\, \left[ \phi(\vec{x})\pi(\vec{y}) + \pi(\vec{y})\phi(\vec{x})\right]\,.
\end{align*}
Here, $C(|\vec{x}-\vec{y}|)$ is the Fourier transform of $\log\left(\frac{\vec{k}^2 + e^{-2u}m^2}{\Lambda^2 + m^2} \right)$.  The function $C(|\vec{x}-\vec{y}|)$ decays like $\sim \, e^{-e^{-u} m |\vec{x}-\vec{y}|}$, and so has a width of $\sim 1/(e^{-u} m)$ which is the scale of the inverse renormalized mass.  The function $D(|\vec{x}-\vec{y}|)$ is the Fourier transform of $\theta'(1-|\vec{k}|/\Lambda)$.  Suppose that the derivative $\theta'(0)$ is $\sim 1$, meaning that the discontinuity of the Heaviside step function has been smoothed over a scale $\sim 1$.  Since $\theta'(1-|\vec{k}|/\Lambda)$ is nearly localized on the sphere $|\vec{k}| = \Lambda$ in Fourier space, its Fourier transform $D(|\vec{x}-\vec{y}|)$ will be localized near the origin and be highly oscillatory with frequency $\sim \Lambda$.  Since $C(|\vec{x}-\vec{y}|)$ has a width of $\sim 1/(e^{-u} m)$ and $D(|\vec{x}-\vec{y}|)$ oscillates with frequency $\sim \Lambda$, when $C(|\vec{x}-\vec{y}|)$ and $D(|\vec{x}-\vec{y}|)$ are integrated against one another the result will be non-negligible only if $1/(e^{-u} m) \lesssim 1/\Lambda$.  This result is intuitive: it means that we can only see the effect of the mass $m$ of the UV theory if we probe distance scales around $\sim 1/m$ or larger.  Otherwise, we are only probing modes which are effectively massless.

\subsection{Perturbative cMERA for scalar $\varphi^4$ theory}
\subsubsection{$1$-loop cMERA circuit}

We again follow the procedure outlined in Section \ref{sec:OutlineProcedure} above.  Recall once more that the Hamiltonian of $\varphi^4$ theory is
\begin{align}
H^\Lambda &= \frac{1}{2} \int^\Lambda d^d \vec{k} \, \left(\widehat{\pi}(\vec{k}) \widehat{\pi}(-\vec{k}) + \widehat{\phi}(\vec{k}) \left(\vec{k}^2 + m^2\right) \widehat{\phi}(-\vec{k})\right) \nonumber \\
& \qquad \qquad \qquad \qquad + \frac{\lambda}{4!} \int^\Lambda d^d \vec{k}_1 \, d^d \vec{k}_2 \, d^d \vec{k}_3 \, \widehat{\phi}(\vec{k}_1) \widehat{\phi}(\vec{k}_2) \widehat{\phi}(\vec{k}_3) \widehat{\phi}(-\vec{k}_1 - \vec{k}_2 - \vec{k}_3)
\end{align} 
where we take $\lambda$ to be a perturbatively small parameter.  First, we need to use an RG scheme to calculate $|\Psi(\Lambda e^u)\rangle$ for the ground state of $\varphi^4$ theory.  Performing the spatial Wilsonian RG procedure explained in Section \ref{sec:cMERAfreerevisited} to first order in the $\phi^4$ coupling $\lambda$ (see Appendix \ref{sec:AppA} for details), we obtain
\begin{align}
H_{1-\text{loop}}^{\Lambda e^u} &= \frac{1}{2} \int^\Lambda d^d \vec{k} \, \left(\widehat{\pi}(\vec{k}) \widehat{\pi}(-\vec{k}) + \widehat{\phi}(\vec{k}) \left(\vec{k}^2 + e^{-2u}\,\widetilde{m}^2\right) \widehat{\phi}(-\vec{k})\right) \nonumber \\
& \qquad \qquad \qquad \qquad + \frac{e^{(d-3)u}\lambda}{4!} \int^\Lambda d^d \vec{k}_1 \, d^d \vec{k}_2 \, d^d \vec{k}_3 \, \widehat{\phi}(\vec{k}_1) \widehat{\phi}(\vec{k}_2) \widehat{\phi}(\vec{k}_3) \widehat{\phi}(-\vec{k}_1 - \vec{k}_2 - \vec{k}_3)
\end{align}
where
\begin{align}
\widetilde{m}^2 &=: m^2 + \delta m^2\,.
\end{align}
The explicit form of $\delta m^2$ is given in Appendix \ref{sec:AppA} in Eqn.~\eqref{eq:deltam2}.  The ground state wavefunctional of $H_{1-\text{loop}}^{\Lambda e^u}$ is given by
\begin{align}
\label{phi4gs1second}
\langle \phi | \Psi(\Lambda e^u)\rangle &= \mathcal{N} \, \exp\left(- G[\phi] - e^{-2u} \delta m^2 \, R_1[\phi] - e^{(d-3)u} \lambda \, R_2[\phi] \right) \,\, + \mathcal{O}(\lambda^2)
\end{align}
where
\begin{align}
G[\phi] &= \frac{1}{2}\int^\Lambda d^d \vec{k} \, \phi(\vec{k}) \, \omega_k \, \phi(-\vec{k}) \\ \nonumber \\
R_1[\phi]&=\frac{1}{4}\int^\Lambda d^d \vec{k}\, \frac{1}{\omega_k}\,\phi(\vec{k}) \phi(-\vec{k}) \\ \nonumber \\
\begin{split}
R_2[\phi] &=\frac{1}{16}\int^\Lambda d^d \vec{k}\, \frac{1}{\omega_k}\,\left(\int \frac{d^d \vec{q}}{(2\pi)^{d}} \frac{1}{\omega_k + \omega_q} \right)\,\phi(\vec{k}) \phi(-\vec{k})
\\
&+ \frac{1}{24}\frac{1}{(2\pi)^d} \int^\Lambda \frac{d^d \vec{k}_1 \, d^d \vec{k}_2 \, d^d \vec{k}_3}{\omega_{k_1} + \omega_{k_2} + \omega_{k_3} + \omega_{-\vec{k}_1 - \vec{k}_2 - \vec{k}_3}} \, \phi(\vec{k}_1) \phi(\vec{k}_2) \phi(\vec{k}_3) \phi(-\vec{k}_1 - \vec{k}_2 - \vec{k}_3)
\end{split}
\end{align}
and where $\omega_k = \sqrt{\vec{k}^2 + e^{-2u} m^2}$.  The renormalized Hamiltonian and ground state were previously mentioned in Section \ref{sec:phi4circuit} above.

Next, we want to pick some simple IR state $|\Omega\rangle$ and find a unitary $U(u)$ such that $\langle \phi(\vec{p})|\Psi(\Lambda e^u)\rangle = \langle \phi(\vec{p})| U(u) |\Omega\rangle$ for $\phi(\vec{p})$ for which $\phi(\vec{p}) = 0$ when $|\vec{p}| \geq \Lambda$.  In other words, $|\Psi(\Lambda e^u)\rangle$ equals $U(u) |\Omega\rangle$ for modes with $|\vec{p}| \leq \Lambda$.  In Section \ref{sec:cMERAfreerevisited}, we chose $|\Omega\rangle$ to be a Gaussian state (see Eqn.~\eqref{IRstate1}).  However, since we are now working with $\varphi^4$ theory and our state of interest in Eqn.~\eqref{phi4gs1second} is Gaussian with quadratic and quartic corrections at first order on $\lambda$, we may instead desire a more general IR state $|\widetilde{\Omega}\rangle$ of the form
\begin{align}
\label{IRstate2}
\langle \phi | \widetilde{\Omega}\rangle &= \mathcal{N} \, \exp\bigg(- \frac{1}{2} \int d^d \vec{k} \, \phi(\vec{k}) \, M \, \phi(-\vec{k}) - \frac{\lambda}{2} \int d^d \vec{k} \, \phi(\vec{k}) \, M_2(\vec{k}) \, \phi(-\vec{k}) \nonumber \\
& \qquad \qquad \qquad \qquad - \frac{\lambda}{4} \int d^d \textbf{k} \, \delta^{(d)}(\textbf{k}) \, M_4(\vec{k}_1,\vec{k}_2,\vec{k}_3,\vec{k}_4) \, \phi(\vec{k}_1) \phi(\vec{k}_2) \phi(\vec{k}_3) \phi(\vec{k}_4) \bigg)\,,
\end{align}
where we assume that $M_2(\vec{k}) = M_2(|\vec{k}|)$ to preserve translation and rotation invariance, and $M_4(\vec{k}_1, \vec{k}_2, \vec{k}_3, \vec{k}_4)$ is symmetric in its four inputs since our state of interest $|\Psi(\Lambda e^u)\rangle$ also has this form.

For the state $|\widetilde{\Omega}\rangle$ to be scale-invariant, we need to put some constraints on the functions $M_2(\vec{k})$ and $M_4(\vec{k}_1,\vec{k}_2,\vec{k}_3,\vec{k}_4)$.  To derive these constraints, it is convenient to use an equivalent definition of $|\widetilde{\Omega}\rangle$, namely as a state satisfying
\begin{align}
&\bigg( (M + \lambda \, M_2(\vec{k})) \, \widehat{\phi}(\vec{k}) \nonumber \\
& \qquad + \lambda \int d^d \vec{k}_2 \, d^d \vec{k}_3 \, d^d \vec{k}_4 \, \delta(-\vec{k} + \vec{k}_2 + \vec{k}_3 + \vec{k}_4) \, M_4(-\vec{k},\vec{k}_2,\vec{k}_3,\vec{k}_4) \, \widehat{\phi}(\vec{k}_2)\widehat{\phi}(\vec{k}_3)\widehat{\phi}(\vec{k}_4) + i\,\widehat{\pi}(\vec{k}) \bigg) |\widetilde{\Omega}\rangle = 0 \nonumber \\ \nonumber \\ 
& \qquad \qquad \qquad \qquad \qquad \qquad \qquad \qquad \qquad \qquad \qquad \qquad \qquad \qquad \qquad \qquad \qquad \qquad \text{for all } \vec{k} \in \mathbb{R}^d\,. \nonumber \\ \nonumber \\
\end{align}
This definition of $|\widetilde{\Omega}\rangle$ uses the same construction explained in Eqn.'s~\eqref{Ocondition1} and~\eqref{scaleinvcond1}.  Then, according to Eqn.~\eqref{Ocondition1}, we need
\begin{align}
\label{M2eq1}
M_2(e^{-u}|\vec{k}|) &= M_2(|\vec{k}|) \\ \nonumber \\
M_4(e^{-u}\vec{k}_1,e^{-u}\vec{k}_2,e^{-u}\vec{k}_3,e^{-u}\vec{k}_4) &= e^{3 d u} M_4(\vec{k}_1,\vec{k}_2,\vec{k}_3, \vec{k}_4)\,.
\end{align}
The first of the two equations forces $M_2$ to be a constant.  The second equation leaves room for $M_4$ to be a broad class of functions. 

Having defined our state $|\widetilde{\Omega}\rangle$, we must now find the unitary $U(u)$.  From Section \ref{sec:phi4circuit}, we know it suffices for the unitary to take the form
\begin{equation}
\label{QFTUnitaryAgain}
U = \exp\left(i \, K_{2,0} + i \, \lambda \left(K_{2,1} + K_4 \right) \right)\,,
\end{equation}
where we define
\begin{align}
K_{2,0} &= - \int d^d \vec{k}_1 \, d^d \vec{k}_2 \, \delta^{(d)}(\vec{k}_1 + \vec{k}_2)\,g_{2,0}(\vec{k}_1 \, ; \, u) \, S_2^{(1)}(\vec{k}_1, \vec{k}_2) \\
K_{2,1} &= - \int d^d \vec{k}_1 \, d^d \vec{k}_2 \,\delta^{(d)}(\vec{k}_1 + \vec{k}_2)\, g_{2,1}(\vec{k}_1 \, ; \, u) \, S_2^{(1)}(\vec{k}_1, \vec{k}_2) \\
K_{4} &= \int d^d \vec{k}_1 \, d^d \vec{k}_2 \, d^d \vec{k}_3 \, d^d \vec{k}_4 \, \delta^{(d)}(\vec{k}_1 + \vec{k}_2 + \vec{k}_3 + \vec{k}_4) \bigg( g_{4}^{(1)}(\vec{k}_1,\vec{k}_2,\vec{k}_3,\vec{k}_4 \, ; \, u) \, S_4^{(1)}(\vec{k}_1, \vec{k}_2, \vec{k}_3, \vec{k}_4) \nonumber \\
& \qquad \qquad \qquad \qquad \qquad \qquad \qquad \qquad \qquad \qquad \qquad +  g_{4}^{(3)}(\vec{k}_1,\vec{k}_2, \vec{k}_3, \vec{k}_4 \, ; \, u) \, S_4^{(3)}(\vec{k}_1, \vec{k}_2, \vec{k}_3, \vec{k}_4) \bigg)\,.
\end{align}

Performing essentially the same analysis as in Section \ref{sec:phi4circuit}, we can solve for the unknown functions $g_{2,0}(\vec{k}_1 \, ; \, u)$, $g_{2,1}(\vec{k}_1 \, ; \, u)$, $g_{4}^{(1)}(\vec{k}_1,\vec{k}_2, \vec{k}_3, \vec{k}_4 \, ; \, u)$, $g_{4}^{(3)}(\vec{k}_1,\vec{k}_2, \vec{k}_3, \vec{k}_4 \, ; \, u)$, as well as $M$, $M_2$ and $M_4(\vec{k}_1,\vec{k}_2, \vec{k}_3, \vec{k}_4 \, ; \, u)$.  As in Section \ref{sec:phi4circuit}, letting
\begin{align}
\mathcal{G}_1(\vec{k}_1,\vec{k}_2,\vec{k}_3,\vec{k}_4 \, ; \, u) &:= 2\big(g_{2,0}(\vec{k}_1 \, ; \, u) - g_{2,0}(\vec{k}_2 \, ; \, u) - g_{2,0}(\vec{k}_3 \, ; \, u) - g_{2,0}(\vec{k}_4 \, ; \, u)\big) \\
\mathcal{G}_3(\vec{k}_1,\vec{k}_2,\vec{k}_3,\vec{k}_4 \, ; \, u) &:= 2\big(g_{2,0}(\vec{k}_1 \, ; \, u) + g_{2,0}(\vec{k}_2 \, ; \, u) + g_{2,0}(\vec{k}_3 \, ; \, u) - g_{2,0}(\vec{k}_4 \, ; \, u)\big)\,.
\end{align}
and defining
\begin{align}
\widetilde{g}_4^{(j)}(\vec{k}_1, \vec{k}_2, \vec{k}_3, \vec{k}_4 \, ; \, u) &:= \frac{e^{- \mathcal{G}_j(\vec{k}_1,\vec{k}_2,\vec{k}_3,\vec{k}_4 \, ; \, u)}-1}{\mathcal{G}_j(\vec{k}_1,\vec{k}_2,\vec{k}_3,\vec{k}_4 \, ; \, u)} \, g_4^{(j)}(\vec{k}_1, \vec{k}_2, \vec{k}_3, \vec{k}_4 \, ; \, u)
\end{align}
with $j=1,3$, the functions in $U(u)$ and $|\widetilde{\Omega}\rangle$ above are given by
\begin{align}
\label{eq:defg20}
g_{2,0}(k \, ; \, u) &= \frac{1}{4} \log\left(\frac{M}{\omega_{k}} \right)\\ \nonumber \\
\label{eq:defg21}
g_{2,1}(k \, ; \, u) &= - \frac{1}{32}\bigg\{\frac{1}{(2\pi)^d}\int^\Lambda d^d \vec{q} \, \left(\frac{e^{(d-3) u}}{\omega_{k}^2(\omega_k+\omega_q)} - \frac{1}{M^2(M+\sqrt{q^2+m^2})}\right) \\  \nonumber \\
& \qquad \qquad \qquad \qquad \qquad  \qquad \qquad \qquad + 4 \, (\delta m^2/\lambda) \left(\frac{e^{-2u}}{\omega_k^2} - \frac{1}{M^2} \right) \bigg\}\nonumber \\
\label{eq:defg41}
\widetilde{g}_{4}^{(1)}(\vec{k}_1,\vec{k}_2,\vec{k}_3,\vec{k}_4 \, ; \, u) &= \frac{1}{96}\frac{e^{(d-3)u}}{(2\pi)^d}\,\frac{1}{\omega_{k_2} \omega_{k_3} \omega_{k_4}(\omega_{k_1} + \omega_{k_2} + \omega_{k_3} + \omega_{k_4})} \\ \nonumber \\
\label{eq:defg43}
\widetilde{g}_{4}^{(3)}(\vec{k}_1,\vec{k}_2,\vec{k}_3,\vec{k}_4 \, ; \, u) &= \frac{1}{32}\frac{e^{(d-3)u}}{(2\pi)^d}\,\frac{1}{\omega_{k_4}(\omega_{k_1} + \omega_{k_2} + \omega_{k_3} + \omega_{k_4})}\,,
\end{align}
where $\omega_k := \sqrt{\vec{k}^2 + e^{-2u} m^2}$, and
\begin{align}
M &=\sqrt{\Lambda^2 + m^2} \\ \nonumber \\
M_2 &= - \frac{4}{(2\pi)^d}\int^\Lambda d^d \vec{q} \, \frac{1}{M(M+\sqrt{q^2+m^2})} + \frac{1}{2}\,(\delta m^2/\lambda) \, \frac{1}{M} \\ \nonumber \\
M_4(\vec{k}_1, \vec{k}_2, \vec{k}_3, \vec{k}_4) &= 0\,.
\end{align}
We point out that matching $|\Psi(\Lambda e^u)\rangle$ and $U(u) |\widetilde{\Omega}\rangle$ forces $M_4(\vec{k}_1, \vec{k}_2, \vec{k}_3, \vec{k}_4)$ to be zero, but allows us to choose $M$ and $M_2$.  Above, we have chosen values of $M$ and $M_2$ so that the kernel for the quadratic term in the exponential of the wavefunctional is continuous at $|\vec{k}| = \Lambda$, akin to Eqn.~\eqref{whatwewant1} and the comments which immediately follow it.  However, we cannot force the kernel for the quartic term in the exponential of the wavefunction to be continuous at $|\vec{k}| = \Lambda$, but such continuity is not essential and instead is aesthetic. 

For the final step of the cMERA procedure outlined in Section \ref{sec:OutlineProcedure}, we need to find a $K(s)$ such that
\begin{equation}
\mathcal{P}_s \, e^{-i \int_{-\infty}^u ds \, (K(s) + L)}|\widetilde{\Omega}\rangle = U(u) |\widetilde{\Omega}\rangle\,.
\end{equation}
As per Eqn.'s~\eqref{rewrite1},~\eqref{rewrite2} and~\eqref{rewrite3}, we write
\begin{equation}
\label{rewrite1v2}
\mathcal{P}_s \, e^{-i \int_{-\infty}^u ds \, (K(s) + L)}|\widetilde{\Omega}\rangle = e^{-i u L}\, \mathcal{P}_s \, e^{-i \int_{-\infty}^u ds \, \widetilde{K}(s)}|\widetilde{\Omega}\rangle
\end{equation}
where
\begin{equation}
\label{rewrite2v2}
\widetilde{K}(s) = e^{i s L} \, K(s) \, e^{- i s L}\,.
\end{equation}
Since $|\widetilde{\Omega}\rangle = e^{i u L} |\widetilde{\Omega}\rangle$, we can write
\begin{equation}
\label{rewrite3v2}
e^{-iuL} \left(\mathcal{P}_s e^{-i \int_{-\infty}^u ds \, \widetilde{K}(s)}\right) \, e^{iuL} \, |\widetilde{\Omega}\rangle = U(u) \, |\widetilde{\Omega}\rangle \,.
\end{equation}
Suppose that $\widetilde{K}(s)$ has the form
\begin{align}
\label{tildeKform}
\widetilde{K}(s) =& \int d^d \vec{k}_1 \, d^d \vec{k}_2 \, \delta^{(d)}(\vec{k}_1 + \vec{k}_2)\,f_{2,0}(e^{-s}\vec{k}_1 \, ; \, s) \, S_2^{(1)}(\vec{k}_1, \vec{k}_2) \nonumber \\
& + \lambda \int d^d \vec{k}_1 \, d^d \vec{k}_2 \, \delta^{(d)}(\vec{k}_1 + \vec{k}_2)\,f_{2,1}(e^{-s}\vec{k}_1 \, ; \, s) \, S_2^{(1)}(\vec{k}_1, \vec{k}_2) \nonumber \\
& + \lambda \int d^d \textbf{k} \, \delta^{(d)}(\textbf{k}) \bigg( f_{4}^{(1)}(e^{-s}\vec{k}_1,e^{-s}\vec{k}_2,e^{-s}\vec{k}_3,e^{-s}\vec{k}_4 \, ; \, s) \, S_4^{(1)}(\vec{k}_1, \vec{k}_2, \vec{k}_3, \vec{k}_4) \nonumber \\
& \qquad \qquad \qquad \qquad \qquad +  f_{4}^{(3)}(e^{-s}\vec{k}_1,e^{-s}\vec{k}_2, e^{-s}\vec{k}_3, e^{-s}\vec{k}_4 \, ; \, s) \, S_4^{(3)}(\vec{k}_1, \vec{k}_2, \vec{k}_3, \vec{k}_4) \bigg)\,.
\end{align}
Next, we perturbatively expand the path-ordered exponential $\mathcal{P}_s \, e^{-i \int_{-\infty}^u ds \, \widetilde{K}(s)}$ and arrange the terms as
\begin{align}
\label{pathorderExpand1}
&\mathcal{P}_s \, e^{-i \int_{-\infty}^u ds \, \widetilde{K}(s)} \nonumber \\
&= \bigg\{1 - i \lambda \int_{-\infty}^u ds \int d^d \vec{k}_1 d^d \vec{k}_2 \, \delta^{(d)}(\vec{k}_1 + \vec{k}_2) \, f_{2,1}(e^{-s}\vec{k}_1 \, ; \, s)\,\, S_2^{(1)}(\vec{k}_1, \vec{k}_2) \nonumber \\
& \, \qquad \,\,\, + i \lambda \int_{-\infty}^u ds \, \int d^d \textbf{k} \, \delta^{(d)}(\textbf{k}) \bigg( f_{4}^{(1)}(e^{-s}\vec{k}_1,e^{-s}\vec{k}_2,e^{-s}\vec{k}_3,e^{-s}\vec{k}_4 \, ; \, s) \, e^{\mathcal{F}_1(s,u)}\, S_4^{(1)}(\vec{k}_1, \vec{k}_2, \vec{k}_3, \vec{k}_4) \nonumber \\
& \qquad \qquad \qquad \qquad \qquad \qquad \qquad + f_{4}^{(3)}(e^{-s}\vec{k}_1,e^{-s}\vec{k}_2,e^{-s}\vec{k}_3,e^{-s}\vec{k}_4 \, ; \, s) \, e^{\mathcal{F}_3}(s,u) \, S_4^{(3)}(\vec{k}_1, \vec{k}_2, \vec{k}_3, \vec{k}_4) \bigg)\bigg\} \nonumber \\
& \qquad \qquad \qquad \qquad \qquad \qquad \qquad \qquad \qquad \qquad \quad \times e^{-i \int_{-\infty}^u ds \, \int d^d \vec{k}_1 \, d^d \vec{k}_2 \, \delta^{(d)}(\vec{k}_1 + \vec{k}_2)\,f_{2,0}(e^{-s}\vec{k}_1 \, ; \, s) \, S_2^{(1)}(\vec{k}_1, \vec{k}_2)} \nonumber \\
& \qquad \qquad \qquad \qquad \qquad \qquad \qquad \qquad \qquad \qquad \qquad \qquad \qquad \qquad \qquad \qquad \qquad \qquad \qquad \quad + \mathcal{O}(\lambda^2)\,, \nonumber \\
\end{align}
where $\mathcal{F}_1(s,u)$ and $\mathcal{F}_3(s,u)$ are defined by
\begin{align}
\mathcal{F}_1(s,u) &:= 2 \int_s^u dt \, \left(f_{2,0}(e^{-t}\vec{k}_1 \, ; \, t) - f_{2,0}(e^{-t}\vec{k}_2\, ; \, t) - f_{2,0}(e^{-t}\vec{k}_3\, ; \, t) - f_{2,0}(e^{-t}\vec{k}_4\, ; \, t) \right) \\ \nonumber \\
\mathcal{F}_3(s,u) &:= 2 \int_s^u dt \, \left(f_{2,0}(e^{-t}\vec{k}_1\, ; \, t) + f_{2,0}(e^{-t}\vec{k}_2\, ; \, t) + f_{2,0}(e^{-t}\vec{k}_3\, ; \, t) - f_{2,0}(e^{-t}\vec{k}_4\, ; \, t) \right) \,.
\end{align}
Next, we write $e^{i u L} U(u) e^{-i u L}$ in the same form, as
\begin{align}
\label{pathorderExpandToMatch1}
&e^{i u L} U(u) e^{-i u L} \nonumber \\
&= \bigg\{1 - i \lambda  \int d^d \vec{k}_1 d^d \vec{k}_2 \, \delta^{(d)}(\vec{k}_1 + \vec{k}_2) \, g_{2,1}(e^{-u}\vec{k}_1 \, ; \, u)\,\, S_2^{(1)}(\vec{k}_1, \vec{k}_2) \nonumber \\
& \, \qquad \,\,\, - i \lambda  \, \int d^d \textbf{k}\hspace{1mm} e^{-du}\, \delta^{(d)}(\textbf{k}) \bigg( \widetilde{g}_{4}^{(1)}(e^{-u}\vec{k}_1,e^{-u}\vec{k}_2,e^{-u}\vec{k}_3,e^{-u}\vec{k}_4 \, ; \, u) \, S_4^{(1)}(\vec{k}_1, \vec{k}_2, \vec{k}_3, \vec{k}_4) \, \nonumber \\
&  \qquad \qquad \qquad \qquad \qquad \qquad \quad + \widetilde{g}_{4}^{(3)}(e^{-u}\vec{k}_1,e^{-u}\vec{k}_2,e^{-u}\vec{k}_3,e^{-u}\vec{k}_4 \, ; \, u) \, S_4^{(3)}(\vec{k}_1, \vec{k}_2, \vec{k}_3, \vec{k}_4) \bigg)\bigg\} \nonumber \\
& \qquad \qquad \qquad \qquad \qquad \qquad \qquad \qquad \qquad \quad \times e^{-i  \, \int d^d \vec{k}_1 \, d^d \vec{k}_2 \, \delta^{(d)}(\vec{k}_1 + \vec{k}_2)\,g_{2,0}(e^{-u}\vec{k}_1 \, ; \, u) \, S_2^{(1)}(\vec{k}_1, \vec{k}_2)} \nonumber \\
& \qquad \qquad \qquad \qquad \qquad \qquad \qquad \qquad \qquad \qquad \qquad \qquad \qquad \qquad \qquad \qquad \qquad \qquad \quad + \mathcal{O}(\lambda^2)\,. \nonumber \\
\end{align}
Notice that both Eqn's~\eqref{pathorderExpand1} and~\eqref{pathorderExpandToMatch1} have the form
\begin{align}
\left(1 + \lambda (\, \cdots \, ) S_{2}^{(1)} + \lambda (\, \cdots \, ) S_{4}^{(1)} + \lambda (\, \cdots \, ) S_{4}^{(3)} \right) \, e^{i (\, \cdots \,) S_2^{(1)}}\,.
\end{align}
Therefore, we can match Eqn's~\eqref{pathorderExpand1} and~\eqref{pathorderExpandToMatch1} term by term to determine $f_{2,0}$, $f_{2,1}$, $f_{4}^{(1)}$ and $f_{4}^{(3)}$.  Matching the equations, we can extract $f_{2,0}$ and $f_{2,1}$ by differentiating with respect to $u$ :
\begin{align}
\label{f20sol1}
f_{2,i}(e^{-u}\vec{k}\,;\,u) &= \frac{d}{du} \left[\theta(1-|\vec{k}|/\Lambda e^u)  \, g_{2,0}(e^{-u}\vec{k} \, ; \, u)\right]\,, \;\;\;\;\;\;\;\;\;\;i=0,1\,.
\end{align}
To obtain $f_{4}^{(1)}$ and $f_{4}^{(3)}$, we note that the matching gives us
\begin{align}
\label{f41match1}
&\int_{-\infty}^u ds \, e^{\mathcal{F}_j(s,u)} \, f_4^{(j)}(e^{-s}\vec{k}_1, e^{-s}\vec{k}_2, e^{-s}\vec{k}_3, e^{-s}\vec{k}_4 \, ; \, s) \nonumber \\
& \qquad \qquad \qquad \qquad = - e^{-du} \widetilde{g}_4^{(j)}(e^{-s}\vec{k}_1, e^{-s}\vec{k}_2, e^{-s}\vec{k}_3, e^{-s}\vec{k}_4 \, ; \, s) \, \prod_{p=1}^4 \theta(1-|\vec{k}_p|/\Lambda e^u)
\end{align}
with $j=1,3$. Differentiating both sides of the above two equations with respect to $u$, we obtain
\begin{align}
\label{f41diff1}
\begin{split}
& f_{4}^{(j)}(e^{-u}\vec{k}_1, e^{-u}\vec{k}_2, e^{-u}\vec{k}_3, e^{-u}\vec{k}_4 \, ; \, u) \\
& \qquad = e^{-(d+1)u}\bigg\{- \widetilde{g}_4^{(j)}(e^{-u}\vec{k}_1, e^{-u}\vec{k}_2, e^{-u}\vec{k}_3, e^{-u}\vec{k}_4 \, ; \, u) \sum_{\ell=1}^4 \frac{|\vec{k}_\ell|}{\Lambda} \frac{\theta'(1-|\vec{k}_\ell|/\Lambda e^u)}{\theta(1-|\vec{k}_\ell|/\Lambda e^u)} \\
& \qquad \qquad + d \, e^u \, \widetilde{g}_4^{(j)}(e^{-u}\vec{k}_1, e^{-u}\vec{k}_2, e^{-u}\vec{k}_3, e^{-u}\vec{k}_4 \, ; \, u) \\
& \qquad \qquad \qquad + e^u \, \frac{d}{du} \, \widetilde{g}_4^{(j)}(e^{-u}\vec{k}_1, e^{-u}\vec{k}_2, e^{-u}\vec{k}_3, e^{-u}\vec{k}_4 \, ; \, u)\bigg\} \prod_{p=1}^4 \theta(1-|\vec{k}_\ell|/\Lambda e^u) \\
& \qquad \qquad \qquad \qquad \,\,\,\, - \int_{-\infty}^u ds \, e^{\mathcal{F}_j(s,u)} \frac{\partial \mathcal{F}_j(s,u)}{\partial u} \, f_4^{(j)}(e^{-s}\vec{k}_1, e^{-s}\vec{k}_2, e^{-s}\vec{k}_3, e^{-s}\vec{k}_4 \, ; \, s)
\end{split}
\end{align}
again with $j=1,3$.
Notice that Eqn~\eqref{f41diff1}
does not immediately allow us to solve for $f_4^{(j)}$, $j=1,3$.  For instance, in Eqn.~\eqref{f41diff1}, $f_4^{(j)}$ appears both as the left-hand side, and in an integral on the right-hand side.  Noting that
\begin{align}
\frac{\partial}{\partial u} \mathcal{F}_1(s,u) &=  2 \left(f_{2,0}(e^{-u}\vec{k}_1 \, ; \, u) - f_{2,0}(e^{-u}\vec{k}_2\, ; \, u) - f_{2,0}(e^{-u}\vec{k}_3\, ; \, u) - f_{2,0}(e^{-u}\vec{k}_4\, ; \, u) \right)\\ \nonumber \\
\frac{\partial}{\partial u} \mathcal{F}_3(s,u) &= 2 \left(f_{2,0}(e^{-u}\vec{k}_1 \, ; \, u) + f_{2,0}(e^{-u}\vec{k}_2\, ; \, u) + f_{2,0}(e^{-u}\vec{k}_3\, ; \, u) - f_{2,0}(e^{-u}\vec{k}_4\, ; \, u) \right)\,,
\end{align}
we can plug Eqn.'s~\eqref{f41match1}
back into Eqn.'s~\eqref{f41diff1}
and solve for $f_4^{(1)}$ and $f_4^{(3)}$.  We find the solutions 
\begin{align}
\begin{split}
\label{f41sol1}
& f_4^{(j)}(e^{-u}\vec{k}_1, e^{-u}\vec{k}_2, e^{-u}\vec{k}_3, e^{-u}\vec{k}_4 \, ; \, u) \\
& \qquad = e^{-(d+1)u}\bigg\{e^u\left(d - \frac{\partial \mathcal{F}_j(s,u)}{\partial u}\right) \, \widetilde{g}_4^{(j)}(e^{-u}\vec{k}_1, e^{-u}\vec{k}_2, e^{-u}\vec{k}_3, e^{-u}\vec{k}_4 \, ; \, u) \\
& \qquad \qquad + e^u \frac{d}{du} \widetilde{g}_4^{(j)}(e^{-u}\vec{k}_1, e^{-u}\vec{k}_2, e^{-u}\vec{k}_3, e^{-u}\vec{k}_4 \, ; \, u) \\
& \qquad \qquad \qquad- \widetilde{g}_{4}^{(j)}(e^{-u}\vec{k}_1, e^{-u}\vec{k}_2, e^{-u}\vec{k}_3, e^{-u}\vec{k}_4 \, ; \, u) \, \sum_{\ell=1}^4 \frac{|\vec{k}_\ell|}{\Lambda} \frac{\theta'(1-|\vec{k}_p|/\Lambda e^u)}{\theta(1-|\vec{k}_\ell|/\Lambda e^u)}\bigg\} \\
&  \qquad \qquad \qquad \qquad \qquad \qquad \qquad \qquad \qquad \qquad \qquad \qquad \qquad \times \prod_{p=1}^4 \theta(1-|\vec{k}_p|/\Lambda e^u)
\end{split}
\end{align}
for $j=1,3$.
Having solved for $f_{2,0}$, $f_{2,1}$, $f_4^{(1)}$ and $f_4^{(3)}$ in Eqn.'s~\eqref{f20sol1} and~\eqref{f41sol1}, we have thus determined the dilated entangler $\widetilde{K}(u)$ given in Eqn.~\eqref{tildeKform} and hence the entangler $K(u)$ itself, to first order in perturbation theory.  We summarize the result below, and write down an explicit form for $K(u)$. \\

\noindent \textbf{Summary of 1-loop cMERA circuit for the ground state of scalar $\varphi^4$ theory}
\\ \\
Consider the cMERA circuit
\begin{equation}
\mathcal{P}_s \, e^{-i \int_{-\infty}^u ds \, (K(s) + L)}|\widetilde{\Omega}\rangle = U(u) |\widetilde{\Omega}\rangle\,.
\end{equation}
where
\begin{align}
\label{IRstate2copy}
\langle \phi | \widetilde{\Omega}\rangle &= \mathcal{N} \, \exp\bigg(- \frac{1}{2} \int d^d \vec{k} \, \phi(\vec{k}) \, M \, \phi(-\vec{k}) - \frac{\lambda}{2} \int d^d \vec{k} \, \phi(\vec{k}) \, M_2(\vec{k}) \, \phi(-\vec{k}) \bigg)\,,
\end{align}
with
\begin{align}
M &=\sqrt{\Lambda^2 + m^2} \\ \nonumber \\
M_2 &= - \frac{4}{(2\pi)^d}\int^\Lambda d^d \vec{q} \, \frac{1}{M(M+\sqrt{q^2+m^2})} + \frac{1}{2}\,(\delta m^2/\lambda) \, \frac{1}{M} \,.
\end{align}
Let the entangler $K(s)$ be
\begin{align}
\label{untildeKform}
K(s) =& \int d^d \vec{k}_1 \, d^d \vec{k}_2 \, \delta^{(d)}(\vec{k}_1 + \vec{k}_2)\,f_{2,0}(\vec{k}_1 \, ; \, s) \, S_2^{(1)}(\vec{k}_1, \vec{k}_2) \nonumber \\
& + \lambda \, \int d^d \vec{k}_1 \, d^d \vec{k}_2 \, \delta^{(d)}(\vec{k}_1 + \vec{k}_2)\,f_{2,1}(\vec{k}_1 \, ; \, s) \, S_2^{(1)}(\vec{k}_1, \vec{k}_2) \nonumber \\
& + \lambda \, e^{ds} \int d^d \textbf{k} \, \delta^{(d)}(\textbf{k}) \bigg( f_{4}^{(1)}(\vec{k}_1,\vec{k}_2,\vec{k}_3,\vec{k}_4 \, ; \, s) \, S_4^{(1)}(\vec{k}_1, \vec{k}_2, \vec{k}_3, \vec{k}_4) \nonumber \\
& \qquad \qquad \qquad \qquad \qquad +  f_{4}^{(3)}(\vec{k}_1, \vec{k}_2, \vec{k}_3, \vec{k}_4 \, ; \, s) \, S_4^{(3)}(\vec{k}_1, \vec{k}_2, \vec{k}_3, \vec{k}_4) \bigg)
\end{align}
where
\begin{align}
\label{eq:f2i}
f_{2,i}(e^{-u}\vec{k}\,;\,u) &= \frac{d}{du} \left[\theta(1-|\vec{k}|/\Lambda e^u)  \, g_{2,i}(e^{-u}\vec{k} \, ; \, u)\right]\,,\;\;\;\;\;\;\;\;\;\;i=0,1
\end{align}
\begin{align}
\label{f41sol2}
\begin{split}
& f_4^{(j)}(\vec{k}_1, \vec{k}_2, \vec{k}_3, \vec{k}_4 \, ; \, u) \\
& \qquad = e^{-(d+1)u}\Bigg\{e^u\left(d - \frac{\partial \mathcal{F}_j(s,u)}{\partial u}\right) \, \widetilde{g}_4^{(j)}(\vec{k}_1, \vec{k}_2, \vec{k}_3, \vec{k}_4 \, ; \, u) \\
&  \qquad \qquad + e^u \left(- \sum_{j=1}^4 \vec{k}_j \cdot \frac{\partial}{\partial \vec{k}_j}\, \widetilde{g}_4^{(j)}(\vec{k}_1, \vec{k}_2, \vec{k}_3, \vec{k}_4 \, ; \, u) + \frac{\partial}{\partial u} \widetilde{g}_4^{(j)}(\vec{k}_1, \vec{k}_2, \vec{k}_3, \vec{k}_4 \, ; \, u) \right)  \\
& \qquad \qquad \qquad 
- \widetilde{g}_{4}^{(j)} (\vec{k}_1, \vec{k}_2, \vec{k}_3, \vec{k}_4 \, ; \, u) \, \sum_{\ell=1}^4 \frac{|\vec{k}_\ell|}{\Lambda} \frac{\theta'(1-|\vec{k}_p|/\Lambda)}{\theta(1-|\vec{k}_\ell|/\Lambda)}\Bigg\} 
 \prod_{p=1}^4 \theta(1-|\vec{k}_p|/\Lambda)
\end{split}
\end{align}
with $j=1,3$ 
and $\mathcal{F}_1(s,u)$ and $\mathcal{F}_3(s,u)$ defined by
\begin{align}
\mathcal{F}_1(s,u) &:= 2 \int_s^u dt \, \left(f_{2,0}(e^{-t}\vec{k}_1 \, ; \, t) - f_{2,0}(e^{-t}\vec{k}_2\, ; \, t) - f_{2,0}(e^{-t}\vec{k}_3\, ; \, t) - f_{2,0}(e^{-t}\vec{k}_4\, ; \, t) \right) \\ \nonumber \\
\mathcal{F}_3(s,u) &:= 2 \int_s^u dt \, \left(f_{2,0}(e^{-t}\vec{k}_1\, ; \, t) + f_{2,0}(e^{-t}\vec{k}_2\, ; \, t) + f_{2,0}(e^{-t}\vec{k}_3\, ; \, t) - f_{2,0}(e^{-t}\vec{k}_4\, ; \, t) \right) \,.
\end{align}
The functions $g_{2,0}$, $g_{2,1}$, $\widetilde{g}_{4}^{(1)}$ and $\widetilde{g}_4^{(3)}$ are given in Eqn.'s~\eqref{eq:defg20}--\!~\eqref{eq:defg43} above.
Then we have
$$\langle \phi(\vec{p})|\Psi(\Lambda e^u)\rangle = \langle \phi(\vec{p})| \mathcal{P}_s \, e^{-i \int_{-\infty}^u ds \, (K(s) + L)} |\Omega\rangle + \mathcal{O}(\lambda^2)$$
for $\phi(\vec{p})$ such that $\phi(\vec{p}) = 0$ for $|\vec{p}| \geq \Lambda$.  In words, $|\Psi(\Lambda e^u)\rangle$ equals $\mathcal{P}_s \, e^{-i \int_{-\infty}^u ds \, (K(s) + L)} |\Omega\rangle$ up to $\mathcal{O}(\lambda^2)$ corrections for modes with $|\vec{p}| \leq \Lambda$.

\newpage
The explicit form of $f^{(1)}_4(\vec{k}_1,\vec{k}_2,\vec{k}_3,\vec{k}_4,u)$ is
\begin{align}
\begin{split}
\hspace{-30mm}
f^{(1)}_4&(\vec{k}_1,\vec{k}_2,\vec{k}_3,\vec{k}_4,u)=
\\&
\frac{e^{-3 u}}{24 (2 \pi )^d}
\frac{\left(\vec{k}_1^2+ e^{-2 u}m^2\right)\sqrt{\Lambda^2+m^2}\left[1-\left(\dfrac{\prod_{i=2}^4\left(\vec{k}_i^2+e^{-2 u} m^2\right)}{\left(\Lambda ^2+m^2\right)^2 \left(\vec{k}_1^2+e^{-2 u} m^2\right)}\right)^\frac{1}{4}\right]
}{\prod_{i=1}^4 \left(e^{2 u} \vec{k}_i^2+m^2\right)^\frac{3}{4} \left(\sum_{i=1}^4\left(e^{2 u} \vec{k}_i^2+ m^2\right)^\frac{1}{2}\right)
\log \left(\dfrac{\prod_{i=2}^4\left(\vec{k}_i^2+e^{-2 u} m^2\right)}{ \left(\Lambda^2+m^2\right)^2 \left(\vec{k}_1^2+e^{-2 u} m^2\right)}\right)}\times
\\&
\qquad \Bigg\{
d-\sum_{\ell=1}^4 \frac{|\vec{k}_\ell|}{\Lambda} \frac{\theta'(1-|\vec{k}_\ell|/\Lambda)}{\theta(1-|\vec{k}_\ell|/\Lambda)}  \prod_{p=1}^4 \theta(1-|\vec{k}_p|/\Lambda) 
\\&
\qquad \qquad 
-\frac{1}{4}\frac{e^{-u}|\vec{k}_1|}{\Lambda}\log\left(\frac{e^{-2u}(\vec{k}_1^2+m^2)}{\Lambda^2+m^2}\right)\theta'\left(1-\frac{e^{-u}|\vec{k}_1|}{\Lambda}\right)
+\frac{1}{2}\frac{m^2}{\vec{k}_1^2+m^2}\theta\left(1-\frac{e^{-u}|\vec{k}_1|}{\Lambda}\right)
\\&
\qquad \qquad 
+\frac{1}{4}\sum_{i=2}^4\frac{e^{-u}|\vec{k}_i|}{\Lambda}\log\left(\frac{e^{-2u}(\vec{k}_i^2+m^2)}{\Lambda^2+m^2}\right)\theta'\left(1-\frac{e^{-u}|\vec{k}_i|}{\Lambda}\right)
+\frac{1}{2}\sum_{i=2}^4\frac{m^2}{\vec{k}_i^2+m^2}\theta\left(1-\frac{e^{-u}|\vec{k}_i|}{\Lambda}\right)
\Bigg\}
\\&
+\frac{e^{-3 u}}{24 (2 \pi )^d}
\frac{\left(\vec{k}_1^2+ e^{-2 u}m^2\right)\sqrt{\Lambda^2+m^2}\left[1-\left(\dfrac{\prod_{i=2}^4\left(\vec{k}_i^2+e^{-2 u} m^2\right)}{\left(\Lambda ^2+m^2\right)^2 \left(\vec{k}_1^2+e^{-2 u} m^2\right)}\right)^\frac{1}{4}\right]
}{\prod_{i=1}^4 \left(e^{2 u} \vec{k}_i^2+m^2\right)^\frac{3}{2} \left(\sum_{i=1}^4\left(e^{2 u} \vec{k}_i^2+ m^2\right)^\frac{1}{2}\right)^2
\log^2 \left(\dfrac{\prod_{i=2}^4\left(\vec{k}_i^2+e^{-2 u} m^2\right)}{ \left(\Lambda ^2+m^2\right)^2 \left(\vec{k}_1^2+e^{-2 u} m^2\right)}\right)}\times
\\&
\qquad \Bigg\{
2\frac{\left(\sum_{i=1}^4\left(e^{2 u} \vec{k}_i^2+ m^2\right)^\frac{1}{2}\right)}{\prod_{i=1}^4 \left(e^{2 u} \vec{k}_i^2+m^2\right)^\frac{1}{4}}
\Big[e^{-2 u} m^2 \Big(\vec{k}_1^2 \left(\vec{k}_2^2 \vec{k}_3^2+\vec{k}_2^2\vec{k}_4^2+\vec{k}_3^2 \vec{k}_4^2\right)-\vec{k}_2^2 \vec{k}_3^2 \vec{k}_4^2
\\&
\qquad \qquad \qquad \qquad 
+2 \vec{k}_1^2 \left(\vec{k}_2^2+\vec{k}_3^2+\vec{k}_4^2\right)e^{-2 u} m^2+\left(3 \vec{k}_1^2+\vec{k}_2^2+\vec{k}_3^2+\vec{k}_4^2\right)m^4 e^{-4 u}+2 m^6 e^{-6 u}\Big)\Big]
\\&
\qquad \qquad 
+e^{-2u} m^2\prod_{i=1}^4 \left(e^{2 u} \vec{k}_i^2+m^2\right)^\frac{3}{4}\left(\sum_{i=1}^4\frac{1}{\left(\vec{k}_i^2+e^{-2u} m^2\right)^\frac{1}{2}}\right)
\log \left(\dfrac{\prod_{i=2}^4\left(\vec{k}_i^2+e^{-2 u} m^2\right)}{ \left(\Lambda ^2+m^2\right)^2 \left(\vec{k}_1^2+e^{-2 u} m^2\right)}\right)
\\&
\qquad \qquad 
+\prod_{i=1}^4 \left(e^{2 u} \vec{k}_i^2+m^2\right)^\frac{3}{4}\left(\sum_{i=1}^4\left(e^{2 u} \vec{k}_i^2+ m^2\right)^\frac{1}{2}\right)
\left(d-3+\sum_{i=2}^4\frac{e^{-2u} m^2}{\vec{k}_i^2+e^{-2u} m^2}\right)\times
\\&
\qquad \qquad \qquad \qquad \qquad \qquad \qquad \qquad  \qquad \qquad \qquad \qquad 
\log \left(\dfrac{\prod_{i=2}^4\left(\vec{k}_i^2+e^{-2 u} m^2\right)}{ \left(\Lambda ^2+m^2\right)^2 \left(\vec{k}_1^2+e^{-2 u} m^2\right)}\right)
\Bigg\}
\\&
+\frac{e^{-3 u}}{48 (2 \pi )^d}
\frac{\left(\vec{k}_1^2+ e^{-2 u}m^2\right)\sqrt{\Lambda^2+m^2}}{\prod_{i=1}^4 \left(e^{2 u} \vec{k}_i^2+m^2\right)^\frac{5}{2} \left(\sum_{i=1}^4\left(e^{2 u} \vec{k}_i^2+ m^2\right)^\frac{1}{2}\right)
\log\left(\dfrac{\prod_{i=2}^4\left(\vec{k}_i^2+e^{-2 u} m^2\right)}{ \left(\Lambda ^2+m^2\right)^2 \left(\vec{k}_1^2+e^{-2 u} m^2\right)}\right)}\times
\\&
\qquad \qquad 
\Big[e^{-2 u} m^2 \Big(\vec{k}_1^2 \left(\vec{k}_2^2 \vec{k}_3^2+\vec{k}_2^2\vec{k}_4^2+\vec{k}_3^2 \vec{k}_4^2\right)-\vec{k}_2^2 \vec{k}_3^2 \vec{k}_4^2
\\&
\qquad \qquad \qquad \qquad 
+2 \vec{k}_1^2 \left(\vec{k}_2^2+\vec{k}_3^2+\vec{k}_4^2\right)e^{-2 u} m^2+\left(3 \vec{k}_1^2+\vec{k}_2^2+\vec{k}_3^2+\vec{k}_4^2\right)m^4 e^{-4 u}+2 m^6 e^{-6 u}\Big)\Big]\,,
\end{split}
\end{align}
and the explicit form of $f^{(3)}_4(\vec{k}_1,\vec{k}_2,\vec{k}_3,\vec{k}_4,u)$ is 
\begin{align}
\begin{split}
\hspace{-30mm}
f^{(3)}_4&(\vec{k}_1,\vec{k}_2,\vec{k}_3,\vec{k}_4,u)=
\\&
\frac{e^{-3u}}{8(2 \pi )^d}
\frac{\left[1-\left(\dfrac{\prod_{i=1}^3\left(\vec{k}_i^2+e^{-2 u} m^2\right)}{\left(\Lambda ^2+m^2\right)^2 \left(\vec{k}_4^2+e^{-2 u} m^2\right)}\right)^\frac{1}{4}\right]
}{\left(\vec{k}_4^2+e^{-2 u} m^2\right)^\frac{1}{2}
\left(\sum_{i=1}^4 \left(\vec{k}_i^2+ e^{-2 u} m^2\right)^\frac{1}{2}\right)
\log \left(\dfrac{\prod_{i=1}^3\left(\vec{k}_i^2+e^{-2 u} m^2\right)}{\left(\Lambda ^2+m^2\right)^2 \left(\vec{k}_4^2+e^{-2 u} m^2\right)}\right)}\times
\\&
\qquad \Bigg\{
d-\sum_{\ell=1}^4 \frac{|\vec{k}_\ell|}{\Lambda} \frac{\theta'(1-|\vec{k}_\ell|/\Lambda)}{\theta(1-|\vec{k}_\ell|/\Lambda)}  \prod_{p=1}^4 \theta(1-|\vec{k}_p|/\Lambda) 
\\&
\qquad \qquad
+\frac{1}{4}\frac{e^{-u}|\vec{k}_4|}{\Lambda}\log\left(\dfrac{e^{-2u}(\vec{k}_4^2+m^2)}{\Lambda^2+m^2}\right)\theta'\left(1-\frac{e^{-u}|\vec{k}_4|}{\Lambda}\right)
-\frac{1}{2}\frac{m^2}{\vec{k}_4^2+m^2}\theta\left(1-\frac{e^{-u}|\vec{k}_4|}{\Lambda}\right)
\\&
\qquad \qquad 
-\frac{1}{4}\sum_{i=1}^3\frac{e^{-u}|\vec{k}_i|}{\Lambda}\log\left(\frac{e^{-2u}(\vec{k}_i^2+m^2)}{\Lambda^2+m^2}\right)\theta'\left(1-\frac{e^{-u}|\vec{k}_i|}{\Lambda}\right)
+\frac{1}{2}\sum_{i=2}^4\frac{m^2}{\vec{k}_i^2+m^2}\theta\left(1-\frac{e^{-u}|\vec{k}_i|}{\Lambda}\right)
\Bigg\}
\\&
+\frac{e^{-3u}}{8(2 \pi )^d}\frac{\left[1-\left(\dfrac{\prod_{i=1}^3\left(\vec{k}_i^2+e^{-2 u} m^2\right)}{\left(\Lambda ^2+m^2\right)^2 \left(\vec{k}_4^2+e^{-2 u} m^2\right)}\right)^\frac{1}{4}\right]
}{\left(\vec{k}_4^2+e^{-2 u} m^2\right)^\frac{3}{2} \left(\sum_{i=1}^4 \left(\vec{k}_i^2+ e^{-2 u} m^2\right)^\frac{1}{2}\right)^2
\log^2 \left(\dfrac{\prod_{i=1}^3\left(\vec{k}_i^2+e^{-2 u} m^2\right)}{\left(\Lambda ^2+m^2\right)^2 \left(\vec{k}_4^2+e^{-2 u} m^2\right)}\right)}\times
\\&
\qquad \Bigg\{
2e^{-2u} m^2\left(\vec{k}_4^2+e^{-2 u} m^2\right)\left(\sum_{i=1}^4 \left(\vec{k}_i^2+ e^{-2 u} m^2\right)^\frac{1}{2}\right)\left(\sum_{i=1}^3\frac{1}{\left(\vec{k}_i^2+e^{-2 u} m^2\right)}-\frac{1}{\left(\vec{k}_4^2+e^{-2 u} m^2\right)}\right)
\\&
\qquad \qquad
+e^{-2u} m^2\left(\sum_{i=1}^4 \left(\vec{k}_i^2+ e^{-2 u} m^2\right)^\frac{1}{2}\right)
\log \left(\dfrac{\prod_{i=1}^3\left(\vec{k}_i^2+e^{-2 u} m^2\right)}{\left(\Lambda ^2+m^2\right)^2 \left(\vec{k}_4^2+e^{-2 u} m^2\right)}\right)
\\&
\qquad \qquad
+(d-3)\left(\vec{k}_4^2+e^{-2 u} m^2\right)\left(\sum_{i=1}^4 \left(\vec{k}_i^2+ e^{-2 u} m^2\right)^\frac{1}{2}\right)\log \left(\dfrac{\prod_{i=1}^3\left(\vec{k}_i^2+e^{-2 u} m^2\right)}{\left(\Lambda ^2+m^2\right)^2 \left(\vec{k}_4^2+e^{-2 u} m^2\right)}\right)
\\&
\qquad \qquad
+e^{-2u} m^2\left(\vec{k}_4^2+e^{-2 u} m^2\right)\left(\sum_{i=1}^4\frac{1}{\left(\vec{k}_i^2+e^{-2 u} m^2\right)}\right)
\log \left(\dfrac{\prod_{i=1}^3\left(\vec{k}_i^2+e^{-2 u} m^2\right)}{\left(\Lambda ^2+m^2\right)^2 \left(\vec{k}_4^2+e^{-2 u} m^2\right)}\right)
\Bigg\}
\\&
+\frac{e^{-3u}}{16(2 \pi )^d}\frac{\left(\dfrac{\prod_{i=1}^3\left(\vec{k}_i^2+e^{-2 u} m^2\right)}{\left(\Lambda ^2+m^2\right)^2 \left(\vec{k}_4^2+e^{-2 u} m^2\right)}\right)^\frac{1}{4}
}{\left(\vec{k}_4^2+e^{-2 u} m^2\right)^\frac{1}{2} \left(\sum_{i=1}^4 \left(\vec{k}_i^2+ e^{-2 u} m^2\right)^\frac{1}{2}\right)
\prod_{i=1}^4 \left(\vec{k}_i^2+ e^{-2 u} m^2\right)
\log \left(\dfrac{\prod_{i=1}^3\left(\vec{k}_i^2+e^{-2 u} m^2\right)}{\left(\Lambda ^2+m^2\right)^2 \left(\vec{k}_4^2+e^{-2 u} m^2\right)}\right)}\times
\\&
\qquad \qquad
\Big[e^{-2u} m^2\Big(\vec{k}_4^2 \left(\vec{k}_1^2 \vec{k}_2^2+\vec{k}_1^2 \vec{k}_3^2+\vec{k}_2^2 \vec{k}_3^2\right)-\vec{k}_1^2 \vec{k}_2^2 \vec{k}_3^2+2 \vec{k}_4^2 e^{-2 u} m^2 \left(\vec{k}_1^2+\vec{k}_2^2+\vec{k}_3^2\right)
\\&
\qquad \qquad \qquad \qquad \qquad \qquad \qquad \qquad \qquad \qquad
+m^4 e^{-4 u} \left(\vec{k}_1^2+\vec{k}_2^2+\vec{k}_3^2+3 \vec{k}_4^2\right)+2 m^6 e^{-6 u}\Big)\Big]\,.
\end{split}
\end{align}


\subsubsection{Features of $1$-loop effects}

\begin{figure}
\centering
\includegraphics[width=8cm]{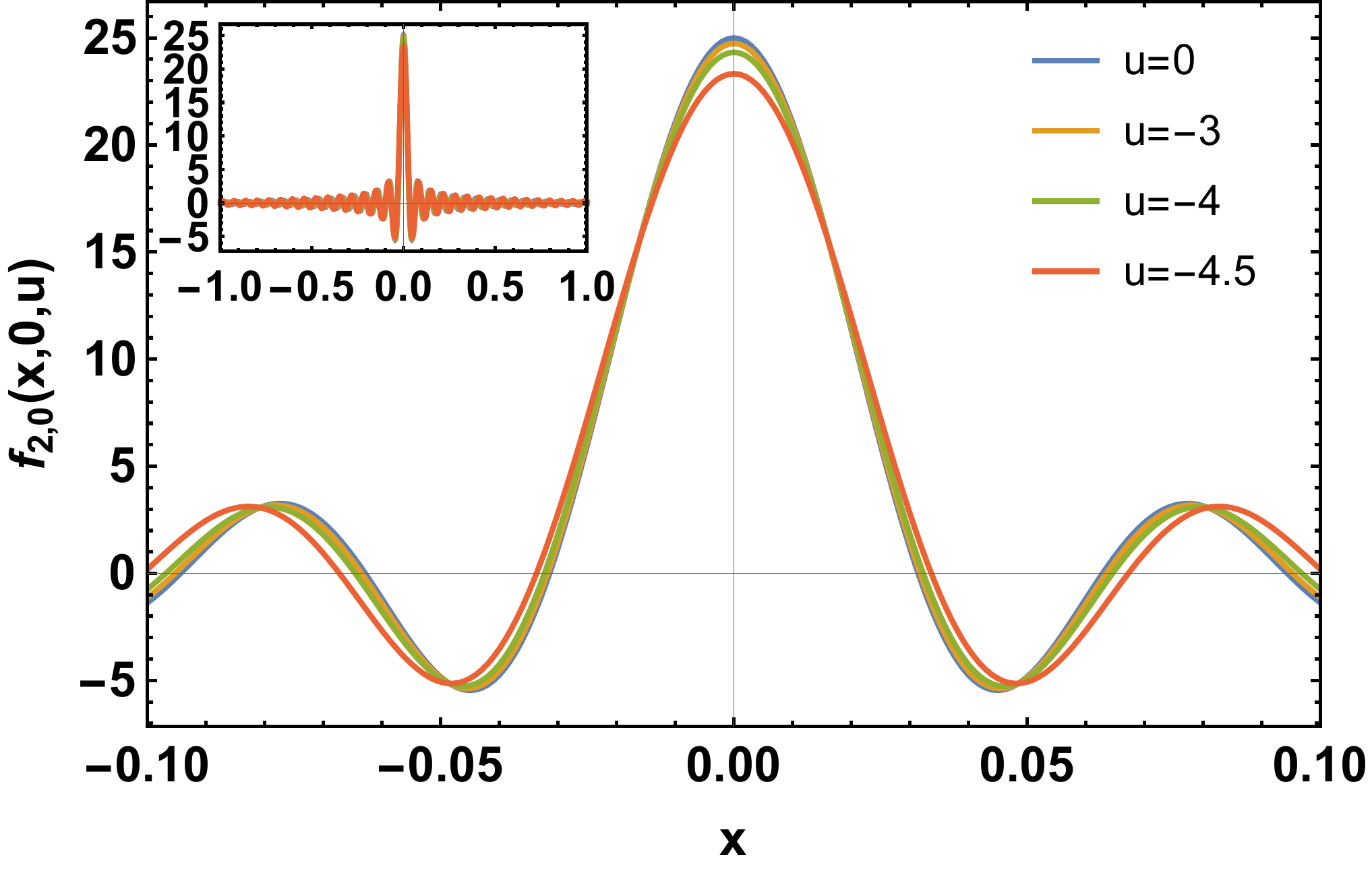}
\hspace{0mm}
\includegraphics[width=8cm]{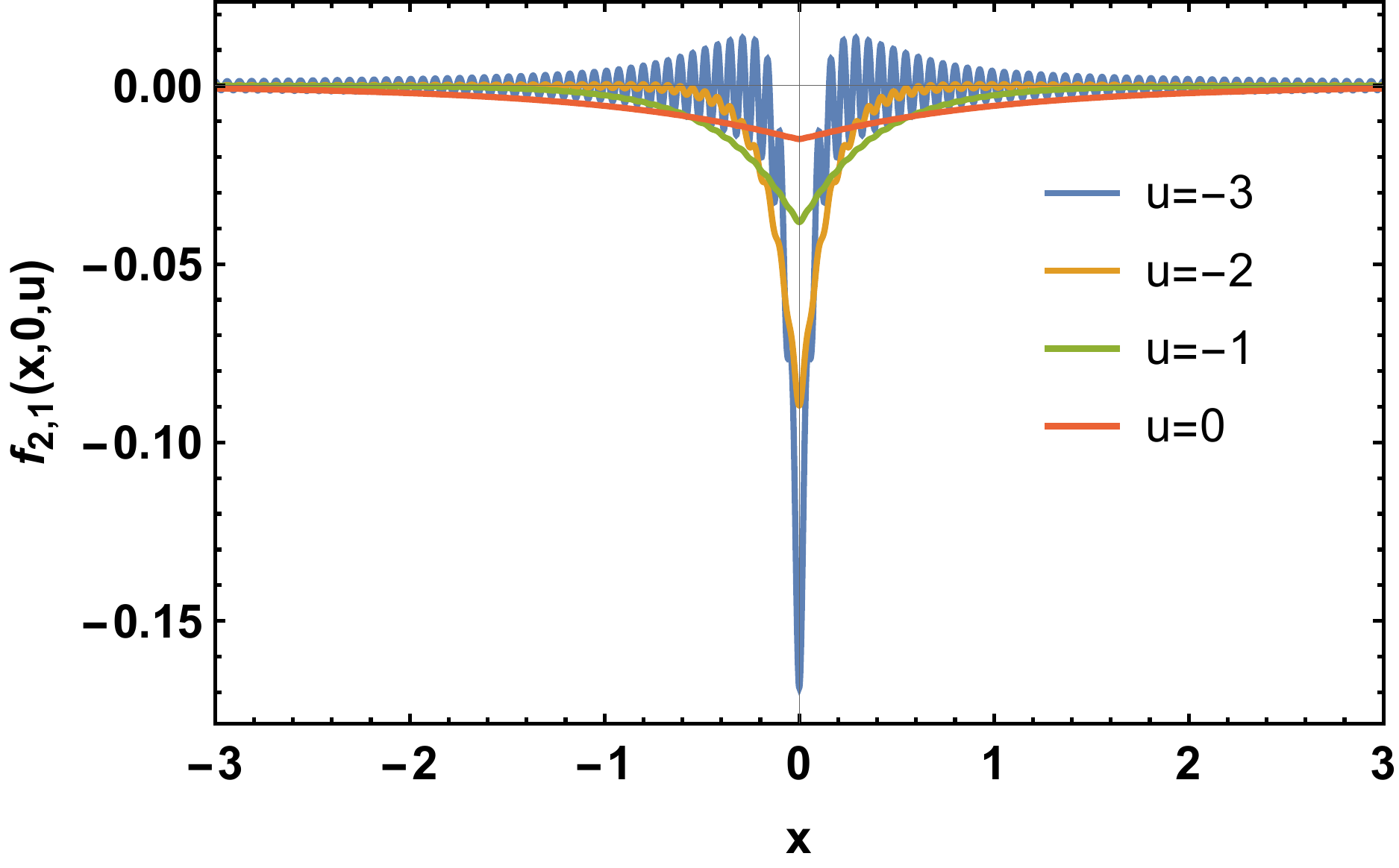}
\\
\hspace{1mm}
\includegraphics[width=7.9cm]{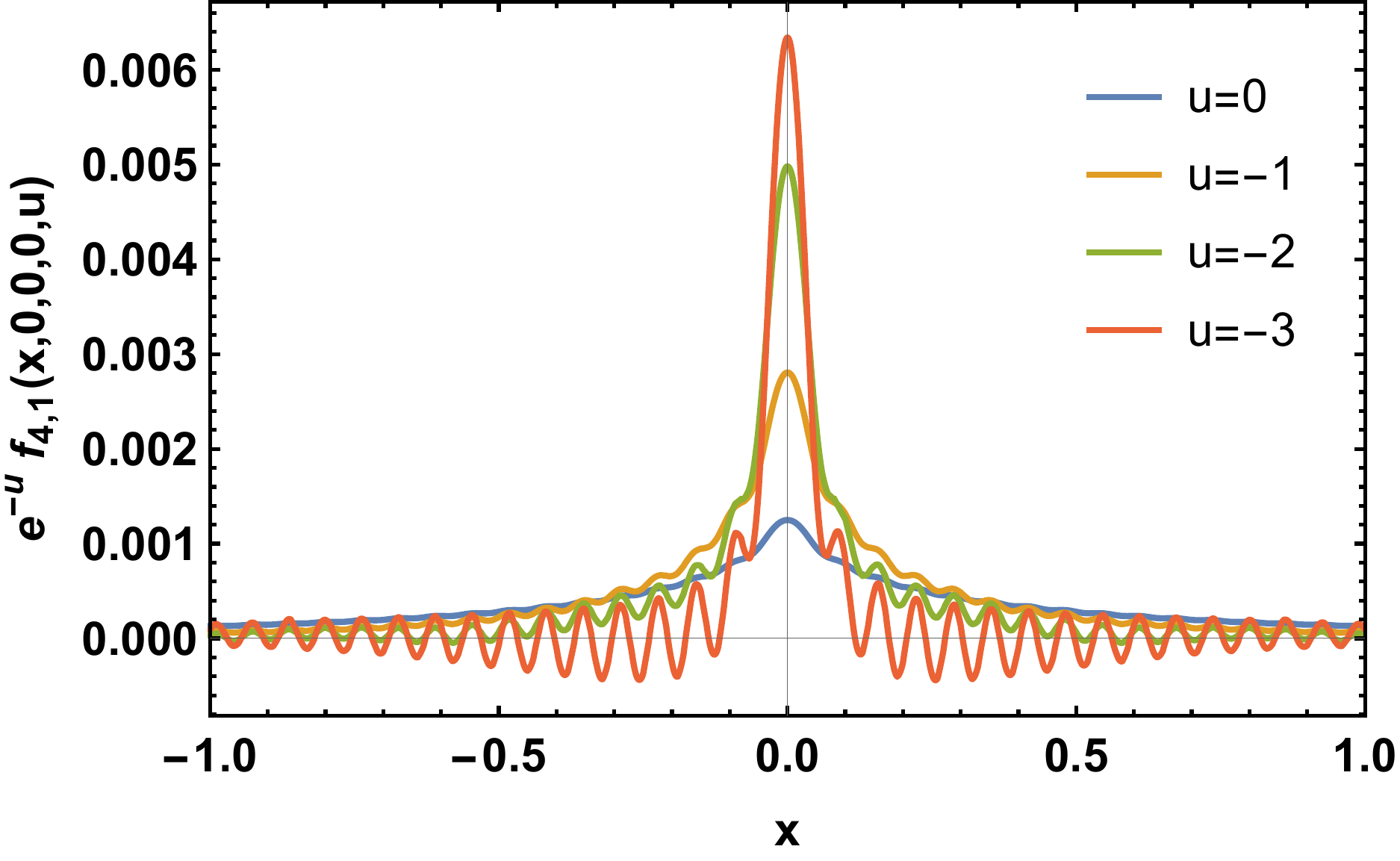}
\hspace{1mm}
\includegraphics[width=7.9cm]{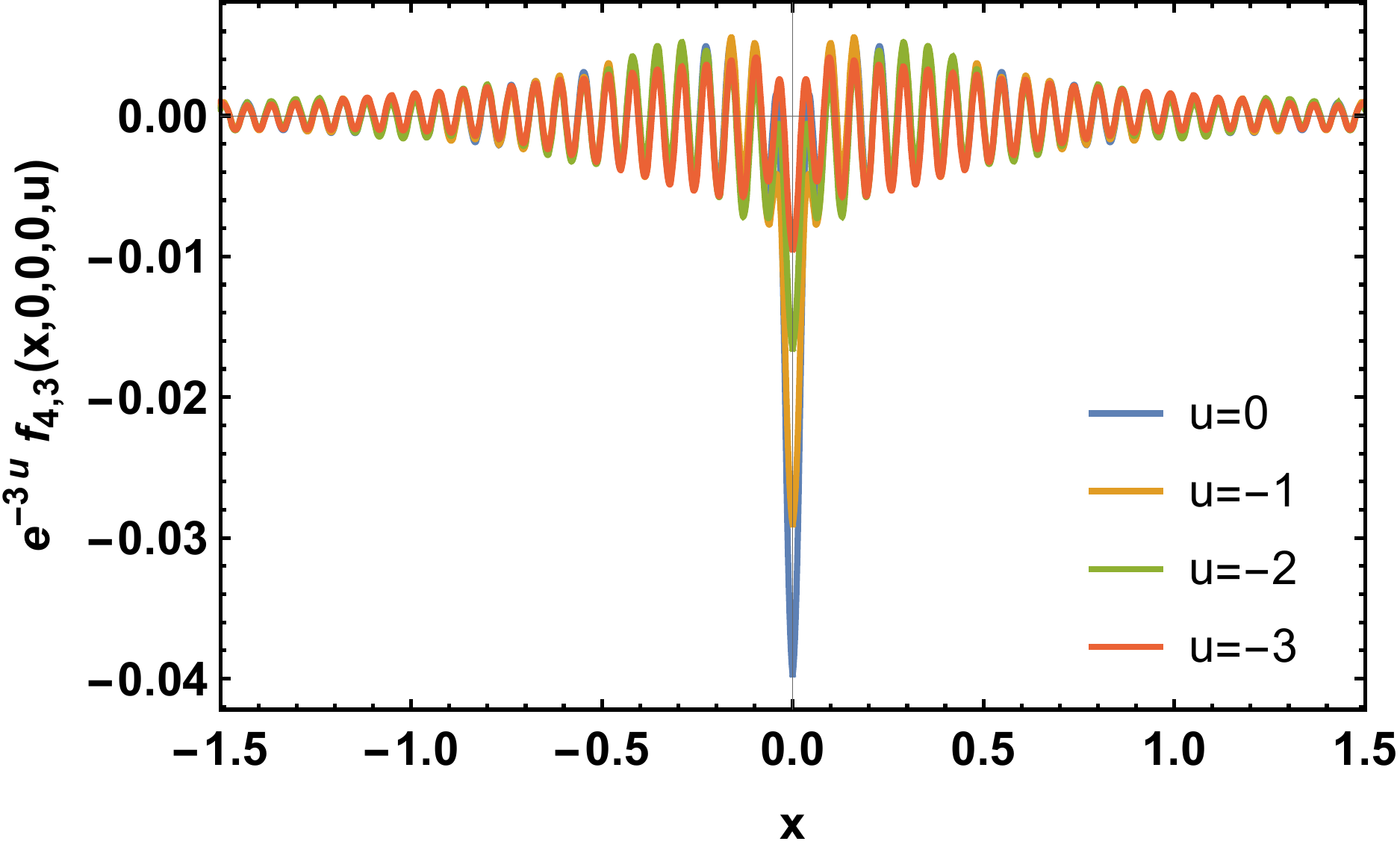}
\caption{The kernels comprising the entangler in position space at different values of $u$, with a soft cutoff (i.e., a sigmoid instead of a Heaviside step function). Here we work in $(1+1)$ dimensions.  The upper left kernel is the free kernel $f_{2,0}(x,0 \, ; u)$ and the upper right kernel is the quadratic kernel $f_{2,1}(x,0 \,; u)$. The momentum space equations corresponding to the upper plots were given in \eqref{eq:f2i}. The lower left and lower right plots are the quartic kernels $f_4^{(1)}(x,0,0,0 \,; u)$ and $f_4^{(3)}(x,0,0,0 \,; u)$, respectively.  The corresponding momentum space equations were given in \eqref{f41sol2}. In all plots we have set $m=1$ and $\Lambda=100$.}
\label{fig:kernels}
\end{figure}

One of the most interesting features of the $1$-loop cMERA kernels is that they are exponentially localized in position space with a characteristic width that goes as $\sim 1/(e^{-u} m)$, the inverse of the renormalized mass scale.  We have numerically plotted the Fourier transform of all of our kernels in Figure \ref{fig:kernels} in order to demonstrate their locality in position space.  We will provide detailed analytic arguments for the locality of the kernels in Appendix \ref{AppLocal}.  The arguments in Appendix \ref{AppLocal} are very general, and should apply to any theory with a mass scale.  The locality means that perturbative spatial Wilsonian RG (i.e., on spatial momentum modes) for the ground state of $\varphi^4$ theory can be exactly recast as a local position space cMERA. In other words, the renormalization group flow from the UV to the IR is perturbatively equivalent to locally disentangling degrees of freedom at progressively larger distance scales. Our example provides a concrete link between momentum space RG and tensor networks.

Considering the state $|\Psi(\Lambda e^u)\rangle$, the unentangled modes above the scale $\Lambda$ act as a large-momentum cutoff for the state, and accordingly, its correlation functions.  The way in which the unentangled modes act as a cutoff was explored in the context of free field theory in \cite{cMERA4}.  The quartic terms in the 1-loop cMERA kernel for $|\Psi(\Lambda e^u)\rangle$ contain cutoffs for each of four momenta $\vec{k}_1,\vec{k}_2,\vec{k}_3,\vec{k}_4$ individually.  Thus each momenta is regulated individually, once it passes above the scale $\Lambda$.

In the IR, we notice that $|\widetilde{\Omega}\rangle$ in Eqn.~\eqref{IRstate2copy} is an unentangled state in position space since $M$ and $M_2$ are constants and $M_4 = 0$.  In particular, we can write the state as
\begin{equation}
\langle \phi | \widetilde{\Omega}\rangle = \mathcal{N} \, \exp\left(- \frac{1}{2} \int d^d \vec{x} \, \phi(\vec{x}) \, (M + \lambda M_2) \, \phi(\vec{x}) \right) = \mathcal{N} \, \prod_{\vec{x}} \, \exp\left(- \frac{1}{2} \,d^d \vec{x} \, \phi(\vec{x}) \, (M + \lambda M_2) \, \phi(\vec{x}) \right)\,.
\end{equation}
Similar to the free massive scalar field case, the renormalized mass of $\varphi^4$ theory diverges in the IR which localizes correlations and causes entanglement to go away.  Thus, it is consistent that the IR state $|\widetilde{\Omega}\rangle$ be unentangled in position space.

\subsubsection{Momentum space RG vs. position space RG}

For both free massive scalar field theory and $\varphi^4$ theory, we have constructed a spatially local cMERA circuit which is perturbatively equivalent to Wilsonian RG on spatial momentum modes.  Turning the statement around: in the cases we studied, Wilsonian RG on spatial momentum modes is perturbatively equivalent to locally entangling spatial degrees of freedom across a hierarchy of distance scales.  The fact that Wilsonian RG can be recast as a cMERA is pleasing, and is harmonious with the underlying architecture of the standard MERA ansatz.

Our approach suggests that perturbative Wilsonian RG for any renormalizable field theory with a non-zero mass scale can be rewritten as a local position space cMERA.  In essence, our locality argument above depends on the momentum space kernels appearing in the perturbative circuit representation of Wilsonian RG having poles at the renormalized mass scale.  This feature is generic for renormalizable massive theories.  The locality argument holds at higher order in perturbation theory as well.

For massless interacting theories, the momentum space kernels appearing in the perturbative circuit representation of Wilsonian RG have poles on the real line, and decay only polynomially in position space.  If the theory in question is not a CFT, then there will be another length scale in the problem which is not the mass scale (since we are supposing the mass is zero).  This length scale will be defined by some perturbative coupling, and as such, cannot generate a pole in the momentum space kernels since said coupling only appears polynomially.

Our findings offer a new perspective on both cMERA and Wilsonian RG.  In particular, cMERA reorganizes the data of Wilsonian RG on spatial momenta to make the RG structure of the ground state wavefunctional more transparent.  For instance, even in the simple case of free massive scalar field theory, recasting Wilsonian RG as a local cMERA provides new insights.  As remarked above, this cMERA circuit does not essentially depend on the mass of the theory when entangling degrees of freedom at shorter distances than the scale of the inverse mass.  The mass becomes important in the cMERA kernel only for distance scales greater than or equal to the scale of the inverse mass.  These features of position space renormalization are perfectly intuitive, but are not immediately transparent from the standard perspective of RG in momentum space.

It would be interesting to study more examples of cMERA kernels for more diverse quantum field theories, and understand their local structure in position space.  For instance, the techniques in this paper are well-suited to studying the Wilson-Fisher fixed point.  A detailed understanding these kernels could enable a quantitative study of the correlation and entanglement across different distance scales of ground states of weakly interacting QFT's.

\subsection{Lessons for numerical applications}

Ultimately, one would like to use cMERA as an efficient \textit{numerical} ansatz for the ground state of an interacting quantum field theory.  Strongly interacting QFT's and gauge theories are of particular interest.  Our perturbative circuit calculations are useful for analytic studies and offering insight into the structure of cMERA circuits for interacting QFT's.  While it is possible to use numerical methods to compute cMERA circuits to higher order in perturbation theory for weakly interacting field theories, this is not a robust strategy for uncovering non-perturbative behavior of strongly interacting systems.  A better approach is to use the structural insights we have learned to construct judicious numerical ansatzes, which we sketch below.  In this section, we will be interested in ansatzes which produce the correct ground state in the UV instead of at all intermediate length scales.  However, it is still important that our ansatzes capture certain features of correlations at intermediate length scales. 

Starting with the usual IR state $|\Omega\rangle$ given by\footnote{As mentioned before, the state $|\Omega\rangle$ is unentangled in position space.  Even though we may consider systems with non-trivial entanglement in the IR, we can still use $|\Omega\rangle$ as our IR state and hierarchically build entanglement on top of it to obtain a good UV ansatz.  Indeed, in the present numerical setting, our desire is to obtain a good UV ansatz.  We can think of the IR state $|\Omega\rangle$ as imposing an IR cutoff on the largest length scale of correlations -- in particular, we will not get correlations in our final UV state across distances greater than $\sim e^{u_{\text{UV}} - u_\text{IR}}/\Lambda$.}
\begin{equation}
\label{IRstateAgain1}
\langle \phi |\Omega\rangle  = \mathcal{N} \, \exp\left(- \int d^d \vec{x} \, \phi(\vec{x}) M \phi(\vec{x}) \right)
\end{equation}
for $M$ some fixed constant, we choose a fixed IR scale $u_{\text{IR}} < 0$.  We construct an entangler
\begin{align}
K(s) &= \sum_{j_1} \int_{1/\Lambda} d^d \vec{x}_1 \, f_{j_1}(\vec{x}_1 \, ; \, u) \, \mathcal{O}_{j_1}(\vec{x}_1) + \sum_{j_2} \int_{1/\Lambda} d^d \vec{x}_1 \, d^d \vec{x}_2 \, f_{j_2}(\vec{x}_1, \vec{x}_2\, ; \, u) \, \mathcal{O}_{j_2}(\vec{x}_1, \vec{x}_2) \nonumber \\
& \qquad \qquad \qquad \qquad \qquad \qquad \qquad + \cdots + \sum_{j_n} \int_{1/\Lambda} d^d \vec{x}_1 \cdots d^d \vec{x}_n \, f_{j_n}(\vec{x}_1,...,\vec{x}_n\, ; \, u) \, \mathcal{O}_{j_n}(\vec{x}_1,...,\vec{x}_n)\,
\end{align} 
whose form we will further specify shortly.  Note that the position space integrals are cutoff from below at scale $1/\Lambda$, which is the distance scale corresponding to the UV.  We imagine that the $f_{i_1},...,f_{i_n}$ functions are of a specified form, but have undetermined parameters that we can tune and optimize.  We approximate the cMERA circuit $\mathcal{P}_s \exp\left(- i \int_{u_{\text{IR}}}^0 ds \, (K(s) + L) \right)$ by
\begin{align}
\left[e^{- i \,\Delta u \,(K(u_{\text{IR}}) + L)} \,\, e^{- i \,\Delta u \,(K(u_{\text{IR}} + \Delta u) + L)}  \, \cdots \, e^{- i \,\Delta u \,(K(0) + L)} \right]_{T}
\end{align}
where $\Delta u := -u_{\text{IR}}/N$ for some positive integer $N$, and $[ \cdots ]_{T}$ denotes that we truncate the terms inside the bracket at order $\mathcal{O}((\Delta u)^T)$.  In other words, $[ \cdots ]_{T}$ instructs us to construct a power series in $\Delta u$ up to order $\mathcal{O}((\Delta u)^T)$.
Then our cMERA ansatz is
\begin{equation}
|\Psi_{\text{cMERA}}\rangle := \left[e^{- i \,\Delta u \,(K(u_{\text{IR}}) + L)} \,\, e^{- i \,\Delta u \,(K(u_{\text{IR}} + \Delta u) + L)}  \, \cdots \, e^{- i \,\Delta u \,(K(0) + L)} \right]_{T}\, |\Omega\rangle
\end{equation}
which depends on the functions $f_{i_1},...,f_{i_n}$.

To utilize this ansatz, we consider a UV Hamiltonian $H_{\text{UV}}$ and perform the numerical minimization
\begin{equation}
\min_{f_{i_1},...,f_{i_n}} \frac{\langle \Psi_{\text{cMERA}}|H_{\text{UV}}|\Psi_{\text{cMERA}}\rangle}{\langle \Psi_{\text{cMERA}} | \Psi_{\text{cMERA}} \rangle}
\end{equation}
where the denominator is required since $|\Psi_{\text{cMERA}}\rangle$ is not normalized as given.  Since the form of $|\Omega\rangle$ is so simple, we can \textit{explicitly} compute correlation functions of the form
$$\langle \Omega|\mathcal{O}_{b_1}^\dagger \cdots \mathcal{O}_{b_q}^\dagger \, \mathcal{O}_{a_p}\cdots \mathcal{O}_{a_1}|\Omega\rangle$$
and
$$\langle \Omega|\mathcal{O}_{b_1}^\dagger \cdots \mathcal{O}_{b_q}^\dagger \,H_{\text{UV}}\, \mathcal{O}_{a_p}\cdots \mathcal{O}_{a_1}|\Omega\rangle\,,$$
and accordingly can drastically simplify both the numerator and denominator of
$$\langle \Psi_{\text{cMERA}}|H_{\text{UV}}|\Psi_{\text{cMERA}}\rangle/\langle \Psi_{\text{cMERA}} | \Psi_{\text{cMERA}} \rangle$$
before minimizing over the parameters of the $f_{i_1},...,f_{i_n}$.

But what form should the $f_{i_1},...,f_{i_n}$ have?  Using our perturbative results as a guide, we list several properties:
\begin{itemize}
\item The functions $f_{i_p}(\vec{x}_1,...,\vec{x}_{p} \, ; \, u)$ should be translation and rotation-invariant, meaning that
$$f_{i_p}(R \, \vec{x}_1 + \vec{a},..., R \, \vec{x}_{p} + \vec{a} \, ; \, u) = f_{i_p}(\vec{x}_1,...,\vec{x}_{p} \, ; \, u)$$
for all $\vec{a}$ and all rotation matrices $R \in SO(d)$.
\item Each $f_{i_p}(\vec{x}_1,...,\vec{x}_{p} \, ; \, u)$ is exponentially local in the distance between any pair of $\vec{x}_i$ and $\vec{x}_j$, for $1 \leq i,j \leq p$.
\item If the theory under consideration is a CFT, then $f_{i_p}(\vec{x}_1,...,\vec{x}_{p} \, ; \, u)$ should \textit{not} depend on $u$.  The function may depend on the values of marginal couplings.
\item If the theory under consideration is \textit{not} a CFT, then each  $f_{i_p}(\vec{x}_1,...,\vec{x}_{p} \, ; \, u)$ can be decomposed into a term which does not depend of $u$, and terms which do depend on $u$.  The term which does not depend on $u$ may depend on the values of marginal couplings.  The terms which do depend on $u$ should be approximately zero for $u$ such that $1/(\Lambda e^u)$ is below the scale of the inverse renormalized mass $1/(e^{-u} m)$ (or if the theory is massless, some other distance scale associated with the most relevant dimensionful coupling).
\end{itemize}
From the above points, it is clear that CFT's are easiest to work with since the kernels in their entanglers do not depend on the scale $u$.  In this case, a nice way of parametrizing the $f_{i_1},...,f_{i_n}$ is in terms of Sine-Gaussian wavelets which only depend on the differences of coordinates $|\vec{x}_i - \vec{x}_j|$.  For instance, we might parametrize a kernel $f(\vec{x}_1,\vec{x}_2)$ by
\begin{equation}
f(\vec{x}_1,\vec{x}_2 \, ; \, \{a_j, b_j, c_j, d_j, \phi_j\}) = \sum_{j} a_j \, e^{- b_j^2 \, |\vec{x}_1 - \vec{x}_2|^2 + c_j \, |\vec{x}_1 - \vec{x}_2|} \cos(d_j \, |\vec{x}_1 - \vec{x}_2| + \phi_j) 
\end{equation}
which is the form of the sum of the real parts of Gabor wavelets.  A kernel $f(\vec{x}_1,\vec{x}_2, \vec{x}_3, \vec{x}_4)$ might be parametrized similarly by
\begin{equation}
\label{paramkern1}
f(\vec{x}_1,\vec{x}_2, \vec{x}_3, \vec{x}_4 \, ; \, \{a_j, \textbf{B}_j, \textbf{c}_j, \textbf{d}_j, \boldsymbol{\phi}_j\}) = \sum_{j} a_j \, e^{- \textbf{x}^T \textbf{B}_j \textbf{x} + \textbf{c}_j \cdot \textbf{x}} \cos(\textbf{d}_j \cdot \textbf{x} + \boldsymbol{\phi}_j) 
\end{equation}
where
\begin{equation}
\label{paramkern2}
\textbf{x} := ( \, |\vec{x}_1 - \vec{x}_2|\,,\,|\vec{x}_1 - \vec{x}_3|\,,\, |\vec{x}_1 - \vec{x}_4|\,,\,|\vec{x}_2 - \vec{x}_3|\,,\,|\vec{x}_2 - \vec{x}_4|\,,\,|\vec{x}_3 - \vec{x}_4|\,)\,.
\end{equation}
Note that $\textbf{B}_j$ is a $6 \times 6$ matrix of parameters, and $\textbf{c}_j, \textbf{d}_j, \boldsymbol{\phi}_j$ are all $6$--dimensional vectors of parameters.  For non-CFT's, the parameters in Eqn.'s~\eqref{paramkern1} and~\eqref{paramkern2} can depend on $u$, and hence have non-trivial dependence on the distance scale.

We envision that by parametrizing the kernels $f_{i_1},...,f_{i_n}$ in terms of appropriate Sine-Gaussian wavelets, it should be possible for cMERA to become a useful variational method for the ground states of CFT's as well as regular QFT's (for which there are additional parametric dependencies in the kernels).  In particular, the integrals and gradient descent procedure required to minimize
$$\langle \Psi_{\text{cMERA}}|H_{\text{UV}}|\Psi_{\text{cMERA}}\rangle/\langle \Psi_{\text{cMERA}} | \Psi_{\text{cMERA}} \rangle$$
over the parameters of Sine-Gaussian wavelets (or similar such wavelets) can be performed efficiently.  There are many opportunities for future numerical studies.

\section{Discussion}

We have introduced new techniques for ``quantum circuit perturbation theory'' for systems in infinite-dimensional Hilbert spaces.  Specifically, we have focused on single-particle quantum mechanics, and scalar fields in $d+1$ dimensions.  Our techniques generalize naturally to $N$--particle quantum mechanics, and also to systems of fermions \cite{fermion}.  A generalization to gauge theories may also be possible.  It would also be interesting to generalize our techniques to many-body spin systems, along the lines of \cite{Bridging1}.  We have used quantum circuit perturbation theory to systematically construct unitaries which map between specified perturbatively non-Gaussian states.  In the context of QFT in particular, the wavefunctional representation is essential, and elucidates the position space or momentum space ``surgery'' required to unitarily map one state to another.

We have further used quantum circuit perturbation theory to develop cMERA for weakly interacting field theories.  Specifically, we have constructed cMERA circuits which generate the perturbative ground states of weakly interacting scalar field theories by locally building up entanglement at increasingly small distance scales starting from a product state.  Interestingly, our cMERA circuits can be made to exactly perform perturbative Wilsonian RG (in terms of the spatial momentum modes) of the ground state of the weakly interacting theories under consideration.  This is distinct from previous studies of cMERA in non-interacting systems in which the cMERA circuit only produces the correct state in the UV.

In other words, perturbative Wilsonian RG of spatial momentum modes can be exactly recast as a cMERA for the examples we studied.  By example, we explicitly showed that in constructing the ground state of scalar $\varphi^4$ theory, the cMERA kernels are exponentially local, and hence perturbative momentum space RG is equivalent to a local position space cMERA.  Since our procedure for perturbatively constructing cMERA's is systematic, a matching between spatial Wilsonian RG and cMERA should exist in general.

The form of our cMERA circuits explicitly encodes the entanglement structure of quantum states across different distance scales.  For instance, when we exactly solve for the cMERA circuit which produces the RG flow of the ground state of a free massive scalar field theory,\footnote{In previous work, the cMERA circuit would only reproduce the correct ground state in the UV.} we can explicitly track how correlations are built up differently at distance scales smaller than and larger than the scale of the inverse renormalized mass.  For the ground state of $\varphi^4$ theory, there is a more intricate interplay between the correlations the cMERA circuit develops at different distance scales.  It would be very interesting to perform a more detailed study of the entanglement properties of these circuits, which we have only commented on in passing.  One possible avenue for probing entanglement was suggested in the context of ``flow equations'' in \cite{Stefan1}.  Understanding the connection of our work to the flow equation approach to RG \cite{Wegner1, Stefan2} is also interesting more broadly.

We have used various lessons from our exploration of cMERA circuits for weakly interacting theories to suggest a refined numerical approach to cMERA, along with viable ansatzes.  We hope that this is a step towards developing robust numerical methods for the exploration of ground states of strongly interacting QFT's in the continuum and infinite-volume limits, as envisioned in the original cMERA paper \cite{cMERA1}.  Our proposed ansatzes seem ripe for numerical exploration in future studies.

In addition to providing new technical tools, our work emphasizes the perspective that complicated correlations and multipartite entanglement in states of interacting field theories can be recast in the language of local circuits which build up the entanglement architecture of the quantum states.  Although this perspective has been held within a certain community over the last several years, our work provides the first concrete instantiation of this view outside of free field theory.

It would be interesting to adapt our perturbative techniques to CFT's, which have a richer algebraic structure than standard field theories.  One could also use our techniques to explore various generalizations of holography, along the lines of \cite{holo1, holo2, holo3, holo4}.  As another application, one could consider exactly solvable theories (for instance, in low dimensions), for which it may be possible to develop exact circuits which build up the interacting ground states.  Our perturbative techniques are also well-suited for exploring circuit complexity for weakly interacting field theories, along the lines of \cite{complexity1, complexity2}.  For quantum circuit perturbation theory and perturbative cMERA, it would be valuable to analyze a wider collection of examples than scalar fields at 1-loop -- for instance, fermions and gauge fields, and also higher-loop calculations.

\section*{Acknowledgements}
We would like to thank Chris Akers, Ignacio Cirac, William Donnelly, Patrick Hayden, Michal Heller, Javier Molina-Vilaplana, Mark Mueller, Tobias Osborne, Daniel Ranard, Tadashi Takayanagi, Frank Verstraete, and Guifr\'{e} Vidal for valuable conversations and feedback.  We thank Felipe Hern\'{a}ndez for many discussions about the locality of the cMERA kernels in position space, for jointly developing the Fourier-analytic argument, and for reviewing the relevant section of the manuscript.  JC is supported by the Fannie and John Hertz Foundation and the Stanford Graduate Fellowship program. AM would like to thank the It from Qubit summer school in Bariloche and ``Complexity Workshop 2018'' at Max Planck Institute for Gravitational Physics where the main results of this paper were presented. AM would like to thank ICTP for hospitality during the last stages of this work. AN has greatly profited from discussions with Farhad Ardalan on the Wilsonian renormalization group and also Vahid Karimipour and Niloofar Vardian on different aspects of tensor networks. AN also would like to thank CERN TH-Division and ICTP for hospitality during some stages of this work.

\appendix
\section{Review of spatial Wilsonian Renormalization Group}

\label{sec:AppA}
In this Appendix, we perform the spatial Wilsonian Renormalization Group to $1$--loop for $\varphi^4$ theory.  (For a good review of standard, Euclidean Wilsonian RG, see \cite{Kardar}.)  In Section \ref{sec:cMERAfreerevisited}, we carried out a similar procedure for the free massive scalar field.  In the free massive scalar field case, we could perform Wilsonian RG \textit{exactly}.  For $\varphi^4$ theory, we need to instead work perturbatively in a weak coupling.  We will use the same notation as Section \ref{sec:cMERAfreerevisited}.

In spatial Wilsonian RG, we integrate out \textit{spatial} momentum modes for $\Lambda e^u \leq |\vec{p}| \leq \Lambda$.  We do not touch the time component of the fields so that the Hamiltonian (and its ground state) are well-defined after the RG procedure.  Using our notation from Section \ref{sec:cMERAfreerevisited}, let us denote the renormalized partition function as scale $\Lambda e^u$ by $Z^{\Lambda e^u}$.  From this partition function, one can read off the renormalized Hamiltonian at the scale $\Lambda e^u$.

The partition function for $\varphi^4$ theory at the UV scale $\Lambda$ is given by 
\begin{align}
\label{partition1int}
&Z^\Lambda[\{J_i(\vec{p})\}] = \nonumber \\
& \qquad \lim_{T \to \infty(1-i \epsilon)}\int \prod_{|\vec{p}| \leq \Lambda}\mathcal{D}\phi(\vec{p},t) \, \mathcal{D}\pi(\vec{p},t) \, e^{i \int_{-T}^T dt \left(\int^\Lambda d^d \vec{p} \, \pi(\vec{p},t) \dot{\phi}(\vec{p},t) - \mathscr{H}^\Lambda\big(\phi(\vec{p},t), \pi(\vec{p},t)\big) \right)} \, e^{- i \int^\Lambda d^d \vec{p} \, \sum_i J_i(\vec{p}) \mathcal{O}_i(\vec{p},0) }
\end{align}
where $\epsilon$ is a positive, infinitesimal parameter, and $\mathscr{H}^\Lambda\big(\phi(\vec{p},t), \pi(\vec{p},t)\big)$ is the Hamiltonian density
\begin{align}
\begin{split}
\mathscr{H}^\Lambda\big(\phi(\vec{p},t), \pi(\vec{p},t)\big) &= \frac{1}{2} \int^\Lambda d^d \vec{k} \, \left(\pi(\vec{k},t) \pi(-\vec{k},t) + \phi(\vec{k},t) \left(\vec{k}^2 + m^2\right) \phi(-\vec{k},t)\right) \nonumber \\
& \qquad \qquad + \frac{\lambda}{4!} \int^\Lambda d^d \vec{k}_1 \, d^d \vec{k}_2 \, d^d \vec{k}_3 \,\phi(\vec{k}_1,t) \phi(\vec{k}_2,t) \phi(\vec{k}_3,t) \phi(-\vec{k}_1 -\vec{k}_2 -\vec{k}_3,t)
\end{split}
\end{align} 
Now we will perform RG on $\varphi^4$ theory down to scale $\Lambda e^u$.  Letting
\begin{align}
\phi(\vec{p},t) &= \begin{cases} 
   \phi_{<}(\vec{p},t) & \text{if } |\vec{p}| \leq \Lambda e^u \\
   \phi_{>}(\vec{p},t) & \text{if } \Lambda e^u < |\vec{p}| \leq \Lambda
  \end{cases} \\ \nonumber \\
\pi(\vec{p},t) &= \begin{cases} 
   \pi_{<}(\vec{p},t) & \text{if } |\vec{p}| \leq \Lambda e^u \\
   \pi_{>}(\vec{p},t) & \text{if } \Lambda e^u < |\vec{p}| \leq \Lambda
  \end{cases}
\end{align}
we can rewrite the partition function $Z^\Lambda[\{J_i(\vec{p})\}]$ as
\begin{align}
\label{partitionintegrated01}
\hspace{-1.3cm}
\begin{split}
& \int \prod_{|\vec{p}| \leq \Lambda e^u}\mathcal{D}\phi_{<}(\vec{p},t) \, \mathcal{D}\pi_{<}(\vec{p},t) \, \prod_{\Lambda e^u < |\vec{p}| \leq \Lambda}\mathcal{D}\phi_{>}(\vec{p},t) \, \mathcal{D}\pi_{>}(\vec{p},t)
\\
& \,\, \times \, \exp\left(i \int_{-T}^T dt \int^{\Lambda e^u} d^d \vec{p} \, \left(\pi_{<}(\vec{p},t) \dot{\phi}_{<}(-\vec{p},t)-\frac{1}{2}\pi_{<}(\vec{p},t)\pi_{<}(-\vec{p},t) - \frac{1}{2}\phi_{<}(\vec{p},t) (\vec{p}^2 + m^2) \phi_{<}(-\vec{p},t)\right)\right)
\\
& \,\, \times \, \exp\left(i \int_{-T}^T dt \int_{\Lambda e^u}^{\Lambda} d^d \vec{p} \, \left(\pi_{>}(\vec{p},t) \dot{\phi}_{>}(-\vec{p},t)-\frac{1}{2}\pi_{>}(\vec{p},t)\pi_{>}(-\vec{p},t) - \frac{1}{2}\phi_{>}(\vec{p},t) (\vec{p}^2 + m^2) \phi_{>}(-\vec{p},t)\right)\right)
\\
& \,\, \times \, \Bigg[1-
i\,\frac{\lambda}{4!}\int_{-T}^T dt \int_{\Lambda e^u}^{\Lambda} \frac{d^d \vec{p}_1d^d \vec{p}_2d^d \vec{p}_3}{(2\pi)^d}
\phi_{>}(\vec{p}_1,t)\phi_{>}(\vec{p}_2,t)\phi_{>}(\vec{p}_3,t)\phi_{>}(-\vec{p}_1-\vec{p}_2-\vec{p}_3,t)
\\&
\qquad \quad 
-
i\,\frac{\lambda}{4!}\int_{-T}^T dt \int^{\Lambda e^u} \frac{d^d \vec{p}_1d^d \vec{p}_2d^d \vec{p}_3}{(2\pi)^d}
\phi_{<}(\vec{p}_1,t)\phi_{<}(\vec{p}_2,t)\phi_{<}(\vec{p}_3,t)\phi_{<}(-\vec{p}_1-\vec{p}_2-\vec{p}_3,t)
\\&
\qquad \quad 
-6
i\,\frac{\lambda}{4!}\int_{-T}^T dt \int_{\Lambda e^u}^{\Lambda} \frac{d^d \vec{p}_1d^d \vec{p}_2}{(2\pi)^{d/2}}\int^{\Lambda e^u} \frac{d^d \vec{p}_3d^d \vec{p}_4}{(2\pi)^{d/2}}\;
\phi_{>}(\vec{p}_1,t)\phi_{>}(\vec{p}_2,t)\phi_{<}(\vec{p}_3,t)\phi_{<}(\vec{p}_4,t)\delta^{(d)}(\vec{p}_1+\vec{p}_2+\vec{p}_3+\vec{p}_4)
\Bigg]\\
& \,\, \times \, \exp\left(i \int^{\Lambda e^u} d^d \vec{p} \, \sum_i J_i(\vec{p}) \mathcal{O}_i(\vec{p},0) \right)+\mathcal{O}(\lambda^2)\,.
\end{split}
\end{align}
Before we take the path integral over the $\phi_>$ and $\pi_>$ fields, we first take an inventory of the type of terms we have.  There are three terms in the path integral at $\mathcal{O}(\lambda)$.  The $\mathcal{O}(\lambda)$ term containing $\phi_>\,\phi_>\,\phi_>\,\phi_>$ will contribute a $\phi_{<}$--independent term which simply renormalizes the partition function.  The $\mathcal{O}(\lambda)$ term containing $\phi_<\,\phi_<\,\phi_<\,\phi_<$ will become the quartic interaction term in the renormalized action.  Finally, the $\mathcal{O}(\lambda)$ term containing $\phi_> \, \phi_> \, \phi_< \, \phi_<$ will contribute a mass renormalization to the renormalized action after we integrate out the $\phi_>$ modes.  Path integrating over the $\phi_>$ and $\pi_>$ fields, we are left with 
\begin{align}
\label{partitionintegrated02}
\begin{split}
\propto & \int \prod_{|\vec{p}| \leq \Lambda e^u}\mathcal{D}\phi_{<}(\vec{p},t) \, \mathcal{D}\pi_{<}(\vec{p},t)
\\
& \,\, \times \, \exp\left(i \int_{-T}^T dt \int^{\Lambda e^u} d^d \vec{p} \, \left(\pi_{<}(\vec{p},t) \dot{\phi}_{<}(-\vec{p},t)-\frac{1}{2}\pi_{<}(\vec{p},t)\pi_{<}(-\vec{p},t) - \frac{1}{2}\phi_{<}(\vec{p},t) (\vec{p}^2 + m^2) \phi_{<}(-\vec{p},t)\right)\right)
\\
& \,\, \times \, \Bigg[1-
i\,\frac{\lambda}{4!}\int_{-T}^T dt \int^{\Lambda e^u} \frac{d^d \vec{p}_1d^d \vec{p}_2d^d \vec{p}_3}{(2\pi)^d}
\phi_{<}(\vec{p}_1,t)\phi_{<}(\vec{p}_2,t)\phi_{<}(\vec{p}_3,t)\phi_{<}(-\vec{p}_1-\vec{p}_2-\vec{p}_3,t)
\\&
\qquad \qquad \qquad \qquad \qquad \qquad \quad 
-
\frac{\lambda}{4}i\int_{-T}^T dt \int_{\Lambda e^u}^{\Lambda} \frac{d^d \vec{p}_1}{(2\pi)^{d}}\frac{1}{\vec{p}_1+m^2}\int^{\Lambda e^u} d^d \vec{p}_3\;
\phi_{<}(\vec{p}_3,t)\phi_{<}(-\vec{p}_3,t)
\Bigg]\\
& \,\, \times \, \exp\left(i \int^{\Lambda e^u} d^d \vec{p} \, \sum_i J_i(\vec{p}) \mathcal{O}_i(\vec{p},0) \right)+\mathcal{O}(\lambda^2)
\end{split}
\end{align}
As per Section \ref{sec:cMERAfreerevisited}, we rescale the momentum and time as $(\vec{p},t) \to (e^{-u} \vec{p}, e^u t)$ and renormalize the $\phi_<$ and $\pi_<$ fields by
\begin{align}
\label{renormalize21}
e^{\frac{d+1}{2}  \cdot u} \, \phi_{<}(e^u \vec{p},e^{-u}t) \quad &\longrightarrow \quad \phi(\vec{p},t) \\
\label{renormalize22}
e^{\frac{d-1}{2}  \cdot u} \, \pi_{<}(e^u \vec{p}, e^{-u} t) \quad &\longrightarrow \quad \pi(\vec{p},t)
\end{align}
(c.f. Eqn.'s~\eqref{renormalize1} and~\eqref{renormalize2}) to obtain
\begin{align}
\label{partitionintegrated03}
Z_{1-\text{loop}}^{\Lambda e^u}[\{J_i(\vec{p})\}] :=& \,\,\lim_{T \to \infty(1-i \epsilon)} C \! \int \prod_{|\vec{p}| \leq \Lambda}\mathcal{D}\phi(\vec{p},t) \, \mathcal{D}\pi(\vec{p},t) \nonumber \\
& \qquad \qquad \times \, \exp\bigg(i \int_{-T}^T dt \int^{\Lambda} d^d \vec{p} \,\big(\pi(\vec{p},t) \dot{\phi}(\vec{p},t) \nonumber \\
&\qquad \qquad \qquad \qquad \qquad \qquad \qquad - \frac{1}{2}\pi(\vec{p},t)\pi(\vec{p},t) - \frac{1}{2}\phi(\vec{p},t) (\vec{p}^2 + e^{-2u} m^2) \phi(\vec{p},t)\big)\bigg) \nonumber \\
\times \, \Bigg[1 &-i\,
e^{(d-3)u}\,\frac{\lambda}{4!}\int_{-T}^{T} dt \int^{\Lambda} \frac{d^d \vec{p}_1d^d \vec{p}_2d^d \vec{p}_3}{(2\pi)^d}
\phi_{<}(\vec{p}_1,t)\phi_{<}(\vec{p}_2,t)\phi_{<}(\vec{p}_3,t)\phi_{<}(-\vec{p}_1-\vec{p}_2-\vec{p}_3,t)
\nonumber \\ &
\qquad \qquad \quad \quad 
-
i\,e^{-2u}\,\frac{\lambda}{4}\int_{-T}^{T} dt \left(\int_{\Lambda e^u}^{\Lambda} \frac{d^d \vec{p}_1}{(2\pi)^{d}}\frac{1}{\vec{p}_1+m^2} \right)\int^{\Lambda} d^d \vec{p}_2\;
\phi_{<}(\vec{p}_2,t)\phi_{<}(-\vec{p}_2,t)
\Bigg] \nonumber \\
& \qquad \qquad \times \, \exp\left(i \int^{\Lambda} d^d \vec{p} \,e^{du}\, \sum_i J_i(\vec{p}) \mathcal{O}_i^{\,'}(\vec{p},0) \right)\,,
\end{align}
where the primed $\mathcal{O}_i^{\,'}$ operators are the same as the unprimed $\mathcal{O}_i$ operators, but written in terms of $\phi,\pi$ instead of $\phi_<\,, \pi_<$ as per Eqn.'s~\eqref{renormalize21} and~\eqref{renormalize22}.  Writing $\mathcal{O}_i\big(\phi_<(\vec{p},0), \pi_<(\vec{p},0)\big)$ to denote the operator's dependence on $\phi_<(\vec{p},0)$ and $\pi_<(\vec{p},0)$, we have
\begin{equation}
\mathcal{O}_i^{\,'}\big(\phi(\vec{p},0), \pi(\vec{p},0)\big) := \mathcal{O}_i\big(e^{\frac{d+2}{2} \cdot u}\phi_<(e^u\vec{p},0), e^{-\frac{d+2}{2} \cdot u} \pi_<(e^u\vec{p},0)\big)\,.
\end{equation}
We can read off that the Hamiltonian renormalized to scale $\Lambda e^u$ is
\begin{align}
H_{1-\text{loop}}^{\Lambda e^u} &= \frac{1}{2} \int^\Lambda d^d \vec{k} \, \left(\widehat{\pi}(\vec{k}) \widehat{\pi}(-\vec{k}) + \widehat{\phi}(\vec{k}) \left(\vec{k}^2 + e^{-2u}\,\widetilde{m}^2\right) \widehat{\phi}(-\vec{k})\right) \nonumber \\
& \qquad \qquad \qquad \qquad + \frac{e^{(d-3)u}\lambda}{4!} \int^\Lambda d^d \vec{k}_1 \, d^d \vec{k}_2 \, d^d \vec{k}_3 \, \widehat{\phi}(\vec{k}_1) \widehat{\phi}(\vec{k}_2) \widehat{\phi}(\vec{k}_3) \widehat{\phi}(-\vec{k}_1 - \vec{k}_2 - \vec{k}_3)
\end{align}
where 
\be
\widetilde{m}^2=m^2+\delta m^2=m^2
+\frac{\lambda}{2}\int_{\Lambda e^u}^{\Lambda} \frac{d^d \vec{p}}{(2\pi)^{d}}\frac{1}{\vec{p}^2+m^2}\,.
\ee
The explicit form of $\delta m^2$ is given by 
\be\label{eq:deltam2}
\hspace{-15mm}
\delta m^2\;=
\begin{cases}
\dfrac{\lambda}{4\pi m} \left(\arctan(\Lambda/m)-\arctan(\Lambda e^u/m)\right)  & ~~~ ~~~ d=1 \\ \\
\dfrac{\lambda}{8\pi} \log\left(\dfrac{\Lambda^2+m^2}{e^{2u} \Lambda^2+m^2}\right) & ~~~ ~~~ d =2 \\ \\
\dfrac{\lambda \cdot \text{Area}(\mathbb{S}^{d-1})\cdot \Lambda^d}{2 \,d \, m^2(2\pi)^d}\left[_2F_1\left(1,\frac{d}{2}\,;\,1+\frac{d}{2}\,;\,-\frac{\Lambda^2}{m^2}\right)-{e^{du}} _2F_1\left(1,\frac{d}{2}\,;\,1+\frac{d}{2}\,;\,-\frac{\Lambda^2e^{2u}}{m^2}\right)\right]  & ~~~ ~~~ d \ge 3
\end{cases}
\ee
where $\text{Area}(\mathbb{S}^{d-1})$ is the area of the $(d-1)$-dimensional unit sphere.

We can carry out the spatial Wilsonian RG procedure to higher orders in perturbation theory by proceeding in a similar fashion.

\section{Some Useful Formulas}
\label{sec:AppC}

Consider a scalar field theory in $d+1$ dimensions, with wavefunctional $|\Psi_0\rangle$, given in $\phi$--space by
\begin{equation}
\langle \phi | \Psi_0\rangle = \mathcal{N} \, e^{- \frac{1}{2} \int d^d \vec{k} \, \phi(\vec{k}) \, \Omega(\vec{k}) \, \phi(-\vec{k})}
\end{equation}
for some kernel $\Omega_0(\vec{k}) = \Omega(|\vec{k}|)$.  Recall that
\begin{align}
S_2^{(1)}(\vec{k}_1, \vec{k}_2) &= \widehat{\phi}(\vec{k}_1) \widehat{\pi}(\vec{k}_2) + \widehat{\pi}(\vec{k}_2) \widehat{\phi}(\vec{k}_1) \\ \nonumber \\
S_4^{(1)}(\vec{k}_1, \vec{k}_2, \vec{k}_3, \vec{k}_4) &= \widehat{\phi}(\vec{k}_1) \widehat{\pi}(\vec{k}_2) \widehat{\pi}(\vec{k}_3) \widehat{\pi}(\vec{k}_4) + \widehat{\pi}(\vec{k}_2) \widehat{\pi}(\vec{k}_3) \widehat{\pi}(\vec{k}_4) \widehat{\phi}(\vec{k}_1) \\ \nonumber \\
S_4^{(3)}(\vec{k}_1, \vec{k}_2, \vec{k}_3, \vec{k}_4) &= \widehat{\phi}(\vec{k}_1) \widehat{\phi}(\vec{k}_2) \widehat{\phi}(\vec{k}_3) \widehat{\pi}(\vec{k}_4) + \widehat{\pi}(\vec{k}_4) \widehat{\phi}(\vec{k}_1) \widehat{\phi}(\vec{k}_2) \widehat{\phi}(\vec{k}_3)\,.
\end{align}
Then we have the useful formulas:
\begin{align}
\langle \phi | S_2^{(1)}(\vec{k}_1, \vec{k}_2) |\Psi_0\rangle &= \left( 2 i \, \phi(\vec{k}_1) \phi(\vec{k}_2) \, \Omega(\vec{k}_2) - i \, \delta^{(d)}(\vec{k}_1 + \vec{k}_2)\right) \, \langle \phi|\Psi_0\rangle \qquad \quad \,\,\,
\end{align}
\begin{align}
&\langle \phi | S_4^{(1)}(\vec{k}_1, \vec{k}_2,\vec{k}_3,\vec{k}_4) |\Psi_0\rangle \nonumber \\
&= \big(- 2 i \, \phi(\vec{k}_1) \phi(\vec{k}_2) \phi(\vec{k}_3) \phi(\vec{k}_4) \, \Omega(\vec{k}_2) \Omega(\vec{k}_3) \Omega(\vec{k}_4) + 2i \, \delta^{(d)}(\vec{k}_2 + \vec{k}_3) \, \phi(\vec{k}_1) \phi(\vec{k}_4) \, \Omega(\vec{k}_3) \Omega(\vec{k}_4) \nonumber \\
&\qquad + 2i \, \delta^{(d)}(\vec{k}_2 + \vec{k}_4) \, \phi(\vec{k}_1) \phi(\vec{k}_3) \, \Omega(\vec{k}_3) \Omega(\vec{k}_4) + 2i \, \delta^{(d)}(\vec{k}_3 + \vec{k}_4) \, \phi(\vec{k}_1) \phi(\vec{k}_2) \, \Omega(\vec{k}_2) \Omega(\vec{k}_4) \nonumber \\
&\qquad +i \delta^{(d)}(\vec{k}_1 + \vec{k}_2) \, \phi(\vec{k}_3) \phi(\vec{k}_4) \, \Omega(\vec{k}_3) \Omega(\vec{k}_4) +i \delta^{(d)}(\vec{k}_1 + \vec{k}_3) \, \phi(\vec{k}_2) \phi(\vec{k}_4) \, \Omega(\vec{k}_2) \Omega(\vec{k}_4) \nonumber \\
&\qquad +i \delta^{(d)}(\vec{k}_1 + \vec{k}_4) \, \phi(\vec{k}_2) \phi(\vec{k}_3) \, \Omega(\vec{k}_2) \Omega(\vec{k}_3) - i \delta^{(d)}(\vec{k}_1 + \vec{k}_2) \delta^{(d)}(\vec{k}_3 + \vec{k}_4) \, \Omega(\vec{k}_4) \nonumber \\
&\qquad - i \delta^{(d)}(\vec{k}_1 + \vec{k}_3) \delta^{(d)}(\vec{k}_2 + \vec{k}_4) \, \Omega(\vec{k}_4) - i \delta^{(d)}(\vec{k}_1 + \vec{k}_4) \delta^{(d)}(\vec{k}_2 + \vec{k}_3) \, \Omega(\vec{k}_3) \big) \, \langle \phi | \Psi_0\rangle
\end{align}
\begin{align}
\langle \phi | S_4^{(3)}(\vec{k}_1, \vec{k}_2,\vec{k}_3,\vec{k}_4) |\Psi_0\rangle &= \big(2i \, \phi(\vec{k}_1) \phi(\vec{k}_2) \phi(\vec{k}_3) \phi(\vec{k}_4) \, \Omega(\vec{k}_4) - i \delta^{(d)}(\vec{k}_1 + \vec{k}_4) \, \phi(\vec{k}_2) \phi(\vec{k}_3) \nonumber\\
& \qquad \quad - i \delta^{(d)}(\vec{k}_2 + \vec{k}_4) \, \phi(\vec{k}_1) \phi(\vec{k}_3) - i \delta^{(d)}(\vec{k}_3 + \vec{k}_4) \, \phi(\vec{k}_1) \phi(\vec{k}_2) \big) \, \langle \phi | \Psi_0\rangle\,.
\end{align}

\section{Spatial locality of entangler kernels for $\varphi^4$ theory}
\label{AppLocal}
In our analysis above, we have expressed $K(s)$ in terms of operators and kernels with arguments in momentum space.  Fourier-transforming $K(s)$ into position space yields kernels:
\begin{align}
\label{posspaceK20}
&f_{2,0}(|\vec{x}_1 - \vec{x}_2|\,;\,u) := \frac{1}{(2 \pi)^{d}} \int d^d \vec{k}_1 \, d^d \vec{k}_2  \, e^{- i \, (\vec{k}_1 \cdot \vec{x}_1 + \vec{k}_2 \cdot \vec{x}_2)} \, \delta^{(d)}(\vec{k}_1 + \vec{k}_2) \, f_{2,0}(\vec{k}_1\,;\,u) \\
\label{posspaceK21}
&f_{2,1}(|\vec{x}_1 - \vec{x}_2|\,;\,u) := \frac{1}{(2 \pi)^{d}} \int d^d \vec{k}_1 \, d^d \vec{k}_2  \, e^{- i \, (\vec{k}_1 \cdot \vec{x}_1 + \vec{k}_2 \cdot \vec{x}_2)} \, \delta^{(d)}(\vec{k}_1 + \vec{k}_2) \, f_{2,1}(\vec{k}_1\,;\,u) \\
\label{posspaceK41}
& f_{4}^{(1)}(|\vec{x}_1 - \vec{x}_2|, |\vec{x}_1 - \vec{x}_3|, |\vec{x}_1 - \vec{x}_4|, |\vec{x}_2 - \vec{x}_3|, |\vec{x}_2 - \vec{x}_4|, |\vec{x}_3 - \vec{x}_4|\,;\,u) & \nonumber \\
& \qquad := \frac{1}{(2 \pi)^{2d}} \int d^d \vec{k}_1 \, d^d \vec{k}_2 \, d^d \vec{k}_3 \, d^d \vec{k}_4  \, e^{- i \, (\vec{k}_1 \cdot \vec{x}_1 + \vec{k}_2 \cdot \vec{x}_2 + \vec{k}_3 \cdot \vec{x}_3 + \vec{k}_4 \cdot \vec{x}_4)} \, \delta^{(d)}(\vec{k}_1 + \vec{k}_2 + \vec{k}_3 + \vec{k}_4) \, f_{4}^{(1)}(\vec{k}_1, \vec{k}_2, \vec{k}_3, \vec{k}_4\,;\,u) \\
\label{posspaceK43}
& f_{4}^{(3)}(|\vec{x}_1 - \vec{x}_2|, |\vec{x}_1 - \vec{x}_3|, |\vec{x}_1 - \vec{x}_4|, |\vec{x}_2 - \vec{x}_3|, |\vec{x}_2 - \vec{x}_4|, |\vec{x}_3 - \vec{x}_4|\,;\,u) & \nonumber \\
& \qquad := \frac{1}{(2 \pi)^{2d}} \int d^d \vec{k}_1 \, d^d \vec{k}_2 \, d^d \vec{k}_3 \, d^d \vec{k}_4  \, e^{- i \, (\vec{k}_1 \cdot \vec{x}_1 + \vec{k}_2 \cdot \vec{x}_2 + \vec{k}_3 \cdot \vec{x}_3 + \vec{k}_4 \cdot \vec{x}_4)} \, \delta^{(d)}(\vec{k}_1 + \vec{k}_2 + \vec{k}_3 + \vec{k}_4) \, f_{4}^{(3)}(\vec{k}_1, \vec{k}_2, \vec{k}_3, \vec{k}_4\,;\,u)\,.
\end{align} 
We demonstrate that if our theory has a soft UV cutoff in momentum space (we will specify precisely what this means shortly), for any spatial dimension $d$ (for which $\varphi^4$ theory is renormalizable, namely $d \leq 3$) and for fixed $u$, these position space kernels are at least exponentially decaying with rate $e^{-u}m$ (i.e., the renormalized mass) with respect to the pairwise differences $|\vec{x}_i-\vec{x}_j|$.  For instance, $f_{2,0}(|\vec{x}_1 - \vec{x}_2| \, ; \, u)$ for fixed $u$ decays at least as fast as $e^{- e^{-u}m |\vec{x}_1 - \vec{x}_2|}$.  

First, we consider the quadratic kernels $f_{2,0}(|\vec{x}_1 - \vec{x}_2|\,;\,u)$ and $f_{2,1}(|\vec{x}_1 - \vec{x}_2|\,;\,u)$ for fixed $u$.  Simplifying Eqn.'s~\eqref{posspaceK20} and~\eqref{posspaceK21} as
\begin{align}
\label{posspaceK20v2}
&f_{2,0}(|\vec{x}_1 - \vec{x}_2|\,;\,u) := \frac{1}{(2 \pi)^{d}} \int d^d \vec{k} \, e^{- i \, (\vec{k} \cdot (\vec{x}_1 - \vec{x}_2))} \, f_{2,0}(\vec{k}\,;\,u) \\
\label{posspaceK21v2}
&f_{2,1}(|\vec{x}_1 - \vec{x}_2|\,;\,u) := \frac{1}{(2 \pi)^{d}} \int d^d \vec{k} \, e^{- i \, (\vec{k} \cdot (\vec{x}_1 - \vec{x}_2))} \, f_{2,1}(\vec{k}\,;\,u)\,,
\end{align}
we see that we need to analyze the $d$--dimensional Fourier transforms of $f_{2,0}(\vec{k}\,;\,u)$ and $f_{2,1}(\vec{k}\,;\,u)$.  These functions have the feature that they only depend on $|\vec{k}|$ and $u$; that is, $f_{2,0}(\vec{k}\,;\,u) = f_{2,0}(|\vec{k}|\,;\,u)$ and $f_{2,1}(\vec{k}\,;\,u) = f_{2,1}(|\vec{k}|\,;\,u)$.  For instance, see Eqn.~\eqref{eq:f2i}.  Given the functions $f_{2,0}(|\vec{k}|\,;\,u)$ and $f_{2,1}(|\vec{k}|\,;\,u)$, we can consider their analytic continuations $f_{2,0}(z\,;\,u)$ and $f_{2,1}(z\,;\,u)$ where $z$ is a complex argument in place of $|\vec{k}|$.  We can choose branch cuts of the logarithms in $f_{2,0}(z\,;\,u)$ and $f_{2,1}(z\,;\,u)$ so that the functions are analytic in the complex $z$--plane on the strip
\begin{equation}
\{ \, z \in \mathbb{C}  \,\, | \, - e^{-u}m < \text{Im}(z) < e^{-u}m \}\,.
\end{equation} 
In particular, the branch cuts of $f_{2,0}(z\,;\,u)$ and $f_{2,1}(z\,;\,u)$ in the complex $z$--plane are shown in Figure 2 below.

\begin{figure}[h!]
	\begin{center}
		\includegraphics[scale=.6]{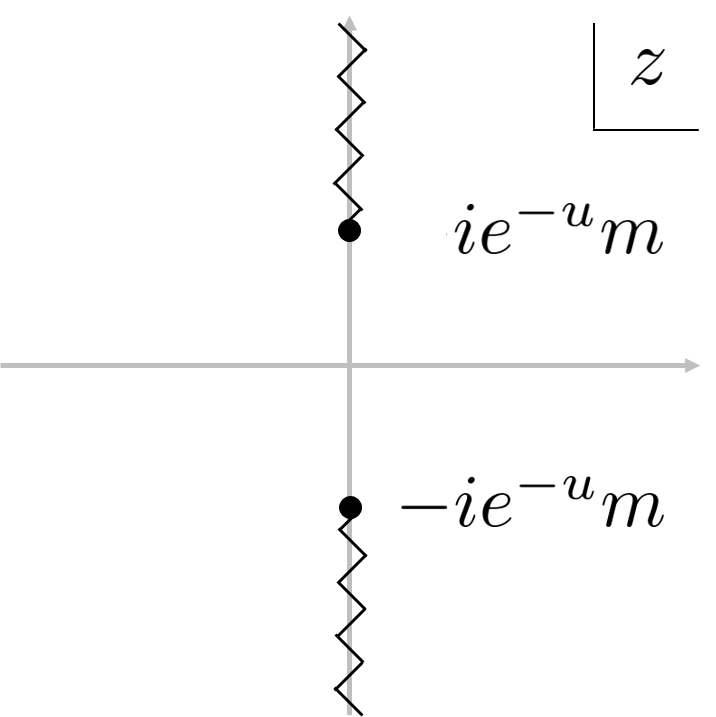}
		\caption{Branch cuts in the complex $z$--plane.}
	\end{center}
\end{figure}
\newpage
To abstract our analysis, suppose we have a complex function $F_2(z)$ with two properties:
\begin{enumerate}
	\item For $\text{Im}(z) = 0$, $F_2(z) = F_2(-z)$.
	\item $z^k \, F_2(z)$ is analytic in the strip $\{ \, z \in \mathbb{C}  \,\, | \, - e^{-u}m < \text{Im}(z) < e^{-u}m \}$ for all $k \in \mathbb{Z}_{\geq 0}$.
\end{enumerate} 
Indeed, both $f_{2,0}(z\,;\,u)$ and $f_{2,1}(z\,;\,u)$ have the above two properties.  We would like to understand the decay properties of
\begin{equation}
\int d^d \vec{k} \, e^{- i \, \vec{k} \cdot \vec{x}} F_2(|\vec{k}|)
\end{equation}
for functions $F_2(z)$ with the properties specified above.  Due to the first property of $F_2(z)$, the Fourier transform $\int d^d \vec{k} \, e^{- i \, \vec{k} \cdot \vec{x}} F_2(|\vec{k}|)$ as a function of $|\vec{x}|$ is strictly real--valued.

Next, suppose $d$ is odd (we will treat the even $d$ case shortly).  Then we can use the formula \cite{Grafakos1, Estrada1, Grafakos2}
\begin{equation}
\label{RadialReduction1}
\int d^d \vec{k} \, e^{- i \, \vec{k} \cdot \vec{x}} F_2(|\vec{k}|) = \left(\sum_{\ell = 1}^{(d-1)/2} \frac{(-1)^\ell (d - \ell - 2)!}{2^{\frac{d-1}{2} - \ell} (\frac{d-1}{2} - \ell)! (\ell - 1)!} \frac{1}{|\vec{x}|^{d - \ell - 2}} \left(\frac{d}{d |\vec{x}|}\right)^\ell \right) \int d |\vec{k}| \, e^{- i \, |\vec{k}|\cdot|\vec{x}|} F_2(|\vec{k}|)\,.
\end{equation}
In words, we can express $\int d^d \vec{k} \, e^{- i \, \vec{k} \cdot \vec{x}} F_2(|\vec{k}|)$ as a differential operator acting on the \textit{radial} Fourier transform $\int d |\vec{k}| \, e^{- i \, |\vec{k}|\cdot|\vec{x}|} F_2(|\vec{k}|)$.  Since $z^k \, F_2(z)$ (for any $k \in \mathbb{Z}_{\geq 0}$) is analytic on the strip $\{ \, z \in \mathbb{C}  \,\, | \, - e^{-u}m < \text{Im}(z) < e^{-u}m \}$ by the second property above, it follows that
\begin{equation}
\label{FourierBound1}
\frac{d^k}{d|\vec{x}|^k}\int_{\text{Im}(z) = 0} d z \, e^{- i \, z\cdot|\vec{x}|} F_2(z) = \int_{\text{Im}(z) = 0} d z \, e^{- i \, z\cdot|\vec{x}|} \, (-i z)^k\, F_2(z) \leq C_k \, e^{- (e^{-u} m - \delta) |\vec{x}|}
\end{equation}
for all $k \in \mathbb{Z}_{\geq 0}$\,, for some constants $C_k \sim \text{poly}(e^{-u} m)$ depending on $k$, and an arbitrarily small $\delta > 0$.  We achieve this bound since we can deform the integration contour $\text{Im}(z) = 0$ by pushing the contour down to $\text{Im}(z) = - e^{-u} m + \delta$ since $|\vec{x}| \geq 0$.  By utilizing the bound in Eqn.~\eqref{FourierBound1} in conjunction with the formula in Eqn.~\eqref{RadialReduction1}, we find that
\begin{equation}
\int d^d \vec{k} \, e^{- i \, \vec{k} \cdot (\vec{x}_1 - \vec{x}_2)} F_2(|\vec{k}|) \leq C \, e^{- (e^{-u} m - \delta) |\vec{x}_1 - \vec{x}_2|}
\end{equation}
for some (different) constant $C \sim \text{poly}(e^{-u} m)$ and an arbitrarily small $\delta > 0$.  Finally, it follows that
\begin{align}
\label{posspaceK20bound}
&f_{2,0}(|\vec{x}_1 - \vec{x}_2|\,;\,u) \leq C_1 \, e^{- (e^{-u} m - \delta) |\vec{x}_1 - \vec{x}_2|} \\
\label{posspaceK21bound}
&f_{2,1}(|\vec{x}_1 - \vec{x}_2|\,;\,u) \leq C_2 \, e^{- (e^{-u} m - \delta) |\vec{x}_1 - \vec{x}_2|}
\end{align}
for constants $C_1, C_2 \sim \text{poly}(e^{-u} m)$ and an arbitrarily small $\delta > 0$.

If $d$ is even, Eqn.~\eqref{RadialReduction1} does not immediately help us: it only relates the radial Fourier transform $\int d |\vec{k}| \, e^{- i \, |\vec{k}|\cdot|\vec{x}|} F_2(|\vec{k}|)$ to the radially symmetric Fourier transform $\int d^{d'} \vec{k} \, e^{-i \vec{k} \cdot \vec{x}}F_2(|\vec{k}|)$ in $d'$ dimensions, if $d'$ is odd.  Since we are now considering $d$ even, we let $d' = d-1$ which is odd.  Then using Eqn.~\eqref{RadialReduction1} and the arguments above, we find that
\begin{equation}
\widetilde{F}_{2,d-1}(|\vec{x}_1-\vec{x}_2|):= \int d^{d-1} \vec{k} \, e^{-i \vec{k} \cdot \vec{x}}F_2(|\vec{k}|) \leq C' \, e^{- (e^{-u} m - \delta) |\vec{x}_1 - \vec{x}_2|}
\end{equation}
for some $C'$.  We have a similar bound on the derivatives of $\widetilde{F}_{2,d-1}(|\vec{x}_1-\vec{x}_2|)$ with respect to $|\vec{x}_1-\vec{x}_2|$.  However, our goal is to bound
\begin{equation}
\widetilde{F}_{2,d}(|\vec{x}_1-\vec{x}_2|):= \int d^{d} \vec{k} \, e^{-i \vec{k} \cdot \vec{x}}F_2(|\vec{k}|)\,.
\end{equation}
Now we need to relate $\widetilde{F}_{2,d-1}(|\vec{x}_1-\vec{x}_2|)$ and $\widetilde{F}_{2,d}(|\vec{x}_1-\vec{x}_2|)$.  Using \cite{Estrada1, Grafakos2}, we find
\begin{align}
\widetilde{F}_{2,d}(|\vec{x}_1-\vec{x}_2|) &= - \frac{2}{\sqrt{2\pi}}\frac{1}{|\vec{x}_1 - \vec{x}_2|} \frac{d}{d|\vec{x}_1-\vec{x}_2|} \int_0^\infty ds\, \widetilde{F}_{2,d}(\sqrt{|\vec{x}_1-\vec{x}_2|^2 +s^2}) \\ \nonumber \\
&=  - \frac{2}{\sqrt{2\pi}} \int_0^\infty ds\, \frac{\widetilde{F}_{2,d}'(\sqrt{|\vec{x}_1-\vec{x}_2|^2 +s^2})}{\sqrt{|\vec{x}_1-\vec{x}_2|^2 +s^2})} \\ \nonumber \\
&\leq C'' \int_0^\infty ds \, e^{-(e^{-u} m - \delta)\sqrt{|\vec{x}_1 - \vec{x}_2|^2 + s^2}} \\ \nonumber \\
&\leq C'' e^{(e^{-u} m - \delta)|\vec{x}_1 - \vec{x}_2|}
\end{align}
where $C''$ is some constant.  It follows that Eqn.'s~\eqref{posspaceK20bound} and~\eqref{posspaceK21bound} also hold for $d$ even as well.

Now we will consider both even and odd $d$ at the same time, and derive similar bounds for the position space kernels $f_{4}^{(1)}$ and $f_{4}^{(3)}$, for fixed $u$.  We notice that the momentum space kernels $f_{4}^{(1)}(\vec{k}_1,\vec{k}_2,\vec{k}_3, \vec{k}_4 \, ; \, u)$ and $f_{4}^{(3)}(\vec{k}_1,\vec{k}_2,\vec{k}_3, \vec{k}_4 \, ; \, u)$ are only even functions of $|\vec{k}_1|, |\vec{k}_2|, |\vec{k}_3|, |\vec{k}_4|$.  See, for instance, Eqn.~\eqref{f41sol2}.  If we replace $|\vec{k}_1|, |\vec{k}_2|, |\vec{k}_3|, |\vec{k}_4|$ by the complex variables $z_1, z_2, z_3, z_4$, then $f_{4}^{(1)}(z_1,z_2,z_3,z_4\, ; \, u)$ and $f_{4}^{(3)}(z_1,z_2,z_3,z_4\, ; \, u)$ are even with respect to each of their complex arguments when restricted to $\text{Im}(z_1) = \text{Im}(z_2) = \text{Im}(z_3) = \text{Im}(z_4) = 0$.  Furthermore, the functions are analytic on the strip
\begin{equation}
\{ (z_1, z_2, z_3, z_4) \in \mathbb{C}^{4} \,\, | \, - e^{-u} m < \text{Im}(z_1),\text{Im}(z_2),\text{Im}(z_3),\text{Im}(z_4) < e^{-u} m \}\,.
\end{equation}
Utilizing the formula in Eqn.~\eqref{RadialReduction1} for each $|\vec{k}_1|, |\vec{k}_2|, |\vec{k}_3|, |\vec{k}_4|$ individually, we can derive the analogous bound
\begin{align}
\label{PrelimBound41}
&\frac{1}{(2 \pi)^{2d}} \int d^d \vec{k}_1 \, d^d \vec{k}_2 \, d^d \vec{k}_3 \, d^d \vec{k}_4  \, e^{- i \, (\vec{k}_1 \cdot \vec{x}_1 + \vec{k}_2 \cdot \vec{x}_2 + \vec{k}_3 \cdot \vec{x}_3 + \vec{k}_4 \cdot \vec{x}_4)} \, f_{4}^{(1)}(\vec{k}_1, \vec{k}_2, \vec{k}_3, \vec{k}_4\,;\,u) \leq C \, e^{-e^{-u} m (|\vec{x}_1| + |\vec{x}_2| + |\vec{x}_3| + |\vec{x}_4|)} \\
\label{PrelimBound43}
&\frac{1}{(2 \pi)^{2d}} \int d^d \vec{k}_1 \, d^d \vec{k}_2 \, d^d \vec{k}_3 \, d^d \vec{k}_4  \, e^{- i \, (\vec{k}_1 \cdot \vec{x}_1 + \vec{k}_2 \cdot \vec{x}_2 + \vec{k}_3 \cdot \vec{x}_3 + \vec{k}_4 \cdot \vec{x}_4)} \, f_{4}^{(3)}(\vec{k}_1, \vec{k}_2, \vec{k}_3, \vec{k}_4\,;\,u) \leq C' \, e^{-e^{-u} m (|\vec{x}_1| + |\vec{x}_2| + |\vec{x}_3| + |\vec{x}_4|)}
\end{align}
for some constants $C,C' \sim \text{poly}(e^{-u} m)$ and an arbitrarily small $\delta > 0$.  However, these are \textit{not exactly} the bounds we want.  Specifically, the left-hand sides of Eqn.'s~\eqref{PrelimBound41} and~\eqref{PrelimBound43} differ from the right-hand sides of Eqn.'s~\eqref{posspaceK41} and~\eqref{posspaceK43} by a delta function $\delta^{(d)}(\vec{k}_1 + \vec{k}_2 + \vec{k}_3 + \vec{k}_4)$ in the integrand.  However, this delta function is easy to account for.  Defining
\begin{align}
&F_{4}^{(1)}(\vec{x}_1, \vec{x}_2, \vec{x}_3, \vec{x}_4 \, ; \, u) := \frac{1}{(2 \pi)^{2d}} \int d^d \vec{k}_1 \, d^d \vec{k}_2 \, d^d \vec{k}_3 \, d^d \vec{k}_4  \, e^{- i \, (\vec{k}_1 \cdot \vec{x}_1 + \vec{k}_2 \cdot \vec{x}_2 + \vec{k}_3 \cdot \vec{x}_3 + \vec{k}_4 \cdot \vec{x}_4)} \, f_{4}^{(1)}(\vec{k}_1, \vec{k}_2, \vec{k}_3, \vec{k}_4\,;\,u) \\
&F_{4}^{(3)}(\vec{x}_1, \vec{x}_2, \vec{x}_3, \vec{x}_4 \, ; \, u) := \frac{1}{(2 \pi)^{2d}} \int d^d \vec{k}_1 \, d^d \vec{k}_2 \, d^d \vec{k}_3 \, d^d \vec{k}_4  \, e^{- i \, (\vec{k}_1 \cdot \vec{x}_1 + \vec{k}_2 \cdot \vec{x}_2 + \vec{k}_3 \cdot \vec{x}_3 + \vec{k}_4 \cdot \vec{x}_4)} \, f_{4}^{(3)}(\vec{k}_1, \vec{k}_2, \vec{k}_3, \vec{k}_4\,;\,u)
\end{align}
it is easy to check that
\begin{align}
&\int d^d \vec{v} \,\, F_{4}^{(1)}(\vec{x}_1 + \vec{v}, \vec{x}_2 + \vec{v}, \vec{x}_3 + \vec{v}, \vec{x}_4 + \vec{v} \, ; \, u) \nonumber \\
& \qquad = \frac{1}{(2 \pi)^{2d}} \int d^d \vec{k}_1 \, d^d \vec{k}_2 \, d^d \vec{k}_3 \, d^d \vec{k}_4  \, e^{- i \, (\vec{k}_1 \cdot \vec{x}_1 + \vec{k}_2 \cdot \vec{x}_2 + \vec{k}_3 \cdot \vec{x}_3 + \vec{k}_4 \cdot \vec{x}_4)} \, \delta^{(d)}(\vec{k}_1 + \vec{k}_2 + \vec{k}_3 + \vec{k}_4) \, f_{4}^{(1)}(\vec{k}_1, \vec{k}_2, \vec{k}_3, \vec{k}_4\,;\,u) \\
&\int d^d \vec{v} \,\, F_{4}^{(3)}(\vec{x}_1 + \vec{v}, \vec{x}_2 + \vec{v}, \vec{x}_3 + \vec{v}, \vec{x}_4 + \vec{v} \, ; \, u) \nonumber \\
& \qquad = \frac{1}{(2 \pi)^{2d}} \int d^d \vec{k}_1 \, d^d \vec{k}_2 \, d^d \vec{k}_3 \, d^d \vec{k}_4  \, e^{- i \, (\vec{k}_1 \cdot \vec{x}_1 + \vec{k}_2 \cdot \vec{x}_2 + \vec{k}_3 \cdot \vec{x}_3 + \vec{k}_4 \cdot \vec{x}_4)} \, \delta^{(d)}(\vec{k}_1 + \vec{k}_2 + \vec{k}_3 + \vec{k}_4) \, f_{4}^{(3)}(\vec{k}_1, \vec{k}_2, \vec{k}_3, \vec{k}_4\,;\,u)
\end{align}
and hence
\begin{align}
\label{fFrelation1}
& f_{4}^{(1)}(|\vec{x}_1 - \vec{x}_2|, |\vec{x}_1 - \vec{x}_3|, |\vec{x}_1 - \vec{x}_4|, |\vec{x}_2 - \vec{x}_3|, |\vec{x}_2 - \vec{x}_4|, |\vec{x}_3 - \vec{x}_4|\,;\,u) \nonumber \\
& \qquad \qquad \qquad \qquad \qquad \qquad \qquad \qquad \qquad \qquad = \int d^d \vec{v} \,\, F_{4}^{(1)}(\vec{x}_1 + \vec{v}, \vec{x}_2 + \vec{v}, \vec{x}_3 + \vec{v}, \vec{x}_4 + \vec{v} \, ; \, u) \\
\label{fFrelation2}
& f_{4}^{(3)}(|\vec{x}_1 - \vec{x}_2|, |\vec{x}_1 - \vec{x}_3|, |\vec{x}_1 - \vec{x}_4|, |\vec{x}_2 - \vec{x}_3|, |\vec{x}_2 - \vec{x}_4|, |\vec{x}_3 - \vec{x}_4|\,;\,u) \nonumber \\
& \qquad \qquad \qquad \qquad \qquad \qquad \qquad \qquad \qquad \qquad = \int d^d \vec{v} \,\, F_{4}^{(3)}(\vec{x}_1 + \vec{v}, \vec{x}_2 + \vec{v}, \vec{x}_3 + \vec{v}, \vec{x}_4 + \vec{v} \, ; \, u)\,.
\end{align}
Since Eqn.'s~\eqref{PrelimBound41} and~\eqref{PrelimBound43} give us the bounds 
\begin{align}
F_{4}^{(1)}(\vec{x}_1, \vec{x}_2, \vec{x}_3, \vec{x}_4 \, ; \, u) \leq C \, e^{-e^{-u} m (|\vec{x}_1| + |\vec{x}_2| + |\vec{x}_3| + |\vec{x}_4|)}  \\
F_{4}^{(3)}(\vec{x}_1, \vec{x}_2, \vec{x}_3, \vec{x}_4 \, ; \, u) \leq C' \, e^{-e^{-u} m (|\vec{x}_1| + |\vec{x}_2| + |\vec{x}_3| + |\vec{x}_4|)} 
\end{align}
for some constants $C,C' \sim \text{poly}(e^{-u} m)$ and an arbitrarily small $\delta > 0$, Eqn.'s~\eqref{fFrelation1} and~\eqref{fFrelation2} imply that
\begin{align}
& f_{4}^{(1)}(|\vec{x}_1 - \vec{x}_2|, |\vec{x}_1 - \vec{x}_3|, |\vec{x}_1 - \vec{x}_4|, |\vec{x}_2 - \vec{x}_3|, |\vec{x}_2 - \vec{x}_4|, |\vec{x}_3 - \vec{x}_4|\,;\,u) \nonumber \\
& \qquad \qquad \qquad \qquad \qquad \qquad \qquad \leq C_3 \, e^{-e^{-u} m (|\vec{x}_1 - \vec{x}_2|+|\vec{x}_1 - \vec{x}_3|+|\vec{x}_1 - \vec{x}_4|+|\vec{x}_2 - \vec{x}_3|+|\vec{x}_2 - \vec{x}_4|+|\vec{x}_3 - \vec{x}_4|)} \\ \nonumber \\
& f_{4}^{(3)}(|\vec{x}_1 - \vec{x}_2|, |\vec{x}_1 - \vec{x}_3|, |\vec{x}_1 - \vec{x}_4|, |\vec{x}_2 - \vec{x}_3|, |\vec{x}_2 - \vec{x}_4|, |\vec{x}_3 - \vec{x}_4|\,;\,u) \nonumber \\
& \qquad \qquad \qquad \qquad \qquad \qquad \qquad \leq C_4 \, e^{-e^{-u} m (|\vec{x}_1 - \vec{x}_2|+|\vec{x}_1 - \vec{x}_3|+|\vec{x}_1 - \vec{x}_4|+|\vec{x}_2 - \vec{x}_3|+|\vec{x}_2 - \vec{x}_4|+|\vec{x}_3 - \vec{x}_4|)}
\end{align}
for some constants $C_3, C_4 \sim \text{poly}(e^{-u} m)$ and an arbitrarily small $\delta > 0$.  These are the desired inequalities.

In summary, we have shown that
\begin{align}
\label{posspaceK20boundv2}
&f_{2,0}(|\vec{x}_1 - \vec{x}_2|\,;\,u) \leq C_1 \, e^{- (e^{-u} m - \delta) |\vec{x}_1 - \vec{x}_2|} \\ \nonumber \\
\label{posspaceK21boundv2}
&f_{2,1}(|\vec{x}_1 - \vec{x}_2|\,;\,u) \leq C_2 \, e^{- (e^{-u} m - \delta) |\vec{x}_1 - \vec{x}_2|} \\ \nonumber \\
\label{posspaceK41boundv2}
& f_{4}^{(1)}(|\vec{x}_1 - \vec{x}_2|, |\vec{x}_1 - \vec{x}_3|, |\vec{x}_1 - \vec{x}_4|, |\vec{x}_2 - \vec{x}_3|, |\vec{x}_2 - \vec{x}_4|, |\vec{x}_3 - \vec{x}_4|\,;\,u) \nonumber \\
& \qquad \qquad \qquad \qquad \qquad \qquad \qquad \leq C_3 \, e^{-e^{-u} m (|\vec{x}_1 - \vec{x}_2|+|\vec{x}_1 - \vec{x}_3|+|\vec{x}_1 - \vec{x}_4|+|\vec{x}_2 - \vec{x}_3|+|\vec{x}_2 - \vec{x}_4|+|\vec{x}_3 - \vec{x}_4|)} \\ \nonumber \\
\label{posspaceK43boundv2}
& f_{4}^{(3)}(|\vec{x}_1 - \vec{x}_2|, |\vec{x}_1 - \vec{x}_3|, |\vec{x}_1 - \vec{x}_4|, |\vec{x}_2 - \vec{x}_3|, |\vec{x}_2 - \vec{x}_4|, |\vec{x}_3 - \vec{x}_4|\,;\,u) \nonumber \\
& \qquad \qquad \qquad \qquad \qquad \qquad \qquad \leq C_4 \, e^{-e^{-u} m (|\vec{x}_1 - \vec{x}_2|+|\vec{x}_1 - \vec{x}_3|+|\vec{x}_1 - \vec{x}_4|+|\vec{x}_2 - \vec{x}_3|+|\vec{x}_2 - \vec{x}_4|+|\vec{x}_3 - \vec{x}_4|)}
\end{align}
for some constants $C_1, C_2, C_3, C_4 \sim \text{poly}(e^{-u} m)$ and an arbitrarily small $\delta > 0$.  Therefore, the kernels are exponentially local in position space with decay rate $\sim e^{-u} m$ with respect to the pairwise differences $|\vec{x}_i - \vec{x}_j|$.


\end{document}